\documentclass{sigchi}

\usepackage{balance}       % to better equalize the last page
\usepackage{graphics}      % for EPS, load graphicx instead 
\usepackage[T1]{fontenc}   % for umlauts and other diaeresis
\usepackage{txfonts}
\usepackage{mathptmx}
\usepackage[pdflang={en-US},pdftex]{hyperref}
\usepackage{color}
\usepackage{booktabs}
\usepackage{textcomp}
\usepackage{xspace}
\usepackage{comment}

\usepackage{enumitem}

\usepackage{microtype}        % Improved Tracking and Kerning
\usepackage{ccicons}          % Cite your images correctly!

\def\plaintitle{How to organize an in-person, online or hybrid hackathon -- A revised planning kit}

\def\emptyauthor{}
\def\plainkeywords{Hackathon; Planning kit; Organizers; Collaboration; Social computing}

\makeatletter
\def\url@leostyle{%
  \@ifundefined{selectfont}{
    \def\UrlFont{\sf}
  }{
    \def\UrlFont{\small\bf\ttfamily}
  }}
\makeatother
\def\pprw{8.5in}
\def\pprh{11in}

\definecolor{linkColor}{RGB}{6,125,233}
\hypersetup{%
  pdftitle={\plaintitle},
  pdfauthor={\emptyauthor},
  pdfkeywords={\plainkeywords},
  pdfdisplaydoctitle=true, % For Accessibility
  bookmarksnumbered,
  pdfstartview={FitH},
  colorlinks,
  citecolor=black,
  filecolor=black,
  linkcolor=black,
  urlcolor=linkColor,
  breaklinks=true,
  hypertexnames=false
}

\makeatletter
\def\@copyrightspace{\relax}
\makeatother

\newcommand{\GH}{{\sc GitHub}\xspace}

\usepackage{titlesec}

\titlespacing{\subsubsection}
    {0pt}{3pt}{0pt}
\newcommand{\forceindent}{\leavevmode{\parindent=1em\indent}}

\begin{document}
\pagenumbering{arabic}

\title{\plaintitle}

\numberofauthors{9}
\author{%
  \alignauthor{Abasi-Amefon Obot Affia-Jomants\\
    \affaddr{University of Tartu, Estonia}\\
    \email{amefon.affia@ut.ee}}\\
  \alignauthor{Kiev Gama\\
    \affaddr{Federal University of Pernambuco - UFPE, Recife, PE, Brazil}\\
    \email{kiev@cin.ufpe.br}}\\
  \alignauthor{James D. Herbsleb\\
    \affaddr{Carnegie Mellon University, Pittsburgh, PA, USA}\\
    \email{jdh@cs.cmu.edu}}\\
  \alignauthor{Alexander Nolte\\
    \affaddr{Eindhoven University of Technology, The Netherlands}\\
    \affaddr{Carnegie Mellon University, Pittsburgh, PA, USA}\\
    \email{a.u.nolte@tue.nl}}\\
}

\maketitle

\begin{abstract}
Hackathons and similar time-bounded events are a global phenomenon. Their proliferation in various domains and their usefulness for a variety of goals has led to the emergence of different formats. While there are a multitude of guidelines available on how to prepare and run a hackathon, most of them focus on a particular format that was created for a specific purpose within a domain for a certain type of participant. This makes it difficult, in particular, for novice organizers to decide how to run an event that fits their needs. To address this gap we developed the original version of this planning kit in 2020 which focused on in-person events that were the dominant form of hackathons then ~\cite{nolte2020organize}. That planning kit was organized around \textbf{12 key decisions} that organizers need to take when preparing for, running, and following up on a hackathon. Fast forward to 2025, after going through a global pandemic that forced all events to move online, we now see different forms of events -- in-person, online, and hybrid -- taking place across the globe, and while they can be all valuable, they have different affordances and require different considerations when planning. To account for these differences, we decided to update the original planning kit by adding a section that discusses the affordances and requirements of in-person, online, and hybrid events to each of the 12 decisions. In addition, we modified the original example timelines to include different forms and types of events. We also updated the planning kit in general based on insights we gained through continuing to organize and study hackathons. The main planning kit is available online while this report is meant to be a downloadable and citable resource.
\end{abstract}

% Author Keywords
\keywords{\plainkeywords}

\section{Introduction}
\label{sec:intro}
Hackathons are time-bounded events during which participants with different backgrounds and expertise form teams to collaborate on a project and create an artefact~\cite{falk2024future}. They are a global phenomenon~\cite{taylor2018everybody} with thousands of events taking place every year\footnote{The largest hackathon database Devpost (\url{https://devpost.com/}) alone lists more than 1000 annual events.} in various domains including entrepreneurship~\cite{irani2015hackathons}, corporations~\cite{pe2022corporate}, (scientific) communities~\cite{huppenkothen2018hack}, education~\cite{schulten2024we}, civic engagement~\cite{maillart2024computational} and others~\cite{falk202010}. These events are organized to foster various goals including the development of (innovative) technology~\cite{gama2023developers,falk2022supporting,briscoe2014digital}, supporting learning ~\cite{holmen2024facultyhack,chounta2023re,affia2020developing,porras2019code,gama2018hackathon}, tackling civic and environmental issues~\cite{maillart2024computational,zapico2013hacking,baccarne2014urban,berger2017karachi,porter2017reappropriating} and building new or expanding existing communities ~\cite{mahmoud2023exploratory,nolte2020support,trainer2014community,moller2014community}. The versatility of hackathons has subsequently led to a large variety of different event formats, making it difficult, in particular, for inexperienced organizers to decide how to run an event that fits their needs.

\vspace{0.15cm}
\forceindent We aimed to address this issue in 2020 when we released the first version of this planning kit~\cite{nolte2020organize}. It was organized around \textbf{12 key decision} that we found to be crucial to consider when organizing a hackathon. We intentionally did not go into the details of how to organize an event in general, such as how to secure a room, book catering, and ensure that participants have WiFi access. Instead, we focused on the specifics that organizers need to consider when planning a hackathon that aims to foster a specific goal for a specific audience in a specific domain.

\vspace{0.15cm}
\forceindent Fast forward to 2025, a lot has changed. We went through a global pandemic that forced all events to move online. Through this move it quickly became clear that online events can not only still be valuable but can also expand participation opportunities to individuals who might otherwise not have been able to join due to resources, travel constraints or other commitments~\cite{paganini2023opportunities,mendes2022socio,schulten2022participants,powell2021organizing}. Based on these insights, many organizers decided to run hybrid events after the pandemic had ended. These events combine in-person participation with the option to join online as well. Today, we see all different types of events -- in-person, online, and hybrid -- taking place across the globe, and while they can be all valuable, they have different affordances and require different considerations when planning. To account for these differences, we decided to create and share this new version of the original planning kit~\cite{nolte2020organize}. It is based on our insights from studying and organizing hackathons for the past ten years. We retain large parts of the original version, but we updated them to account for new insights we gained after the original version was published. We also added a section that discusses the affordances and requirements of in-person, online, and hybrid events to each of the 12 decisions. In addition, we modified the original example timelines to include a corporate in-person hackathon, an entrepreneurial in-person hackathon, an educational online hackathon, and a community-building hybrid hackathon. Moreover, we also added resources to study hackathons, including a post-hackathon survey template ~\cite{nolte2025survey} and instrumentation to study hybrid events ~\cite{affia2025hybrid}.

\vspace{0.15cm}
\forceindent The main planning kit is available online\footnote{\url{https://hackathon-planning-kit.org/}} while this report is meant as a downloadable and citable reference. We would thus like to encourage the reader to consult and use the online version since we will continue updating it in the future.

\vspace{0.15cm}
\forceindent The remainder of this report is structured as follows. We will first provide example timelines for four specific types of hackathons before elaborating on the 12 aforementioned decisions. The example timelines are mainly aimed at first-time organizers. They walk the reader through the different decision points and show examples of how and why they could decide on one or the other option. Afterward, we will present the 12 decisions in detail.

\vspace{0.15cm}
\forceindent We would also like to encourage everyone who uses this planning kit or who has experience organizing, supporting, or researching hackathons to provide feedback or suggestions on how to improve the planning kit. Please feel free to contact us\footnote{hackathon.planning.kit@gmail.com} or directly propose changes on \GH\footnote{\url{https://github.com/herbsleb-group/herbsleb-group.github.io}}.

\section{Example timelines}
\label{sec:timelines}
The following example timelines show an idealized procedure for the organization of four different types of hackathons, which focus on different goals (product innovation, entrepreneurship, education, and community building) and are organized as in-person, online, and hybrid events. These timelines indicate the timing and outcome of key decisions.

\subsection{Corporate in-person hackathon}
\label{sec:timeline:corporate}
This timeline shows an example of a medium-sized in-person hackathon (between 50 and 80 participants) that aimed to promote product innovation at a large IT company.

\vspace{0.3cm}
\textbf{4 months before the hackathon:}
\begin{itemize}[leftmargin=0.5cm]
    \item \textit{Goal:} Identification of innovative ideas that can turn into company products
    \item \textit{Theme:} Expanding the user base of the main product lines
    \item \textit{Competition / cooperation:} Decision for a competitive event where the top 3 teams can win gift cards and material prizes, and the first place team will receive support to continue their project with the goal to eventually roll it out as a product. 
    \item \textit{Duration / breaks:} Decision for a 48-hour event over 3 days starting on Wednesday afternoon.
    \item \textit{Agenda:} Decision about a tentative agenda that includes a checkpoint, a final presentation, a jury session, and an award ceremony.
    \item \textit{Stakeholder involvement:} Discussion with leadership from different departments about their involvement in the event as mentors and judges.
    \item \textit{Participant recruitment:} Decision to run an internal event to which all employees of the company are invited.
    \item \textit{Mentoring:} Decision to recruit company stakeholders as mentors. Focus on product and leadership experience.
    \item \textit{Ideation:} Decision to run a pitch session at the beginning of the event.
\end{itemize}

\vspace{0.3cm}
\textbf{3 months before the hackathon:}
\begin{itemize}[leftmargin=0.5cm]
    \item \textit{Competition / cooperation:} Invitation of judges and sharing of judging criteria.
    \item \textit{Participant recruitment:} Invitation email to all employees of the company including a registration form.
    \item \textit{Mentoring:} Invitation email to selected individuals including information about mentor role and responsibilities.
\end{itemize}

\vspace{0.3cm}
\textbf{1 month before the hackathon:}
\begin{itemize}[leftmargin=0.5cm]
    \item \textit{Agenda:} Finalizing event agenda and sharing it across the company.
    \item \textit{Competition / cooperation:} Decision that teams need to show a working demo as their final presentation.
    \item \textit{Team formation:} Decision that teams will form around ideas, need to have between 3 and 6 members, and can be formed before or at the hackathon.
\end{itemize}

\vspace{0.3cm}
\textbf{1 week before the hackathon:}
\begin{itemize}[leftmargin=0.5cm]
    \item \textit{Competition / cooperation:} Sharing of judging criteria with registered individuals.
    \item \textit{Ideation:} Sharing of pitch procedure with registered individuals.
    \item \textit{Team formation:} Sharing of team formation criteria and procedure with registered individuals.
\end{itemize}

\vspace{0.3cm}
\textbf{Hackathon day 1:}
\begin{itemize}[leftmargin=0.5cm]
    \item \textit{Agenda:} Welcoming words by the organizers, presentation of the code of conduct, hackathon agenda, ideation and team formation procedure, mandatory checkpoint, expected final presentation, judging criteria, and prizes.
    \item \textit{Mentoring:} Introduction of mentors, their expertise, and their role during the hackathon.
    \item \textit{Ideation:} Individuals pitch their project idea and state if they seek additional team members.
    \item \textit{Team formation:} Participants that did not pitch ideas talk to idea proposers, discuss their expertise, and voice their interest. Idea proposers select suitable team members. Ideas that do not gain sufficient interest from other participants are abandoned, and the proposers of these ideas have the option to join other teams.
    \item \textit{Agenda:} Idea proposers present their teams. Quick check by the organizers if the teams meet the criteria.
    \item \textit{Competition / cooperation:} If teams do not meet the criteria, organizers can decide to split them up or suggest participants to join a different team.
    \item \textit{Duration / breaks:} Dinner at the end of the day.
\end{itemize}

\vspace{0.3cm}
\textbf{Hackathon day 2:}
\begin{itemize}[leftmargin=0.5cm]
    \item \textit{Agenda:} Teams meet, and idea proposers present their progress in front of organizers and mentors and receive feedback.
    \item \textit{Duration / breaks:} Lunch and dinner breaks.
\end{itemize}

\vspace{0.3cm}
\textbf{Hackathon day 3:}
\begin{itemize}[leftmargin=0.5cm]
    \item \textit{Duration / breaks:} Lunch break.
    \item \textit{Agenda:} Final demo presentation of idea proposers in front of all participants, jury, organizers, mentors, and interested company employees.
    \item \textit{Competition / cooperation:} Jury decision and award ceremony.
    \item \textit{Duration / breaks:} Dinner at the end of the hackathon.
\end{itemize}

\vspace{0.3cm}
\textbf{After the hackathon:}
\begin{itemize}[leftmargin=0.5cm]
    \item \textit{Competition / cooperation:} Distribution of awards to the winners.
    \item \textit{Continuity planning:} Post-hackathon feedback session for all teams. Discussion with leadership of winning team about resources and timeline to complete the project.
\end{itemize}

\subsection{Entrepreneurial in-person hackathon}
\label{sec:timeline:entrepreneur}
This timeline shows an example for a medium-sized in-person hackathon (between 100 and 150 participants), which aimed to attract entrepreneurs and foster innovative projects that can become successful businesses.

\vspace{0.3cm}
\textbf{4 months before the hackathon:}
\begin{itemize}[leftmargin=0.5cm]
    \item \textit{Goal:} Fostering the regional development of the start-up ecosystem related to the cyber security domain
    \item \textit{Theme:} Cyber security
    \item \textit{Competition / cooperation:} Decision for a competition style event. Teams can win prizes ranging from tech gadgets to start-up coaching and participation in accelerator programs.
    \item \textit{Duration / breaks:} Discussion about and decision for a tentative date for a 48-hour event starting in the afternoon on Friday.
    \item \textit{Agenda:} Discussion about and decision for a tentative agenda that includes daily checkpoints, a final pitch presentation, and an award ceremony.
    \item \textit{Participant recruitment:} Creation of an information hub. Contacting local universities, start-up hubs, tech companies, accelerator programs, and government agencies to spread news about the event through their networks. Start of the social-media campaign.
    \item \textit{Stakeholder involvement:} Discussion with representatives of aforementioned groups about their interest in the event and invitation to participate as mentors, give thematic talks, and provide sponsorship and prizes.
\end{itemize}

\vspace{0.3cm}
\textbf{3 months before the hackathon:}
\begin{itemize}[leftmargin=0.5cm]
    \item \textit{Participant recruitment:} Creation of an online form that covers participants’ contact details, their current profession, and their projected role during the hackathon. Registration requires the payment of a small nominal fee that will be refunded after participation.
    \item \textit{Ideation:} Decision for a pitch-style ideation process at the hackathon. Participants can indicate if they have a project idea for the hackathon and provide a short description as part of the registration form.
    \item \textit{Mentoring:} Identification and invitation of a diverse group of individuals who can provide mentorship related to cyber security, various programming languages, design, entrepreneurship, marketing, and others. Decision for a combination of mentor teams and individual on-demand support.
\end{itemize}

\vspace{0.3cm}
\textbf{1 month before the hackathon:}
\begin{itemize}[leftmargin=0.5cm]
    \item \textit{Team formation:} Teams will form around ideas. They cannot have less than 3 members, have to be of similar size, and include individuals with diverse expertise and interests, including cyber security, programming, design, and entrepreneurship.
    \item \textit{Stakeholder involvement:} Finalization of sponsor agreements, including prizes and talks at the hackathon.
    \item \textit{Participant recruitment:} 1-day competitive ideation events in three cities close to the main hackathon location, during which participants can start working on ideas and form teams. Winners receive travel support for the main hackathon.
\end{itemize}

\vspace{0.3cm}
\textbf{1 week before the hackathon:}
\begin{itemize}[leftmargin=0.5cm]
    \item \textit{Agenda:} Adding final event agenda including thematic talks, trainings, and talks by sponsors during each day to the information hub.
    \item \textit{Ideation:} Adding information about the pitch procedure to the information hub.
    \item \textit{Mentoring:} Introduction of mentors on the information hub.
\end{itemize}

\vspace{0.3cm}
\textbf{Hackathon day 1:}
\begin{itemize}[leftmargin=0.5cm]
    \item \textit{Agenda:} Welcoming words by the organizers, presentation of the hackathon agenda including idea pitches, mandatory checkpoints for idea proposers, talks and trainings, expected outcome (pitch presentation), and jury. Reiteration of information hub and contact details for organizers and mentors.
    \item \textit{Stakeholder involvement:} Introduction of sponsors and supporting individuals and institutions.
    \item \textit{Mentoring:} Introduction of mentors, their area of expertise,and their role during the hackathon.
    \item \textit{Ideation:} Participants pitch ideas in front of organizers, mentors, and other participants, including information about which expertise they perceive to be required. Everyone can pitch. Not only participants who submitted ideas through the registration form.
    \item \textit{Team formation:} Ideas are written on large sheets of paper that idea proposers hang on the walls in the foyer of the hackathon venue. Participants who did not pitch ideas go around and talk to idea proposers, discuss their expertise, and voice their interest. Idea proposers select suitable team members based on interest and expertise. Ideas that do not gain sufficient interest from other participants are abandoned, and the proposers of these ideas have the option to join other teams.
    \item \textit{Agenda:} Idea proposers present their teams. Quick check by the organizers if the teams are of roughly equal size and if all teams have sufficient expertise to start working on their projects.
    \item \textit{Mentoring:} Mentors meet and form teams with diverse expertise. Each mentor team is assigned to a group of hackathon teams that they support during the hackathon. Mentors focus on their teams but also support others if necessary.
    \item \textit{Duration / breaks:} Common dinner.
\end{itemize}

\vspace{0.3cm}
\textbf{Hackathon day 2:}
\begin{itemize}[leftmargin=0.5cm]
    \item \textit{Duration / breaks:} Common breakfast.
    \item \textit{Agenda:} Idea proposers present their progress in front of organizers and mentors at the beginning of the day.
    \item \textit{Mentoring:} Mentors meet, discuss potential difficulties that certain teams face, and decide on mentors with related expertise to support them.
    \item \textit{Stakeholder involvement:} Thematic talk before lunch time.
    \item \textit{Duration / breaks:} Lunch break.
    \item \textit{Agenda:} First pitch training for idea proposers shortly the before next checkpoint.
    \item \textit{Agenda:} Idea proposers present their progress in front of organizers and mentors at the end of the day.
    \item \textit{Mentoring:} Mentors meet, discuss potential difficulties that certain teams face, and decide on mentors with related expertise to support them.
    \item \textit{Duration / breaks:} Common dinner.
\end{itemize}

\vspace{0.3cm}
\textbf{Hackathon day 3:}
\begin{itemize}[leftmargin=0.5cm]
    \item \textit{Duration / breaks:} Common breakfast.
    \item \textit{Agenda:} Idea proposers present their progress in front of organizers and mentors at the beginning of the day.
    \item \textit{Mentoring:} Mentors meet, discuss potential difficulties that certain teams face, and decide on mentors with related expertise to support them.
    \item \textit{Agenda:} Second pitch training for idea proposers before lunch.
    \item \textit{Duration / breaks:} Lunch break.
    \item \textit{Agenda:} Third and final pitch training for idea proposers a few hours before the final pitches.
    \item \textit{Agenda:} Final pitches of idea proposers in front of all participants, jury, organizers, mentors, and online audience (live stream).
    \item \textit{Competition / cooperation:} Online voting for audience favorite, jury decision, and award ceremony.
    \item \textit{Duration / breaks:} Group pictures, networking, end of the hackathon, and departure.
\end{itemize}

\vspace{0.3cm}
\textbf{After the hackathon:}
\begin{itemize}[leftmargin=0.5cm]
    \item \textit{Competition / cooperation:} Distribution of prizes to winners.
    \item \textit{Continuity planning:} Organizers share a summary of the hackathon on the information hub, connect interested teams with stakeholders, and periodically contact winning teams about their progress.
\end{itemize}

\subsection{Educational online hackathon}
\label{sec:timeline:education}
This timeline provides an example of a small-scale online hackathon (< 30 participants) that aimed to provide an opportunity for students to learn about using high-performance computing (HPC) resources. The event took place in conjunction with a major conference in the HPC field.

\vspace{0.3cm}
\textbf{6 months before the hackathon:}
\begin{itemize}[leftmargin=0.5cm]
    \item \textit{Goal:} Teaching students about how to use HPC resources
    \item \textit{Theme:} HPC to support a local community
    \item \textit{Competition / cooperation:} Decision for a competitive event where teams can win cash prizes.
    \item \textit{Duration / breaks:} Discussion about and decision for a 5-day event before the conference starting on Thursday afternoon.
    \item \textit{Agenda:} Decision about a tentative agenda that includes 2 daily checkpoints (morning and evening), a final presentation, a jury session, and an award ceremony.
    \item \textit{Stakeholder involvement:} Discussion with conference organizers, HPC and local community members about their involvement in the event as mentors, judges, and sponsors.
    \item \textit{Participant recruitment:} Discussion with educators as well as HPC and local community members to support participant recruitment through their networks.
    \item \textit{Mentoring:} Decision to recruit HPC and local community members as mentors. Focus on prior teaching experience, diverse interests, and backgrounds.
    \item \textit{Ideation:} Decision for a multi-stage process where mentors provide challenges and teams decide how to address them.
\end{itemize}

\vspace{0.3cm}
\textbf{4 months before the hackathon:}
\begin{itemize}[leftmargin=0.5cm]
    \item \textit{Competition / cooperation:} Decision for three award categories. Discussion with organizers to recognize winning teams during the award ceremony of the main conference. Discussion with sponsors about prizes.
    \item \textit{Specialized preparation:} Creation of a Github organization and page, which includes the preliminary schedule, organizer contact information, code of conduct, and links to resources. The page serves as the primary information hub for the hackathon. Creation of a Discord server, including public announcements, troubleshooting, and introduction channels, as well as private organizer and mentor channels.
    \item \textit{Mentoring:} Decisions for mentor pairs. Invitation email including a link to the Github page and a registration form that covers contact information, prior experience, interests, and availability during the event.
    \item \textit{Ideation:} Mentors propose challenges through the registration form.
    \item \textit{Participant recruitment:} Invitation email including a link to the Github page and a registration form that covers contact information, demographics, and prior experience related to programming and HPC.
\end{itemize}

\vspace{0.3cm}
\textbf{3 months before the hackathon:}
\begin{itemize}[leftmargin=0.5cm]
    \item \textit{Competition / cooperation:} Decision for judging criteria and voting procedure. Invitation email for judges including a link to the Github page, information about the judging procedure and pre-event jury training, and a registration form including contact information and area of expertise.
    \item \textit{Specialized preparation:} Decision for pre-hackathon trainings for participants covering the use of Github and Cloud resources.
    \item \textit{Agenda:} Adding a more detailed agenda to the Github page, including pre-event trainings for participants.
    \item \textit{Participant recruitment:} Invitation of registered participants to pre-hackathon trainings.
    \item \textit{Mentoring:} Invitation to pre-hackathon training. Mentors prepare slides to present their challenge.
    \item \textit{Continuity planning:} Invitation of registered participants, mentors, and judges to the Discord server and Github organization.
\end{itemize}

\vspace{0.3cm}
\textbf{1 month before the hackathon:}
\begin{itemize}[leftmargin=0.5cm]
    \item \textit{Competition / cooperation:} Finalization of sponsor agreements and creation of judging and popular vote forms.
    \item \textit{Specialized preparation:} Securing of Cloud Credits that participants can use for their projects during the hackathon.
    \item \textit{Agenda:} Adding calendar invite links and Zoom links for all training and hackathon sessions to Github page.
\end{itemize}

\vspace{0.3cm}
\textbf{2 weeks before the hackathon:}
\begin{itemize}[leftmargin=0.5cm]
    \item \textit{Specialized preparation:} First pre-hackathon training session for participants.
    \item \textit{Mentoring:} Pre-hackathon mentor training session. Organizers introduce the mentoring process and ensure that all mentors are on Discord and in the respective channels. Mentors present their challenges with organizers providing feedback. Forming of mentor pairs based on challenge similarity, availability, and expertise.
    \item \textit{Continuity planning:} Recording of the participant training session is shared on the Github page.
\end{itemize}

\vspace{0.3cm}
\textbf{1 week before the hackathon:}
\begin{itemize}[leftmargin=0.5cm]
    \item \textit{Specialized preparation:} Second pre-hackathon training session for participants.
    \item \textit{Stakeholder involvement:} Securing final confirmation by local community members for short talks during the event.
    \item \textit{Agenda:} Adding sponsor talks to the agenda.
    \item \textit{Continuity planning:} Recording of the participant training session is shared on the Github page.
\end{itemize}

\vspace{0.3cm}
\textbf{Hackathon day 1:}
\begin{itemize}[leftmargin=0.5cm]
    \item \textit{Agenda:} Welcoming words by the organizers, presentation of the code of conduct, hackathon agenda, expected final submission through Github, judging criteria, and prizes. Reminder to join the Discord server and Github organization. Hint to refer to the Github page and contact organizers or mentors if participants get stuck.
    \item \textit{Stakeholder involvement:} Introduction of sponsors and supporting individuals and institutions.
    \item \textit{Mentoring:} Introduction of mentors, their expertise, and their role during the hackathon.
    \item \textit{Ideation:} Mentors present their challenges.
    \item \textit{Agenda:} Organizers explain the team formation procedure and ask each team to create an introduction slide for the first checkpoint the next morning. The slides includes the team name, the names of the team members, the challenge they aim to address, and their initial approach for addressing it. Hint for participants to inform each other about potential parallel activities during the event (e.g., classes or social and work commitments).
    \item \textit{Continuity planning:} Organizers ask participants to create a Github repository for their team within the hackathon Github organization.
    \item \textit{Team formation:} Organizers create one breakout room for each presented challenge. Participants join different breakout rooms, and discuss with those present about the challenge and decide which challenge they want to contribute to.
    \item \textit{Competition / cooperation:} If teams are too large, organizers can decide to split them up or suggest participants to join a different team.
    \item \textit{Mentoring:} Mentors whose challenges were not chosen can decide to join a specific team or provide on-demand support.
\end{itemize}

\vspace{0.3cm}
\textbf{Hackathon day 2:}
\begin{itemize}[leftmargin=0.5cm]
    \item \textit{Agenda:} Organizers introduce the agenda for the day and reiterate the expected final submission. Teams present themselves, their project idea, and their plan until the next checkpoint.
    \item \textit{Specialized preparation:} Creation of one dedicated Discord channel per team. Invitation of team members and team mentors to their respective channel.
    \item \textit{Continuity planning:} Organizers add links to the teams' Github repositories on the main hackathon Github page.
    \item \textit{Stakeholder involvement:} First sponsor talk.
    \item \textit{Duration / breaks:} Suggested off-screen break at noon.
    \item \textit{Mentoring:} Mentors meet with organizers, discuss potential difficulties that teams might face, and decide on mentors with related expertise to support them.
    \item \textit{Stakeholder involvement:} Second sponsor talk and webinar for participants.
    \item \textit{Agenda:} At the end of the day, teams present their progress, problems they faced, and their plan until the next checkpoint. They also receive feedback from organizers, mentors, and other teams.
\end{itemize}

\vspace{0.3cm}
\textbf{Hackathon day 3:}
\begin{itemize}[leftmargin=0.5cm]
    \item \textit{Agenda:} Organizers introduce the agenda for the day. Teams present their progress, problems they faced, and their plan until the next checkpoint. They also receive feedback from organizers, mentors, and other teams.
    \item \textit{Duration / breaks:} Suggested off-screen break at noon.
    \item \textit{Mentoring:} Mentors meet with organizers, discuss potential difficulties that teams might face, and decide on mentors with related expertise to support them.
    \item \textit{Agenda:} At the end of the day, teams present their progress, problems they faced, and their plan until the next checkpoint. They also receive feedback from organizers, mentors, and other teams.
\end{itemize}

\vspace{0.3cm}
\textbf{Hackathon day 4:}
\begin{itemize}[leftmargin=0.5cm]
    \item \textit{Continuity planning:} Presentation of internship and funding opportunities.
    \item \textit{Duration / breaks:} Suggested off-screen break at noon.
    \item \textit{Mentoring:} Mentors meet with organizers, discuss potential difficulties that teams might face, and decide on mentors with related expertise to support them.
    \item \textit{Stakeholder involvement:} Third sponsor talk.
    \item \textit{Agenda:} At the end of the day, teams present their progress, problems they faced, and their plan until the next checkpoint. They also receive feedback from organizers, mentors, and other teams.
\end{itemize}

\vspace{0.3cm}
\textbf{Hackathon day 5:}
\begin{itemize}[leftmargin=0.5cm]
    \item \textit{Agenda:} Organizers introduce the agenda for the day. Teams present their progress, problems they faced, and their plan until the final presentation. They also receive feedback from organizers, mentors, and other teams.
    \item \textit{Competition / cooperation:} Training session for judges. Explanation of judging criteria and judging procedure.
    \item \textit{Duration / breaks:} Suggested off-screen break at noon.
    \item \textit{Agenda:} Final team presentations (live streaming). Presentations are recorded.
    \item \textit{Competition / cooperation:} Online voting, judges deliberation, and decision. Announcement of winners and invitation to the conference award ceremony.
\end{itemize}

\vspace{0.3cm}
\textbf{After the hackathon:}
\begin{itemize}[leftmargin=0.5cm]
    \item \textit{Competition / cooperation:} Recognition of winning teams at conference award ceremony. Prizes are sent to team members.
    \item \textit{Continuity planning:} Team Github repositories, talk recordings, final presentation slides and recordings, and information about winners are added to Github page and announced by the hackathon organizers on social media. After participants consent, they are invited to a common LinkedIn group and added to an email list where organizers and other community members can share future events, trainings, internships, and job and study opportunities.
\end{itemize}

\subsection{Hybrid community hackathon}
\label{sec:timeline:community}
This timeline provides an example of a small-scale hybrid hackathon (< 30 participants) that aimed to bring together researchers, students, and practitioners to form a community around a novel resource.

\vspace{0.3cm}
\textbf{6 months before the hackathon:}
\begin{itemize}[leftmargin=0.5cm]
    \item \textit{Goal:} Formation of a community around a novel data resource that contains a virtually complete collection of all open-source projects around the world.
    \item \textit{Theme:} Development of research ideas and initial prototypes that utilize the resource.
    \item \textit{Competition / cooperation:} Decision for a cooperation style event that focuses on joint exploration of the resource.
    \item \textit{Duration / breaks:} Discussion about and decision for a 3-day event that starts on Friday afternoon.
    \item \textit{Agenda:} Decision about a tentative agenda that includes idea pitches, checkpoints, and a final presentation.
    \item \textit{Participant recruitment:} Identification of key individuals in industry, universities, and scientific communities that could benefit from the resource and that can support the recruitment of individuals that would be interested in and would benefit from using the resource.
    \item \textit{Stakeholder involvement:} Discussions with these key individuals about their interest in the resource.
    \item \textit{Ideation:} Decision to ask participants for initial ideas through the registration form and conduct additional ideation at the beginning of the hackathon.
    \item \textit{Specialized preparation:} Identification of a suitable space at the offices of the organization where one of the organizers is affiliated. Criteria for selection included the availability of one larger common room and multiple smaller breakout rooms in close proximity. All rooms are equipped with suitable hardware to hold conference calls (computer connected to area microphones, cameras, and projection hardware).
\end{itemize}

\vspace{0.3cm}
\textbf{4 months before the hackathon:}
\begin{itemize}[leftmargin=0.5cm]
    \item \textit{Specialized preparation:} Creation of a Github organization and page, which includes the location, preliminary schedule, organizer contact information, code of conduct, and links to resources. The page serves as the primary information hub for the hackathon. Creation of a Discord server, including public announcements and troubleshooting channels, as well as a private organizer channel.
    \item \textit{Participant recruitment:} Invitation of potential participants through previously identified key individuals, including a link to the Github page. Registration through an online form that covers the participants' contact details, open source handle, preferred programming languages, interests in the resource, their intention to attend in-person or online, and their potential needs for funding (travel and accommodation support for in-person participants and daily allowance for online participants).
    \item \textit{Ideation:} Ask invitees to propose initial ideas for hackathon projects through the registration form.
\end{itemize}

\vspace{0.3cm}
\textbf{3 months before the hackathon:}
\begin{itemize}[leftmargin=0.5cm]
    \item \textit{Ideation:} Adding proposed ideas to Github page.
    \item \textit{Specialized preparation:} Decision for a pre-hackathon webinar to introduce participants to the resource.
    \item \textit{Mentoring:} Identification and invitation of individuals who are familiar with the resource and relevant technologies to serve as mentors. Invitation email including a link to the Github page and a registration form that covers contact information, prior experience, interests, availability during the event, and intention to attend in person or online.
    \item \textit{Continuity planning:} Invitation of selected participants and key individuals to the Discord server and Github organization.
\end{itemize}

\vspace{0.3cm}
\textbf{1 month before the hackathon:}
\begin{itemize}[leftmargin=0.5cm]
    \item \textit{Specialized preparation:} Development of documentation for the resource, including sample code for selected project ideas that were submitted through the registration form. Sharing of documentation on the Github page and through Discord.
    \item \textit{Agenda:} Adding a first complete agenda to the Github page, including a pre-hackathon webinar for participants, calendar invite links, locations, and Zoom links for all training and hackathon sessions. Connecting each in-person location to a dedicated Zoom link.
    \item \textit{Ideation:} Planning for a pitching session at the beginning of the hackathon.
    \item \textit{Team formation:} Decision that teams will form around ideas and that members should come from different institutions.
\end{itemize}

\vspace{0.3cm}
\textbf{1 week before the hackathon:}
\begin{itemize}[leftmargin=0.5cm]
    \item \textit{Specialized preparation:} Online pre-hackathon webinar to introduce participants to the capabilities and usage of the resource. Interaction during the webinar allows participants to connect to the resource and run code samples.
    \item \textit{Ideation:} Explanation of the pitch procedure at the beginning of the event to webinar participants.
    \item \textit{Mentoring:} Introduction of mentors and their area of expertise at the webinar.
    \item \textit{Continuity planning:} Sharing recording of the webinar on the Github page.
\end{itemize}

\vspace{0.3cm}
\textbf{Hackathon day 1:}
\begin{itemize}[leftmargin=0.5cm]
    \item \textit{Specialized preparation:} Organizers start Zoom in all hackathon rooms and make sure that online participants are visible and audible and that online participants can see and hear in-person participants.
    \item \textit{Agenda:} Meeting in the main room. Welcoming words by the organizers, presentation of the code of conduct, hackathon agenda, and expected final submission through Github (source code, presentation slides, and project report). Reminder to join the Discord server and Github organization. Hint to refer to the Github page and contact organizers or mentors if participants get stuck.
    \item \textit{Stakeholder involvement:} Introduction of supporting individuals and institutions.
    \item \textit{Mentoring:} Introduction of mentors, their area of expertise, and their role during the hackathon.
    \item \textit{Agenda:} Organizers explain the team formation procedure and ask each team to create an introduction slide for the first checkpoint the next morning. The slides should include the team name, the names of the team members, and a concrete project idea. Hint for online participants to inform the others about potential parallel activities during the event (e.g., classes or social and work commitments).
    \item \textit{Continuity planning:} Organizers ask participants to create a Github repository for their team within the hackathon Github organization.
    \item \textit{Ideation:} Participants pitch ideas. Organizers collect pitched ideas on a shared online whiteboard. After the pitches are complete, participants vote for the ideas they would like to work on.
    \item \textit{Team formation:} Organizers create a Zoom breakout room for the most voted ideas and ask proposers to join the breakout room. In-person proposers spread out in the room while the remaining in-person participants walk around and choose a team to join. In parallel, online participants join the breakout rooms of the idea they are interested in. If there is no in-person proposer for a team, an in-person participant who is interested in the idea joins the breakout room instead.
    \item \textit{Competition / cooperation:} If teams are too large, organizers can suggest them to split up or for participants to join a different team.
    \item \textit{Mentoring:} Mentors join teams and support them to connect to the resource, scope their project, and help with technical issues. Mentors focus on their teams but also support others if necessary.
    \item \textit{Duration / breaks:} Social dinner at the end of the day for in-person participants.
\end{itemize}

\vspace{0.3cm}
\textbf{Hackathon day 2:}
\begin{itemize}[leftmargin=0.5cm]
    \item \textit{Specialized preparation:} Organizers start Zoom in all hackathon rooms and make sure that online participants are visible and audible and that online participants can see and hear in-person participants.
    \item \textit{Agenda:} Meeting in the main room. Organizers introduce the agenda for the day and reiterate the expected final submission. Each team presents itself, their project idea, and their plan until the next checkpoint.
    \item \textit{Specialized preparation:} Creation of one dedicated Discord channel per team. Invitation of team members to their respective channel.
    \item \textit{Duration / breaks:} Lunch break for in-person participants and suggested off-screen break for online participants.
    \item \textit{Agenda:} At the end of the day, all participants, mentors and organizers meet in the main room. Teams present their progress, problems they faced, and their plan until the next checkpoint.
\end{itemize}

\vspace{0.3cm}
\textbf{Hackathon day 3:}
\begin{itemize}[leftmargin=0.5cm]
    \item \textit{Specialized preparation:} Organizers start Zoom in all hackathon rooms and make sure that online participants are visible and audible and that online participants can see and hear in-person participants.
    \item \textit{Agenda:} At the beginning of the day, everyone meets in the main room. The organizers lay out the agenda for the day and reiterate the expected final submission. Teams present their progress, problems they faced, and their plan until the final presentation.
    \item \textit{Agenda:} Final presentations of teams in the main room before lunch. Discussions about the content of the presented projects and problems the teams encountered during the hackathon.
    \item \textit{Continuity planning:} Teams share presentations, code repositories, and report through Discord.
    \item \textit{Duration / breaks:} Group pictures and end of the hackathon. Lunch and departure for in-person participants.
\end{itemize}

\vspace{0.3cm}
\textbf{After the hackathon:}
\begin{itemize}[leftmargin=0.5cm]
    \item \textit{Continuity planning:} Team Github repositories are linked on the hackathon Github page. Organizers distribute a summary of the event directly after the hackathon and provide regular updates about the resource through Discord.
    \item \textit{Stakeholder involvement:} Organizers suggest that stakeholders share publications and other outcomes they produce using the resource through Discord.
\end{itemize}

\section{The 12 key decisions}
\label{sec:dec}
In the following, we will outline the twelve key decisions (see Table~\ref{tab:keydecisionssummary} for an overview) that organizers should consider when planning a hackathon.
\begin{table*}[h!]
    \centering
    \caption{Overview of the 12 key decisions organizer should consider when planning a hackathon}
    \resizebox{0.7\textwidth}{!}{
    \begin{tabular}{ll}
        \toprule
        \textbf{Decision} & \textbf{Overall objective} \\
        \midrule
        Goal & What do you want to achieve when organizing your hackathon? \\
        Theme & What should be the overall theme of your hackathon? \\
        Competition/Cooperation & Should teams compete for prizes or work together? \\
        Stakeholder Involvement & How can you integrate externals into your hackathon? \\
        Participant Recruitment & Who would you like to come to your hackathon? \\
        Specialized Preparation & What will be required for teams to participate in your hackathon? \\
        Duration/Breaks & How long and intense will your hackathon be? \\
        Ideation & When and how will teams develop ideas they can work on? \\
        Team Formation & How will likeminded participants find each other? \\
        Agenda & What is going to happen during your hackathon? \\
        Mentoring & How will you support the participants of your hackathon? \\
        Continuity Planning & What will happen after the hackathon is over? \\
        \bottomrule
    \end{tabular}}
    \label{tab:keydecisionssummary}
\end{table*}
 For each decision, we provide information about \textbf{when} organizers should consider making it, \textbf{who} should be involved in the decision, \textbf{how} to make the choice and implement its result, discuss potential \textbf{tradeoffs} among the various options, and discuss \textbf{differences between in-person, online and hybrid events}. The order in which the decisions are presented is deliberate but not strict. Moreover, not all decisions are relevant for each hackathon. Prospective organizers should thus perceive the following list as a suggestion for which decisions can be important but decide for themselves which ones they consider for their specific purpose.

\subsection{Goal}
\label{sec:dec:goal}
Goal setting is key for hackathon design. All parties should be clear about their goals going into an event and consider the attainability and clarity of these goals. A failure to consider them could result in disappointment among participants and organizers.

\subsubsection{When?}
Setting goals for the hackathon should take place \textit{before any planning of the event begins}. A common timeframe is to form goals about 4 months before the start of the event. Before deciding on one or multiple goals, the organizers can consult with projected future participants (section \ref{sec:dec:part}) and potential stakeholders (section \ref{sec:dec:stake}) about the value and feasibility of the projected goals and continue discussions about details after the general direction has been set.

\subsubsection{Who?}
Organizers, stakeholders, and participants are involved in setting goals for their hackathon. In practice, organizers often set initial goals and modify them in collaboration with stakeholders, which is done in the early phase of hackathon planning. Participants are often also asked to express their goals for the event through, e.g., a pre-hackathon survey (section \ref{sec:dec:part}). It is important to note here that the organizers’ goals may not always be aligned with those of the participants. For example, organizers might aim to foster entrepreneurship, whereas participants simply aim to pursue their interests related to a particular topic or project. It is not necessary that organizers and participants have identical goals, but organizers should be aware of potential goal disparities and take necessary actions to maximize the satisfaction of participants with different goals~\cite{medina2019does}. The participants, for example, might have a superset of the organizers’ goals, or the organizers’ goals may be achievable even if only a subset of participants shares them. The organizers should be aware, though, if goals are incompatible (e.g. the organizers want to benefit a charity, while the participants mainly want to start a business) because participants may leave disappointed. It would be wise for organizers to adapt their recruiting approach (section \ref{sec:dec:part}) and their messaging to attract attendees with compatible goals.

\subsubsection{How?}
There are several goals that organizers may aim to achieve when organizing a hackathon. Clarity of goals is strongly related to participant satisfaction and outcome quality~\cite{filippova2017diversity}.

\vspace{0.15cm}
\forceindent The most common goal around which hackathons are organized is the \textbf{production of artifacts or products}. We observe two different strategies that organizers use to achieve this goal. The first strategy involves the recruitment of experts from different fields (section \ref{sec:dec:part}) and the facilitation of brainstorming (section \ref{sec:dec:ideation}) to produce cross-pollinated project ideas. Brainstorming has also been found to be associated with better outcomes for self-identified minorities~\cite{filippova2017diversity}.

\vspace{0.15cm}
\forceindent The organizers using this strategy should devote most of their effort to recruiting stakeholders (section \ref{sec:dec:stake}) from diverse groups that will benefit from meeting and working with each other. For example, they might include local startup communities, volunteers and activists, non-profits, accelerator programs, investors, technology companies, and others. Moreover, organizers should design their events in a way that fosters participants to meet and work with people outside their regular networks and contribute to initiatives outside their usual scope. The sections dedicated to ideation (section \ref{sec:dec:ideation}) and duration/breaks (section \ref{sec:dec:dur}) contain helpful ideas to facilitate networking among participants. One common example are regular checkpoints (section \ref{sec:dec:agenda}) to ensure that participants communicate with each other about their projects several times during the event. For example, a series of hackathons organized by Garage48~\cite{g48cybersec} aims to foster the startup ecosystem by connecting entrepreneurs with domain experts, investors, accelerators, and others. Similarly, BioInnovation Days~\cite{pe2019understanding} gathers together students, researchers, mentors, and startups to develop prototypes in the domain of bio-medicine. The Global Hack~\cite{globalhack} is an example of an online movement where enthusiasts from different countries hosted several hackathons to address the recent COVID-19 global pandemic. It is important to note, though, that having diverse teams might lead to them needing more time to develop a concrete idea to work on. Spending more time on ideation subsequently leaves less time to actually work on an artifact that can be showcased at the end of an event and that can potentially be continued after an event has ended (section \ref{sec:dec:continuity}).

\vspace{0.15cm}
\forceindent The second strategy is teaming up experts and novices as mentors and apprentices, which enables the positive side effects of vertical \textbf{networking and learning}. Hackathons adopting this strategy should focus their efforts on identifying experts in the field and themes of their events, abilities of participants, modularization of projects to support parallelism, and the use of relevant mentoring strategies~\cite{nolte2020support} (section \ref{sec:dec:mentor}). One example of this type of hackathon is Astro Hack Week~\cite{astrohackweek}, where a diverse group of participants ranging from novices to experts in the fields of astronomy, physics, statistics, machine learning, and data science gather together to learn skills necessary for the analysis of astronomical data through tutorials and then solve open problems in the astronomical community. Another example is a series of hackathons organized by HackHPC~\cite{hackhpc} as part of the workforce development initiative of the high-performance computing community where students team up with researchers to learn how to utilize HPC resources in a real-life context~\cite{powell2021organizing}. World of Code~\cite{worldofcode} is another example of a hackathon where students are teamed up with faculty members to learn how to do quantitative research on the behaviors of open source software (OSS) projects through the world’s largest OSS dataset.

\subsubsection{Trade-offs}
\begin{enumerate}[leftmargin=0.5cm]
    \item \textbf{Goal compatibility (Organization goals vs personal goals):} Although organizational goals may not necessarily be compatible with personal goals of the participants~\cite{medina2019does}, the huge disparity among goals will lead to dissatisfaction with the experience and artifacts generated. One essential responsibility of organizers is then to identify any potential misalignment among the goals and ways to mitigate it. This could be done by a clear presentation of organizational goals, selective recruitment of participants (section \ref{sec:dec:part}), upfront negotiation of goals, and adaptation and customization of the hackathon process and outcome based on these goals (section \ref{sec:dec:continuity}).
    \item \textbf{Illustrating a concept vs producing something useful:} Teams can either aim to develop illustrative proofs of concepts, or software that can actually be used to help accomplish a task. In our experience, diverse teams that have not worked together extensively (i.e., flash or ad hoc teams) are likely to spend more time discussing what to build and how to go about it, and how to pitch their projects to potential stakeholders, and they should thus aim to produce \textbf{lightly-engineered demos and videos}, with mock-up data, which require little engineering effort. For teams with less diverse interests or teams that have worked together extensively, it is often desirable to aim for a prototype with \textbf{sufficient functionality that could be used almost immediately in the real context}~\cite{falk2025creativity,pe2022corporate}. Organizers might also consider organizing larger (60 to 80 participants) events if the goal is to produce a creative and useful artifact~\cite{falk2025creativity}.
    \item \textbf{Challenging vs achievable:} Research has shown that groups accomplish more and are more inclined to continue working on a project when they have goals that are challenging~\cite{nolte2020what}. Goals that are too easy fail to motivate them to do their best. On the other hand, goals that are so challenging that they are, or seem to be, impossible to achieve can also be demotivating. Choosing appropriate goals involves an understanding of both the participant’s capabilities and interests, as well as a realistic view of what can be achieved in a short time frame. One useful approach for managing this tradeoff is to set (or encourage each team to set) both a relatively easy goal, and a more difficult “stretch” goal. This gives teams the opportunity to feel that they achieved something meaningful, as well as motivation to go beyond this and attempt something very challenging.
    \item \textbf{Goal clarity:} Although hackathons should ideally be designed to achieve the goals of both organizers and participants, there are circumstances where their goals are not aligned~\cite{filippova2017diversity,medina2019does}. This occurs, for example, when organizers hold a hackathon to connect professionals working across different fields, but participants wish to develop viable product prototypes. It is important for organizers to recognize such potential goal conflicts and plan accordingly to be reasonably able to achieve their intended goals. One way to address this tradeoff is for organizers to clearly state their goals during recruitment (section \ref{sec:dec:part}) so that participants know what to expect.
\end{enumerate}

\subsubsection{Modes of participation}

\begin{itemize}[leftmargin=0.5cm]
    \item \textbf{In-person:} In-person-only participation is suitable if the aim is to focus on local communities, leverage physical resources such as lab spaces or hardware, and prioritize face-to-face networking opportunities. This mode enables organizers to foster direct interactions that promote immediate feedback, collaboration, and a sense of community. 
    
    \forceindent If the goal involves local community building or addressing regional challenges, in-person participation is more suitable because it provides opportunities for direct engagement. However, organizers should consider subsidizing travel or accommodations for underrepresented groups to achieve inclusivity, ensuring equal access for those who might otherwise be excluded~\cite{paganini2023opportunities}. Co-location also limits the risk of individuals disengaging and leaving an event~\cite{mendes2022socio}.
    
    \item \textbf{Online:} Online-only participation is more suitable for organizers who prioritize global knowledge sharing, aim to solve challenges that require diverse expertise across time zones, or try to engage participants who might face logistical barriers to attending in person. This mode allows for broad accessibility, eliminating travel costs and visa issues, and might allow participation even in the case of tight schedules~\cite{mendes2022socio}.
    
    \forceindent For example, if the goal is to foster inclusivity, online participation -- assuming suitable moderation -- can provide safe spaces, particularly in women-focused hackathons, which can reduce gender-related barriers like the discomfort of being ``the only woman in the room''~\cite{paganini2023opportunities}. Online formats can also be suitable for addressing urgent goals, as seen during the COVID-19 pandemic when hackathons like EUvsVirus were able to engage thousands of participants across Europe in a short period of time~\cite{bertello2022open,braune2021interdisciplinary}.  

    \forceindent However, organizers might also face challenges such as participant disengagement or goal misalignment, which are more common in virtual settings. Clear communication of objectives, regular check-ins, and robust collaborative tools can help maintain alignment and motivation~\cite{gama2021online,mendes2022socio} (section~\ref{sec:dec:agenda}). Moreover, social interactions and networking, which typically just happen during in-person events (e.g., when getting coffee or going together for a meal), need to be carefully planned in an online hackathon.
    
    \item \textbf{Hybrid:} Hybrid participation is more suitable if organizer goals combine the need for localized problem-solving with the global reach and inclusivity of online participation. This format allows scaling an event by engaging global participants virtually while partially retaining the benefits of face-to-face collaboration for co-located participants~\cite{powell2021organizing}.

    \forceindent Hybrid formats are particularly effective for goals that aim at both local and global impact. For example, organizers can support localized innovation while inviting diverse perspectives to enrich solutions. Ensuring equity between in-person and online participants is critical, with resources such as virtual mentorship and synchronized event updates helping to bridge gaps.
    
    \forceindent If event goals require flexibility and scalability while maintaining local engagement, hybrid hackathons can provide a balanced approach.
\end{itemize}

\subsection{Theme}
\label{sec:dec:theme}
Most hackathons focus on a specific theme. A theme refers to a specific topic or cause that motivates the organization of a hackathon, which, in turn, sets its boundary. A theme helps organizers to address a particular topical area and devise solutions for problems within this area.

\subsubsection{When?}
Organizing a hackathon commonly starts with a theme. Themes are generally developed in conjunction with the goals of a hackathon (section \ref{sec:dec:goal}), typically 4 months ahead of the event.

\subsubsection{Who?}
While the decision for a specific theme resides with the organizers, it is advisable to discuss it with relevant stakeholders such as non-profits, investors, educational institutions, or others (section \ref{sec:dec:stake}).

\subsubsection{How?}
The organizers should decide on the initial theme of the event and invite potential stakeholders interested in their proposed theme to discuss its feasibility and develop project ideas related to this theme. Themes can range from general to specific, e.g., fighting a global crisis~\cite{globalhack}, mining a large-scale open source data source~\cite{worldofcode}, building cybersecurity solutions and teaching cybersecurity skills~\cite{g48cybersec,affia2020developing}, developing means to teach high-performance computing~\cite{holmen2024facultyhack} and AI literacy~\cite{schulten2024hack}, developing means to learning data science through solving astronomical and geo challenges~\cite{astrohackweek,geohackweek}, computational biology~\cite{openbio}, neuroscience~\cite{brainhack}, environmental issues~\cite{zapico2013hacking}, and high-performance computing~\cite{hackhpc}.

\vspace{0.15cm}
\forceindent The theme focuses potential project ideas on immediate needs or outstanding challenges in the chosen area. One such need is to \textbf{speed up development work}. A series of hackathons by a team of the Space Telescope Science Institute (STScI) that manages science programs and conducts experiments on the Hubble Space Telescope (HST)~\cite{pe2019understanding} is one such example. In this case, events were organized to speed up the conversion of tools to a language, bringing together individuals who are using such tools for their work. Hackathons organized by Garage48~\cite{g48cybersec} are other examples of speeding up the development of cybersecurity solutions. Another need is solving outstanding challenges, which could take the form of technical debt or scientific advancement. A team participating in a corporate hackathon we studied~\cite{oneweek}, for example, utilized the time during that hackathon to implement a tool to automate recurring manual work.

\vspace{0.15cm}
\forceindent Another reason for choosing a specific theme is \textbf{hack to train}. Astro hack week~\cite{astrohackweek} and Geo Hack Week~\cite{geohackweek} are examples of this type, where participants are first trained to apply basic analysis and coding skills and gather domain knowledge before moving on to solve challenges in the respective domains. Singapore’s BrainHack of the Organization for Human Brain Mapping (OHBM) in 2018~\cite{brainhack} is another example. There, the organizers held parallel training and hacking tracks to get participants up to speed with technologies and approaches they might require to work on projects in this domain. In addition, organizers might aim to help educators develop teaching resources. Examples of this are the FacultyHack~\cite{facultyhack2024} and the Hack beyond the Code hackathon~\cite{hackbeyondthecode2024}.

\vspace{0.15cm}
\forceindent A theme might also be chosen to \textbf{build resources for a specific community}. One example of this is the World of Code (WoC) hackathon~\cite{worldofcode} that aimed to bring interested researchers together and identify requirements for a large archive of open-source data. Another example are events organized by the Open Bioinformatics Foundation, which aims at improving and extending existing resources and infrastructure and publishing papers about the accomplished work~\cite{openbio}.

\subsubsection{Trade-offs}
\begin{enumerate}[leftmargin=0.5cm]
    \item \textbf{Speed vs quality:} Hackathons operated under themes related to speed or acceleration could quickly advance the development work in a specific domain with proper implementation. Getting things done fast and quickly, however, does not mean that such hackathons produce quality artifacts, i.e., they might be workaround solutions and need code cleanup or dependency reduction, and thus, follow-up work is usually required to ensure the quality of the developed artifacts.
    \item \textbf{More general vs more specific:} Themes can be very specific, e.g., building an app to monitor occupancy levels in a local shelter, or much more general, e.g., helping local governments support sustainability. Very specific themes can be effective in helping to ensure the work actually serves a useful purpose but runs the risk of not matching up well with the interests of a large group of potential participants. More general themes allow more latitude for participants to develop projects that align with their personal interests but run a greater risk of not matching up with a real need. To address this tradeoff, organizers should make sure they understand the motivations of potential participants. To achieve this, they may want to conduct a survey to assess their attitudes toward different theme variations.
\end{enumerate}

\subsubsection{Modes of participation}
\begin{itemize}[leftmargin=0.5cm]
    \item \textbf{In-person:} In-person participation is more suitable for themes requiring tangible outcomes, physical interaction, or access to specialized resources. For example, themes involving hands-on prototyping or product development, such as makeathons in the ``Garage48'' series~\cite{g48defence}, leverage physical proximity to connect entrepreneurs, domain experts, and investors to create tangible outcomes.

    \forceindent Additionally, when the theme focuses on region-specific challenges, such as local sustainability issues, in-person participation immerses participants in the local context and facilitates direct engagement with local stakeholders. Participants who have experienced similar challenges in other regions can also join to collaborate with local teams, creating a rich environment for problem-solving. Themes that benefit from this level of immersion and physical interaction are naturally aligned with in-person formats, particularly when regional relevance and stakeholder involvement are critical (section~\ref{sec:dec:stake}). An example of this is the HPC in the city hackathon series that aims to address issues specific to the place where the event is organized~\cite{hpcinthecity}.

    \item \textbf{Online:}  Online-only participation aligns best with global themes or themes that leverage diverse, interdisciplinary perspectives and do not necessarily require physical tools or resources. For example, themes addressing global challenges such as sustainability~\cite{mendes2022socio}, innovation~\cite{bertello2022open}, inclusivity in technology~\cite{paganini2023opportunities}, and pandemic response~\cite{braune2021interdisciplinary,mendes2022socio} are well-suited to online formats if they do not depend on physical resources. Online participation allows organizers to respond to emerging global issues and bring together participants from various regions. The EasterHack~\cite{braune2021interdisciplinary} event during the COVID-19 pandemic demonstrates how online hackathons can effectively tackle pressing health challenges by sourcing interdisciplinary collaboration from around the world~\cite{braune2021interdisciplinary}.

    \forceindent However, organizers need to ensure that online themes align with tools and resources available to remote participants. Themes requiring physical materials should have virtual alternatives to maintain inclusivity and practicality.

    \item \textbf{Hybrid:}  Hybrid hackathons enable organizers to draw on the strengths of in-person engagement for local impact while incorporating remote perspectives to address broader challenges or integrate specialized knowledge from other regions. For example, themes tied to global issues~\cite{braune2021interdisciplinary,bertello2022open} can leverage hybrid formats to combine localized prototyping with international ideation.

    \forceindent However, organizers need to consider whether a theme genuinely benefits from remote or global input. Broad themes that thrive on diversity and inclusivity can be enriched by hybrid participation. However, in-person participation might still offer greater relevance and focus for themes deeply rooted in local contexts or requiring significant physical interaction. Organizers must also balance resources, tools, and engagement strategies and ensure that physical requirements for certain themes are supplemented with virtual alternatives.
\end{itemize}

\subsection{Competition / cooperation}
\label{sec:dec:comp}
One key decision that organizers need to make when designing a hackathon is whether to provide external incentives in the form of prizes, which introduces a strong element of competition into an event.

\subsubsection{When?}
The decision on whether to have a competitive or collaborative hackathon should be made several months before an event. Competitive events, in particular, take considerable time to organize. Organizers must find credible judges, decide on suitable award criteria, determine prizes and prize categories, and ensure that prizes can be handed out at the end of an event. If the prizes have considerable monetary value or will involve a time commitment from experts whose time is very limited (e.g., the opportunity to pitch the team’s winning idea to potential investors), there may be considerable work for the organizers to secure sponsorship.

\subsubsection{Who?}
Organizers, sponsors, and other stakeholders (section \ref{sec:dec:stake}) involved in setting goals for a hackathon should consider whether and how to introduce competition to their event. It is also extremely important to consider the goals of participants -- picking a structure that does not further their goals will make recruiting (section \ref{sec:dec:part}) difficult. If participant goals are not clear to the organizers, a brief survey or interviews with a sample of the population of interested participants can be very helpful.

\subsubsection{How?}
\textbf{Cooperative events} are typically structured around a \textbf{common goal} (section \ref{sec:dec:goal}) or \textbf{theme} (section \ref{sec:dec:theme}). An example for cooperative events is a series of cooperative hackathons~\cite{biohack} that was organized to accelerate the development of integrated web services in the field of bioinformatics. Teams there worked on projects involving data standardization and interoperability of tools and services, among others. Other cooperative hackathons aim to perfect an artifact. For example, the Debian Linux project launched the Debcamp event~\cite{debcamp}, a hacking session before the Debian Conference, where software developers whose work depends on the Debian Linux software distribution worked together to promote the redistribution, general availability, and mutual compatibility of software. Another example is a hackathon series organized by STScI’s Hubble Space Telescope team aimed to accelerate the migration of data analysis tools implemented in an old obsolescing language to a contemporary one~\cite{pe2019understanding}.

\vspace{0.15cm}
\forceindent \textbf{Competitive events} focus on teams winning prizes. Winners can be selected by a \textbf{jury}, based on a \textbf{popular vote}, or a combination of the two. Some hackathons also award special prizes to projects that meet specific challenges posed by sponsors or other stakeholders. An example of this is the HPC in the city hackathon series, which awards a prize for the hackathon project that most benefits the local community the hackathon focuses on~\cite{hpcinthecity}. If winners are selected by a jury, it is necessary to invite judges and to specify suitable judging criteria. For judges, the organizers may consider recruiting (domain) experts from a university, community leaders, or representatives from tech companies~\cite{mlh:judges}. Some commonly used judging criteria include appeal to the market, creativity, originality, completeness, polish, and level of difficulty. For a competition to be perceived as fair, it is important that judging criteria are well known in advance and that organizers make sure that participants do not simply turn up with solutions that have been built prior to the event. There are different approaches to awarding prizes based on a popular vote. Voting can, e.g., be limited to on-site participants only, or it can be replaced or augmented by online voting. The latter requires an online voting system and a final presentation sessions to be live-streamed or presentation materials to be distributed online.

\vspace{0.15cm}
\forceindent In order to select the winning teams, the organizers should create a dedicated session at the conclusion of a hackathon to allow each participating team to pitch and demonstrate their idea to the entire audience of an event. This session could take the form of either an \textbf{on-stage presentation} or a \textbf{(science) fair}. In the former approach, each team is typically given a set time to present their prototypes in front of the entire event audience. For this it is important to inform participants in advance how their presentation will be evaluated, the presentation format, and the time limit. The commonly used presentation format starts with a brief introduction of team members and problems that they tried to solve, which is followed by a live demo~\cite{hackplaybook}. It is also possible that teams record a short introductory video and focus on their demo during the presentation. In a (science) fair style presentation session~\cite{nolte2018you}, each team is commonly allocated a booth in a large enough space to set up their presentations. Judges and visitors visit each booth to interact with the teams. This approach facilitates a greater face-to-face interaction between participants, judges, and other attendees. At the Microsoft OneWeek Hackathon, we observed~\cite{nolte2018you,oneweek}, teams could both attend a science fair and upload a video to an online platform where other participants and observers could vote on their favorite projects. This online voting enables individuals who are not able to attend the fair in person to participate in the winner selection.

\vspace{0.15cm}
\forceindent Prizes can vary greatly, with tech gadgets, cash prizes, gamified incentives, such as digital badges and certificates, and opportunities for continued development of winning ideas probably being the most common. The opportunities for further development can take the form of providing additional resources and computing power, freeing up participants’ time to work on their project post-hackathon, or simply the opportunity to pitch their idea to top executives or investors. Awarding cash prizes is not always feasible or recommended~\cite{mlh:prizes1}. Major League Hacking (MLH), for example, provides a few non-cash prize ideas~\cite{mlh:prizes2} such as laptops, headphones, conference tickets, etc., as alternatives. Moreover, hackathon organizers need to decide for \textbf{how many prizes} they offer in relation to the number of participating teams. Offering many prizes at a small event might reduce their perceived value, which in turn can negatively affect participant motivation~\cite{nolte2020what}.

\subsubsection{Trade-offs}
\begin{enumerate}[leftmargin=0.5cm]
    \item \textbf{Competition vs cooperation:} Competition is suitable for hackathons aiming to create innovations because, if teams perceive an event to be competitive, they are more likely to generate unique solutions to differentiate themselves from other competing teams and put more effort into their projects. At the same time, it is important that organizers provide a space where individuals feel safe enough to work on risky and creative ideas. One approach could be to have larger teams in which individuals can feel the backing of their team members~\cite{falk2025creativity}. Moreover, competition discourages communication among teams and hence is not suitable for hackathons aiming to enrich networking among participants beyond their own teams or to engage participants in a common goal. On the other hand, cooperation works well whenever it furthers the organizers’ and participants’ goals, such as promoting a civic cause, providing different pieces of an integrated solution, or learning about programming tools or a particular domain. To reduce the severity of this trade-off, it might be advisable to de-emphasize the prizes in a competitive event so that participants do not over-emphasize competition, e.g., by having prizes that are largely symbolic rather than great cash value. These can be t-shirts, hats, pieces of old hardware, 3D-printed artifacts, or others. Heavily emphasizing prizes might put off many potential participants, especially if they feel that the odds of winning a significant prize are low, that their idea might not work well with the proposed judging criteria, or that judging might not be fair.
    \item \textbf{Jury vs popular vote:} Experts are more appropriate as judges if the desired criteria are technically complex. They will be much more able to determine if a prototype is actually feasible, for example, and if it actually addresses the problem claimed. On the other hand, if the desired result is something that addresses a widely-experienced need or to produce something that seems cool or stylish, a popular vote may be more suitable.
    \item \textbf{High-value vs low value prizes:} High-value prizes create a more serious atmosphere and level of competitiveness, and organizers will have to be very careful about finding skilled judges and applying the pre-specified criteria in a way that can be perceived as fair by hackathon participants. The rules will have to be very clear, e.g., about whether teams can form and start work prior to the hackathon and whether they can bring the code they have written with them to the hackathon. If the highest possible level of professionalism, skill, and innovation is the goal, high-value prizes are a good choice. Low-value prizes are better when organizers and participants have a variety of goals, such as learning, expanding social ties, or attracting newcomers to a community. As the value of prizes approaches zero (e.g., special hats or shirts), competitive hackathons can look much more like collaborative hackathons. With inherently competitive populations, however, even symbolic prizes can make some teams behave very competitively.
\end{enumerate}

\subsubsection{Modes of participation}
\begin{itemize}[leftmargin=0.5cm]
    \item \textbf{In-person:} In-person participation is more suitable if a hackathon emphasizes spontaneous interactions that benefit from face-to-face engagement. Competitions requiring tangible outputs, such as hardware prototypes or physical demonstrations, are suitable for this setting~\cite{liu2022understanding}. On-stage judging or science-fair-style formats can enable judges to evaluate physical prototypes and artifacts up close and interact directly with the team that built them.
    
    \forceindent Shared physical spaces, like co-working areas and maker spaces, create environments where participants naturally engage. Physical proximity can foster camaraderie and accelerate idea exchange, which is often harder to replicate in virtual formats~\cite{kraus2022coworking}.
    \forceindent However, in-person competitive setups may amplify stress for certain groups, such as newcomers or underrepresented groups, potentially creating an exclusionary environment~\cite{paganini2023opportunities}. Additionally, the absence of global participants in purely in-person formats can limit the diversity of perspectives in cooperative settings. To enhance the experience, organizers should prioritize real-time elements, such as live pitching and on-the-spot problem-solving, while creating inclusive environments that balance competitive intensity~\cite{mendes2022socio}.

    \item \textbf{Online:} Online-only participation aligns well with cooperative goals that leverage global collaboration and interdisciplinary perspectives. Online events can foster cooperation across cultural and professional boundaries by connecting participants from diverse geographies. Events like EasterHack exemplify how online formats can enable global collaboration~\cite{braune2021interdisciplinary}. Online formats can also support ``follow-the-sun'' models~\cite{morrison2020challenges}, allowing cooperation across different time zones to maximize productivity.

    \forceindent Competition in online hackathons can be sustained using digital platforms that facilitate real-time milestone tracking and gamified progress updates~\cite{bertello2022open,schulten2022participants}. However, the lack of real-time physical interaction may diminish the intensity of competition~\cite{mendes2022socio}. Organizers can mitigate this by providing clear visibility into progress, using tools such as virtual leaderboards and regular updates. Virtual tools like Discord and Slack can help implement these features. The EUvsVirus Hackathon~\cite{bertello2022open} provides an example of leveraging online platforms to maintain competitive engagement.

    \forceindent Additionally, digital alternatives to on-stage presentations, such as video submissions, virtual judging sessions, and feedback meetings with judges, can aid fairness and consistency across hackathon formats~\cite{bertello2022open}.

    \forceindent Online hackathons may face challenges in balancing competition and cooperation, as remote participants may arrive with pre-made solutions or differing levels of preparation without organizers necessarily being aware. This requires organizers to establish clear guidelines and oversight to ensure a level playing field.

    \item \textbf{Hybrid:} Hybrid participation is beneficial when the aim is to foster global contributions while leveraging the energy of physical spaces for competitive activities. However, in-person participants often dominate interactions, potentially marginalizing remote contributors. To address this, organizers can propose digital platforms that connect online and in-person participants in real time.

    \forceindent Managing competition and cooperation across both modalities demands hybrid-compatible activities, such as simultaneous live and virtual judging, video submissions, and feedback meetings can help create a level playing field for participants in both formats~\cite{bertello2022open}. 

    \forceindent Physical prizes or local perks, such as meals or event merchandise, should be complemented with equivalent digital rewards, like vouchers or online recognition, to ensure fairness for remote participants and maintain the excitement and productivity of competitive and cooperative dynamics~\cite{mendes2022socio}.
\end{itemize}

\subsection{Stakeholder involvement}
\label{sec:dec:stake}
Hackathons commonly focus on a specific theme (section \ref{sec:dec:theme}) or take place in a specific domain. It thus appears reasonable to include stakeholders related to this theme or domain to participate in the organization, execution, and follow-up of a hackathon. They can provide valuable input, help set the stage for an event, make it more engaging and fun for participants, and support the sustainability of hackathon outcomes (section \ref{sec:dec:continuity}). Deciding how and when to involve stakeholders in the planning and execution of a hackathon is thus a crucial decision that organizers have to take because it will fundamentally shape the experience of participants during the event.

\subsubsection{When?}
Organizers should think about which stakeholders to involve early in the planning process because their input might have a considerable impact on the design of the event itself. The way each stakeholder participates in the planning, execution, and follow-up of a hackathon is then subject to individual planning and can potentially happen later in the planning process. Stakeholders and their role in relation to the hackathon should, however, be decided upon and announced prior to the start of the hackathon to be able to include them in information material.

\subsubsection{Who?}
While conducting a traditional stakeholder analysis~\cite{stakeholderanalysis} might be too time-consuming, organizers certainly should think about who might be interested in or affected by the outcomes of the hackathon they plan to organize. Depending on the theme of the event, organizers might want to involve university departments (e.g., for a collegiate event~\cite{steelhacks}); investors, incubators, and customers (e.g., for an entrepreneurial event~\cite{g48cybersec}); managers and executives (e.g., for a corporate event~\cite{nolte2018you}); volunteers and activists (e.g., for a civic event~\cite{chihacknight}); scientists and technical experts (e.g., for a scientific event~\cite{pe2019understanding}). These are just a few examples of potential stakeholders. The decision on who and how to involve them ultimately lies in the hands of the organizers.

\subsubsection{How?}
Much like the decision for which stakeholders to involve, the decision for how to involve them allows many options. In the following, we will outline common examples for stakeholder involvement. These should, however, not be perceived as exhaustive. Organizers should discuss options with stakeholders and decide on a model that fits their particular event.

\vspace{0.15cm}
\forceindent One common way of involving stakeholders in a hackathon is as \textbf{sponsors}. This can include them providing resources in exchange for being mentioned on the hackathon website, in handouts, or on posters at the hackathon site, or them providing specialized equipment or sponsoring awards or specific activities during a hackathon. The website of Major League Hacking provides a good overview of how to attract sponsors, including different sponsorship options~\cite{mlh:sponsorship}. The advantage of this approach is that the outline and organization of a hackathon remain solely in the hands of the organizers while sponsors provide additional resources for the hackathon to take place.

\vspace{0.15cm}
\forceindent Another common way of stakeholder involvement is to invite them as \textbf{speakers}. Similarly, stakeholders can also hold \textbf{training sessions} during an event, e.g., related to specific technologies they are familiar with and that participants might use for their projects. Both approaches allow stakeholders to be present during an event and provide useful context and input for participants that they can utilize when planning and working on their projects. Moreover, it leaves the option for participants to decide whether to use the input for their projects or not.

\vspace{0.15cm}
\forceindent Another common way of involving stakeholders in a hackathon is for them to serve as \textbf{mentors} (section \ref{sec:dec:mentor}) or \textbf{jurors}. Serving as mentors -- in comparison to the aforementioned role as speakers or trainers -- allows stakeholders to directly work with participants, provide targeted feedback, and steer them into a specific direction~\cite{nolte2020support} and foster learning~\cite{affia2020developing}. Utilizing stakeholders as jurors can also be beneficial because they can provide realistic project assessments based on their area of expertise and again provide useful feedback to participants when discussing their verdict. 

\vspace{0.15cm}
\forceindent In addition to serving in the aforementioned roles, stakeholders can also \textbf{provide access to additional resources} in the form of datasets, documentation, and access to interested parties such as potential future customers or domain experts~\cite{codeforpgh}. This requires them to be accessible during the hackathon, but it allows participants to seek input and advice on demand. 

\vspace{0.15cm}
\forceindent For some hackathons, it might also be feasible for stakeholders to \textbf{propose specific challenges} for the participants to address~\cite{nolte2020support}. We observed this model, for example, in scientific hackathons where scientists proposed project areas or challenges that were related to their area of expertise or that would help them in their work. Participants can then choose which challenge or project to address and how to address it. This model can also be useful in an educational context where teachers propose challenges that students can address.

\vspace{0.15cm}
\forceindent Finally, stakeholders can, of course, also attend an event as \textbf{participants} or serve as \textbf{co-organizers}. This is particularly common for corporate hackathons where organizers and participants often are employees of the same company that organizes the event~\cite{oneweek}. Involving stakeholders as participants can, however, be difficult, especially in an open event that welcomes individuals from various domains and backgrounds since stakeholders might be inclined to take over projects and adjust them to fit their ideas or goals.

\subsubsection{Trade-offs}
\begin{enumerate}[leftmargin=0.5cm]
    \item \textbf{Depth of stakeholder involvement:} The main difference between the aforementioned models of stakeholder participation is how much they can influence what happens during a hackathon. In some cases, it might be useful for stakeholders to be deeply embedded, e.g., when a hackathon aims to solve specific issues within a certain domain, such as the development of software artifacts that fit within an existing ecosystem. This might, however, limit interest by projected participants, thus making it difficult to attract and retain participants during a hackathon. Balancing these two sides can be difficult for organizers. One way of addressing this trade-off is for organizers to allow stakeholders to provide input but limit their interaction with participants and avoid their active participation in projects.
    \item \textbf{Open project selection vs selection among proposed challenges:} Most hackathons allow participants to work on any project they want. This approach can foster creativity and interest because it allows participants to work on any theme they are passionate about. It will, however, likely also lead to participants working on projects that might or might not be related to the goals (section \ref{sec:dec:goal}) organizers had in mind when organizing their event. Participants might also work on projects that might not be useful for the domain the hackathon was organized in. Providing specific challenges ensures that participants work on projects that are relevant to individuals connected to the hackathon, which in turn might increase the probability of projects to live on after the hackathon has ended (section \ref{sec:dec:continuity}). It does, however, limit choice for participants and thus might lead to limited interest and frustration. One way to address this trade-off is for organizers to propose larger topic areas or themes that guide participants toward a specific direction but allow them to develop their own ideas related to this direction.
\end{enumerate}

\subsubsection{Modes of participation}
\begin{itemize}[leftmargin=0.5cm]
    \item \textbf{In-person:} In-person participation excels in fostering direct engagement between stakeholders and participants. Live presentations, mentoring sessions, and networking opportunities allow stakeholders to provide immediate feedback and guidance. A Garage48~\cite{g48cybersec} event we studied highlighted the benefits of co-located stakeholders, such as investors and industry experts, who interacted with teams face-to-face. 

    \forceindent However, requiring physical attendance can limit the diversity of the stakeholder pool, as only those able to travel can participate. This can reduce the variety of perspectives and expertises available to participants~\cite{mendes2022socio}. Organizers can introduce virtual alternatives and elements such as virtual judging and mentoring sessions to supplement in-person activities without compromising the benefits of in-person interaction.
    
    \item \textbf{Online:} Online formats are suitable for engaging a larger variety of stakeholders, especially if the goal is to include individuals who might not be able to travel. Time zone differences pose logistical challenges, though, particularly for live sessions where participants and stakeholders alike may struggle to participate, leading to uneven engagement~\cite{mendes2022socio,schulten2022participants}. Organizers should design schedules accommodating multiple time zones and provide asynchronous alternatives, such as recorded talks or pre-scheduled Q\&A sessions as part of the hackathon agenda (section~\ref{sec:dec:agenda}). 

    \item \textbf{Hybrid:} It is difficult in a hybrid setting to ensure equitable stakeholder engagement with participants in their teams attending through both modalities. In-person stakeholders, such as sponsors and judges, often have more visibility and influence due to physical presence, potentially marginalizing remote participants~\cite{powell2021organizing}. Sponsors attending in person can directly interact with teams, showcase their brand, and network organically. Remote sponsors might not have the same opportunities or connections to participants.

    \forceindent Organizers can actively design interventions to support fairness. For instance, sponsors could be offered hybrid breakout rooms or virtual networking sessions to connect with remote participants. Judges can also evaluate all teams -- whether in-person or remote -- through standardized video presentations or online submission platforms to minimize biases caused by live interactions with in-person participants. Organizers can also provide parallel showcases for in-person and online teams, such as live-streaming presentations or virtual exhibition platforms. This approach ensures that virtual teams are well represented and their contributions are visible to stakeholders~\cite{powell2021organizing}.

    \forceindent Additionally, the goals of the hackathon may influence stakeholder involvement in hybrid formats. For instance, events focused on lived experiences, such as accessibility challenges or women-focused hackathons, may benefit from having stakeholders present in person to engage directly with participants and provide contextual insights~\cite{paganini2023opportunities}.
\end{itemize}

\subsection{Participant recruitment}
\label{sec:dec:part}
Participant recruitment is one of the most crucial elements of hackathon design. After defining goals (section \ref{sec:dec:goal}) and themes (section \ref{sec:dec:theme}) for the hackathon, organizers should ask themselves: Who would be the target audience for an event? When should they start recruitment? How to draw interest and attention to an event? We will provide suggestions for those in the following.

\subsubsection{When?}
Participants need to be recruited and have to register before the hackathon. The exact time that recruitment needs to occur varies based on the scope of the event, the degree to which the target audience is known, and the amount of planning needed for potential participants to take part in a hackathon (i.e. location and time). As such the organizers should discuss who they would want to participate at the very beginning of the event organization before they develop or deploy any recruitment strategy. Organizers sometimes recruit participants up to a year before an event as, e.g., in the case of events for specialized communities such as Astro Hack Week~\cite{astrohackweek}, HackHPC~\cite{hackhpc}, and World of Code~\cite{worldofcode}. Other events, such as collegiate hackathons sponsored by Major League Hacking, typically start recruitment about two months before the actual event~\cite{mlh:promote}.

\subsubsection{Who?}
Taking their event goals (section \ref{sec:dec:goal}) and themes (section \ref{sec:dec:theme}) into consideration, the organizers should identify the characteristics of the target audience they aim to recruit. For some events with broad appeal, \textit{college students who have taken a programming class} may be sufficiently specific. For others, e.g., \textit{PhD level astronomy students}, the recruitment efforts will have to be very targeted and provide compelling motivation for that particular population. Stakeholders (section \ref{sec:dec:stake}) that are connected to the targeted audience can support this recruitment process.

\subsubsection{How?}
There are two general strategies for participant recruitment: \textbf{open and closed}. The organizers should decide which strategy to use based on the goals of their hackathon. Open recruitment targets a wide range of participants with the aim of diversifying participation. As such, open recruitment is typically used for hackathons whose main goal is to build a community around the cause or theme. The aforementioned Astro Hack Week~\cite{astrohackweek}, the World of Code hackathon~\cite{worldofcode}, and HackHPC~\cite{hackhpc} used open recruitment inviting anyone with an interest in the theme of the event. Alternatively, the organizers could also use closed recruitment, thus only inviting specific participants who, e.g., are internal to a specific community. Examples of such hackathons are corporate events~\cite{leemet2021utilizing} such as Microsoft’s OneWeek Hackathon~\cite{oneweek} and the STScI hack days~\cite{pe2019understanding} who only recruited among their employees. Many tech companies hold such events to foster innovation, promote a more open and innovative culture, and help create richer and farther-reaching social ties~\cite{pe2019designing}. If the focus is on innovation, then the organizers might want to think about inviting experienced hackathon participants because their experience can help them come up with projects that are doable within the short timeframe of a hackathon~\cite{falk2025creativity}. If the event is a community event, organizers need to identify and invite individuals who might be interested in the particular cause or theme that serves as the glue for the community. Hackathons mostly are recurring events that are sometimes held by groups, such as scientific research communities, who have continuous needs to train members, maintain or implement new features, or work on interoperability issues for shared tools. The HackWeek Toolkit provides detailed suggestions for defining a suitable audience and scope of an event to meet the needs of specific communities~\cite{hackweektoolkit}.

\vspace{0.15cm}
\forceindent After the target audience is identified, organizers should set up a \textbf{website which should contain basic information about the event} such as dates and venue, contact information to enable interested participants to communicate with the organizers, and a registration form. The process of setting up and publishing a website can be streamlined by, e.g., creating a repository on GitHub as in the case of the World of Code (WoC) hackathon~\cite{worldofcode:schedule}. It is helpful for the organizers to decide on a particular hosting platform, e.g., GitHub, before the event. The organizers should then publicize the website to potential interested participants. When using a GitHub project URL organizers can also encourage interested participants to communicate with them via GitHub issues. In addition to providing information, the organizers should also provide contact information and be accessible to potential participants via Email, Discord, Slack, or other suitable tools.

\vspace{0.15cm}
\forceindent The organizers should also create a \textbf{pre-event registration} form using, e.g., Google Forms, which should be accessible from the website. Through this registration form, organizers can also collect additional information about participants’ skills and backgrounds, preferences for projects and areas, and goals and expectations if they decide to do so. Forms can also include open text boxes to allow participants to propose project ideas (section \ref{sec:dec:ideation}) or indicate if they are planning to come as a team (section \ref{sec:dec:team}).

\vspace{0.15cm}
\forceindent \textbf{Promoting and advertising} an event is typically done by disseminating the previously discussed website to potentially interested individuals through various channels such as processional networks, mailing lists, student groups, university departments, personal networks, and social media such as LinkedIn groups depending on the target audience~\cite{mlh:promote}. The organizers could also send direct email invitations to individuals who might be interested in taking part in the hackathon as participants or who might promote the event to potentially interested individuals.

\vspace{0.15cm}
\forceindent While \textbf{open selection} of participants is often preferred because it allows every registrant to take part in the event, some hackathon organizers decide to \textbf{carefully select participants}~\cite{hackweektoolkit:targetaudience}. Reasons for selection might be constraints such as the maximum capacity of a venue, funding, etc. as well as to broaden participation from various communities. As described earlier, the choice of the strategy is very much dependent on the event. For example, if the goal of the event is to broaden participation in a specific software development community and most registrants are predominantly from a single institution or background, the organizers might want to consider deploying other strategies to broaden their reach. For hackathons focused on learning or innovation, the organizers might consider a mix of more and less experienced participants. Likewise, hackathons that aim to foster entrepreneurship or the development of sustainable artifacts might want to attract participants from diverse backgrounds and expertise~\cite{nolte2020what}. One approach that seems promising is organizing a mini-hackathon with participants from the community that they wish to attract prior to the main event. This approach has been successfully deployed by the She Innovates~\cite{sheinnovates} hackathon. This is an all-women hackathon that aims to get women participants familiarized with the hacking process before they move on to events with more diverse participants.

\vspace{0.15cm}
\forceindent Moreover, \textbf{participant recruitment and selection should start as early as possible}, at best right after the event theme and goals are formulated. This gives organizers more time to adjust their recruitment strategy when needed, increasing the chance of attracting a sufficient number of participants that fit their desired profile. For large events, using an online tool can be helpful for the selection process. Entrofy~\cite{entrofy} is an example of such a tool that allows organizers to extract a subset of registered participants based on certain attributes, e.g., gender, career stage, etc., and given value (i.e., the percentage of distribution for each attribute).

\subsubsection{Trade-offs}
\begin{enumerate}[leftmargin=0.5cm]
    \item \textbf{Open vs selective recruitment:} An open recruitment strategy is advisable for hackathons that aim to facilitate networking among participants or hope that they would find new collaborators for future work. However, if the goal of the hackathon is to have a more concrete outcome, e.g., creating a working prototype or learning a specific tool, selective recruitment may help bring in participants who are most able to contribute or benefit. Selective recruitment helps organizers to diversify participation or select a desired mix of skills, abilities and prior experience that best serve the purpose of the hackathon. This, in turn, helps create teams with the skills and expertise to achieve their project goals, which can then foster long-term project continuation~\cite{nolte2020what}. For hackathons with selective recruitment, it is important to make the selection process as transparent as possible, e.g. by letting potential participants know how the selection process works, how participants are selected, if there are more eligible participants for each category, or which attributes are given greater weight than others, etc. This helps ensure that even if they are not selected, participants might feel the fairness of the selection process and not feel discouraged to participate in similar events.
    \item \textbf{Open vs closed hackathon:} In a closed hackathon, only members of a particular organization, group, or community are eligible to participate. This allows the organizers and participants to freely discuss and work on non-public topics or on topics that require significant inside knowledge. It also makes it easier for teams to coordinate their work since members of the same organization, group, or community often share a culture, technical language, and role expectations. On the other hand, open participation allows a much broader mixture of people from different backgrounds, domains, and areas of expertise, which facilitates innovation, learning, and community building.
\end{enumerate}

\subsubsection{Modes of participation} 
\begin{itemize}[leftmargin=0.5cm]
    \item \textbf{In-person:} In-person-only participation is most suitable when the aim is to target specific local communities or institutions. However, the in-person format can limit participant recruitment diversity~\cite{powell2021organizing,paganini2023opportunities}. In-person hackathons may unintentionally perpetuate gender imbalances or exclusionary environments, especially in STEM-focused events~\cite{powell2021organizing,paganini2023opportunities}. 
    
    \forceindent Participants from underrepresented groups may feel unsafe or unsupported in in-person settings, particularly if the event lacks gender balance or diversity~\cite{paganini2023opportunities}. Additionally, participants who are new to hackathons may feel intimidated by the high-energy atmosphere of some in-person events, potentially leading to lower engagement or dropout~\cite{powell2021organizing}.
    
    \forceindent To address such issues, organizers could consider actively creating and advertising intentional safe spaces for underrepresented groups by incorporating gender-balanced teams, women-only mentorship, or separate recruitment drives~\cite{paganini2023opportunities}.
    
    \item \textbf{Online:} Online hackathons allow for recruiting a broader and more diverse participant pool, particularly when the goal is to engage individuals from various regions or backgrounds~\cite{mendes2022socio}. Recruitment for online participation can be particularly effective in attracting underrepresented groups, such as women, as it can alleviate traditional barriers like male-dominated physical spaces, personal safety concerns, and flexibility to attend~\cite{paganini2023opportunities}.

    \forceindent However, online recruitment comes with its own challenges. Online hackathons often face higher rates of non-attendance, as participants may register but fail to show up due to competing personal obligations or the lack of a physical commitment~\cite{mendes2022socio,schulten2022participants}. The lack of a physical, immersive experience may also deter participants -- who are seeking hands-on engagement or face-to-face networking opportunities -- from attending. 

    \forceindent To alleviate these issues, organizers could implement structured pre-event engagement strategies, such as onboarding sessions, to familiarize participants with the opportunities of an event while offering tangible incentives, such as certificates or prizes. Maintaining consistent communication can also help participants feel invested in the event and motivated to attend~\cite{paganini2023opportunities,gama2021online}.

    \item \textbf{Hybrid:} Hybrid hackathons offer the flexibility to recruit participants from both local and global pools. This format allows organizers to leverage the diversity and scale of online events while retaining the engagement and networking opportunities of in-person formats. The option to participate online might also be beneficial for individuals who are located in close proximity to where the hackathon takes place but instead prefer the convenience and flexibility to work from home to, e.g., to avoid taking public transportation~\cite{gama2023comfort}.

    \forceindent However, hybrid recruitment presents unique challenges. Over-reliance on one modality could diminish the intended hybrid benefits. Remote participants, in particular, may feel left out compared to their in-person counterparts, particularly if organizers fail to provide equitable opportunities for both modalities. For instance, remote participants may lack the same opportunities to interact with judges and sponsors.

    \forceindent Hybrid recruitment requires careful planning to ensure a balanced participant pool across modalities through specialized preparation (section~\ref{sec:dec:spec}) for remote participants.
\end{itemize}

\subsection{Specialized preparation}
\label{sec:dec:spec}
Organizers might want to run a hackathon related to a specific theme (section \ref{sec:dec:theme}), in a specific domain, or utilize specific software and hardware during their event that are not commonly available to participants. Such events thus potentially require the organizers to provide training and access to licenses or hardware so that participants can work on projects during this hackathon. Moreover, online or hybrid hackathons require a suitable technical setup to function.

\subsubsection{When?}
Preparation activities can be done remotely, onsite, or both. If the theme (section \ref{sec:dec:theme}) and goals (section \ref{sec:dec:goal}) will likely require specialized technical knowledge (e.g., particular tools, languages, or frameworks) or domain knowledge (e.g., community needs, or a scientific field) it is important to develop ways to bring participants up to speed before (usually 1 to 3 weeks) or very early during the event. The organizers may also want to facilitate team meetings if teams are formed in advance (section \ref{sec:dec:team}) so that they can discuss project scope and plan, assign tasks, and experiment with technologies to be used during the hackathon. Assuming participants have free time and sufficient motivation, this can help the work move along more quickly when the hackathon begins.

\subsubsection{Who?}
The organizers should work with mentors (sections \ref{sec:dec:team} and \ref{sec:dec:mentor},~\cite{nolte2020support}) to coordinate training programs. Mentors will, in fact, often be running those programs since they are typically chosen based on their expertise. For hackathons, when team formation (section \ref{sec:dec:team}) occurs in advance, it is advisable that the team leaders are chosen so they can organize team meetings. Organizers can also encourage projected participants to prepare for the hackathon by, e.g., setting up a code base in advance~\cite{nolte2020what} or studying technologies that they might want to use for their project.

\subsubsection{How?}
Organizers first need to identify \textbf{what technologies and topics are necessary} for people to participate in their event and to what extent participants should know about them before attending. One common way to help achieve this is for organizers to arrange training programs in which participants are taught specific technologies or domain knowledge that they would need to use at the event. These programs can consist of webinars developed by the organizers, datasets or software provided by stakeholders, or pointers to existing resources that are available, e.g., on Youtube, or the Coding Academy website~\cite{codecademy} and that projected participants can use to prepare themselves~\cite{nolte2020what}.

\vspace{0.15cm}
\forceindent The organizers should ensure that the tutorial materials are \textbf{accessible} to all participants. This typically includes posting them on the hackathon website, the collaboration platform through which the hackathon is organized, e.g., GitHub, or other document-sharing tools, e.g., GoogleDrive.

\vspace{0.15cm}
\forceindent \textbf{Tutorials} are often delivered as pre-recorded videos with interactive Q\&A at scheduled times before the hackathon. If the training, for example, is related to the configuration of the development environment, participants can watch a pre-recorded video, replicate the steps shown in the video, and communicate with trainers during the Q\&A session and/or via emails. Alternatively, tutorials can be delivered live by a mentor (section \ref{sec:dec:mentor},~\cite{nolte2020support}), which guides a group of participants through activities interactively. The EasterHack event highlighted such pre-event training~\cite{braune2021interdisciplinary}. 

\vspace{0.15cm}
\forceindent For groups of larger size, it might be advisable to form smaller subgroups of perhaps 5 to 10 participants, each guided by one mentor. In practice, live tutorials can present a scheduling challenge, as it might not always be possible to find a common time for all participants, particularly when they are geographically distributed. Moreover, tutorials may also need to be customized based on the participants’ skill levels, e.g., novices need foundational knowledge first before learning advanced skills, while experienced participants might want to skip such basics (section \ref{sec:dec:part}). In such situations, it is advisable that organizers cluster participants into groups of similar skills and provide appropriate materials for each group. The World of Code (WoC) hackathon~\cite{worldofcode} is an example where mentors trained participants in small group tutorials in real-time a few days before the event in dedicated online training sessions.

\vspace{0.15cm}
\forceindent For hackathons that wish to train participants during the event, we have observed two approaches that can be effective. One related approach is \textbf{train to hack} as done by Astro Hack Week~\cite{astrohackweek} and Geo Hack Week~\cite{geohackweek}. In this approach, participants spend most of their time hacking while also spending a considerable amount of time (e.g., 25-50\%) on training particular skills and domain knowledge required to conduct research work afterward. The second approach is participants \textbf{alternating between hacking and training}. BrainHack 2018 of the Organization for Human Brain Mapping (OHBM) in Singapore~\cite{brainhack} is an example where participants could switch between hacking and attending training sessions during a concurrent conference track.

\vspace{0.15cm}
\forceindent In case the hackathon involves \textbf{specialized hardware}, the organizers might want to ensure that it arrives at the hackathon site or is sent to participants early so that it can be set up before the event begins. Moreover, organizers might want to ensure that a specialist who can help with technical issues is available before and during the event.

\vspace{0.15cm}
\forceindent Additionally, hackathons -- in particular those that involve online participants -- often utilize a mix of tools and communication platforms, such as Discord, Slack, Zoom, and GitHub, to communicate, share information, submit projects for competition, etc. This can create make it difficult for participants to navigate between them, leading to feelings of being overwhelmed~\cite{bertello2022open}. To alleviate this issue, organizers can create a centralized platform that serves as a unified hub for communication, agendas, updates, and notifications and as a starting point for participants if they get lost.

\subsubsection{Trade-offs}
\begin{enumerate}[leftmargin=0.5cm]
    \item \textbf{Pre-recorded videos vs real-time training:} Pre-recorded training videos resemble a traditional mode of instruction that offers limited interaction between the participants and trainers. While questions can be addressed in a live session after the training, spontaneous adaptation of the training to attain better learning outcomes is not easily achievable. Adaptation and customization are possible in real-time interactive training, as human trainers are aware of the difficulties that participants are experiencing and can quickly act to mitigate such difficulties. The latter, however, is not always feasible for larger groups.
    \item \textbf{Training before vs training during the hackathon:} Pre-event training permits more hack time as opposed to training at the event, which requires participants to split their time between hacking and training. Pre-event training, however, demands participants’ willingness to spend some time for training before the event. In contrast, real-time settings demand the concurrent presence of both trainers and participants, which can present scheduling problems.
\end{enumerate}

\subsubsection{Modes of participation}
\begin{itemize}[leftmargin=0.5cm]
    \item \textbf{In-person:} In-person hackathons allow for organizers, mentors, and support staff to interact directly with participants and help them should they encounter any issues with specialized tools, hardware, and other resources. Organizers can also arrange dedicated spaces for hands-on tutorials, such as prototyping labs. Depending on the size of an event or the size of an event space, it might be difficult for organizers to retain an overview and send help to teams in need. In cases like this, it might be helpful to provide technical systems where participants can ask for help, even during an in-person event.

    \forceindent Pre-event workshops conducted at the venue can offer personalized guidance and reduce the intimidation factor, particularly for participants new to hackathons. Such strategies are particularly useful for affirmative actions aiming to attract participants from underrepresented groups (e.g., women, transgender people)~\cite{paganini2020engaging, prado2020trans}.

    \item \textbf{Online:} Online hackathons require careful attention to participant workspace environments. While some participants may have dedicated quiet workspaces, others may struggle with distractions from their surroundings, making concentration easier or more complex, depending on their environment. Organizers should encourage participants to set up ergonomic and quiet workspaces, provide recommendations for noise-canceling tools, and suggest co-working spaces as alternatives for those who may be facing distractions by their surroundings.
    
    \forceindent Online hackathons also typically use multiple tools such as GitHub, Discord, Slack, or Zoom as their digital infrastructure. While these tools are essential for collaboration, their simultaneous use can overwhelm participants, especially those unfamiliar with some platforms~\cite{schulten2022participants}. Organizers can alleviate this issue by prioritizing the use of a centralized platform that consolidates key functionalities such as communication, task management, and updates. For example, the BookDash hackathon utilized Slack as the primary communication platform~\cite{bookdash2024} while the World of Code hackathon utilized a dedicated Discord server for communication combined with a centralized GitHub page that contained links to all other technologies and provided a schedule and contact point for participants to reach the organizers~\cite{worldofcode}. Organizers can reduce the learning curve, foster better engagement, and ensure smoother collaboration in online hackathons by minimizing the number of tools participants need to navigate.

    \forceindent Additionally, well-designed pre-event preparation events such as training webinars or pre-hackathon technical check-ins can help familiarize participants with the platforms they will use during the hackathon. These webinars can be delivered synchronously through live sessions or asynchronously as recorded tutorials~\cite{paganini2023opportunities}, catering to participants in different time zones.
    
    \forceindent For hackathons requiring specialized hardware, organizers need to make sure that it reaches participants in time and that there are guides and technical support available for them to set it up. Clear, centralized technical instructions are essential to reduce onboarding-related frustration, which can lead to disengagement~\cite{schulten2022participants,affia2022integrating}. 
    
    \forceindent Online participants also may face barriers such as poor internet connectivity or incompatible software, which could hinder their participation. While organizers have limited influence on issues related to connectivity, they can provide resources for participants to buy short-term fast Internet connections or rent a desk at a co-working space. Moreover, they can provide technical support through dedicated tools or platforms, as used in EUvsVirus~\cite{bertello2022open}.
    
    \item \textbf{Hybrid:} Hybrid hackathons face the challenge of connecting physical and virtual participants. One aspect that organizers need to consider is the suitability of the on-site setup is suitable. This includes rooms with suitable hardware to hold conference calls. This hardware should include, at best, a dedicated computer that can run video conferencing software such as Zoom or MS Teams, and that is connected to microphones that pick up audio everywhere in the room, cameras that ensure that in-person participants are visible to online participants, and audio and projection hardware which maker online participants visible and audible in the room. In addition, it is advisable to have a dedicated camera and microphone for the speaker and a camera that shows the projection surface so that online participants can be aware of how they are represented in the room. If a room like this is not available, it is also possible to utilize mobile options such as a meeting owl for smaller groups.
    
    \forceindent Another issue that organizers of hybrid meetings have to deal with is noise. Larger halls in which many on-site participants are located can create sufficient background noise that makes it hard or impossible for online participants to engage or even be aware of activities that are taking place on-site. To address this issue, organizers consider providing online participants with volume-controlled audio feeds to filter excessive background noise and improve clarity. Additionally, assigning remote facilitators to act as liaisons can help online participants keep virtual attendees informed and engaged with the physical space. Moreover, organizers should provide separate spaces, such as breakout rooms for hybrid teams that consist of in-person and online participants.
    
    \forceindent Organizers should also consider holding onboarding sessions tailored to each format to ensure all participants are adequately prepared. For instance, local participants can attend in-person sessions to familiarize themselves with on-site tools and resources. In contrast, online participants can join remote sessions and access shared virtual resources. Organizers should also encourage online participants to set up ergonomic and quiet workspaces, as discussed before. In addition, they can provide recorded training videos, guides, and FAQs ahead of time, allowing participants to prepare at their own pace regardless of timezone constraints~\cite{schulten2022participants,braune2021interdisciplinary}. 

    \forceindent Hybrid events also face challenges related to resource parity. Remote participants may feel disadvantaged if they lack access to the specialized hardware or materials available to on-site participants. Organizers can address this by providing remote participants with access to necessary resources (e.g., software licenses, virtual prototyping tools, shipped hardware), providing clear setup instructions and virtual support to attempt to level the playing field with in-person participants, although this is not entirely possible~\cite{paganini2023opportunities,bertello2022open}.
    Technological issues in hybrid events often disproportionately affect online participants. Organizers can mitigate this by providing suitable technical support. 
\end{itemize}

\subsection{Duration / breaks}
\label{sec:dec:dur}
When organizing a hackathon, organizers have to decide when to start, when to end, and when to take breaks in between. These decisions are crucial because they can influence who would be motivated to come, whether attendees can maintain a high level of motivation, how the event will be perceived, how participants engage with each other beyond working on their projects, and whether or not participants decide to leave while an event is still ongoing.

\subsubsection{When?}
The overall timeline of a hackathon needs to be decided on and announced early during the planning process since it serves as a basis for recruitment material (section \ref{sec:dec:part}) and for peoples’ decision to attend the event. The overall timeline should include dates and times (including start, end, and potential overnight breaks) for each day. Other breaks during the hackathon can potentially be decided on and announced later.

\subsubsection{Who?}
The decision for when a hackathon will take place, how long it should be, and how many breaks it will have is commonly taken by the organizers. For this decision, they can consult projected participants (section \ref{sec:dec:part}), mentors (section \ref{sec:dec:mentor},~\cite{nolte2020support}), and other stakeholders (section \ref{sec:dec:stake}). Including external stakeholders is especially advisable when the hackathon focuses on a specific theme (section \ref{sec:dec:theme}), takes place in a specific domain, or aims to attract participants (section \ref{sec:dec:part}) from backgrounds that the organizers are not particularly familiar with.

\subsubsection{How?}
When thinking about a hackathon, most people will probably think about an event that starts on a Friday afternoon, ends on a Sunday, runs overnight, and has little to no breaks in between with teams just tirelessly hacking away on their project~\cite{g48howitworks}. While this is a common hackathon format, it certainly is not the only one. Organizers can decide for their event to take place at any point during the week and have breaks overnight as well as during the day. When deciding about the timing of their particular event, organizers should take the following aspects into account.

\vspace{0.15cm}
\forceindent They should consider the background of their \textbf{projected participants} (section \ref{sec:dec:part}). While it might be ok for students to participate in an event during the week and stay up overnight, this might not be possible for people who have fixed working times or busy family lives. Corporate events we studied often took place during regular working hours, and participants could choose to go home for the night or stay and continue working~\cite{oneweek, leemet2021utilizing}, while civic events often take place in the evening and can be spread out over multiple weeks with breaks during hacking times to allow for participants to network~\cite{chihacknight}.

\vspace{0.15cm}
\forceindent Another aspect to consider when deciding on the duration of an event is the \textbf{context or domain} (section \ref{sec:dec:theme}) in which an event takes place. In a corporate setting, it might be feasible to focus on regular working hours because relevant stakeholders that can, e.g., serve as mentors or provide thematic input might only be available during certain times. These times can, however, be considerably different, e.g., in a civic context where stakeholders may be more likely to be available after regular working hours.

\vspace{0.15cm}
\forceindent It is also important for organizers to consider their \textbf{goals} (section \ref{sec:dec:goal}) for organizing a hackathon when deciding about when to start, when to end, and when to take breaks in between. If their goal is for teams to develop polished prototypes, they might want teams to focus on their project and thus not take too many breaks to not affect their productivity and rhythm. If the organizers’ goals should, however, be for participants to network, they might want to consider regular breaks during which participants can connect.

\vspace{0.15cm}
\forceindent Breaks can also serve as opportunities for organizers to convene and discuss with mentors (section \ref{sec:dec:mentor},~\cite{nolte2020support}) and stakeholders (section \ref{sec:dec:stake}) and potentially alter the course of an event. For example, during a community hackathon we studied, the organizers took time during a break to sort ideas proposed by participants to structure the following team formation process (section \ref{sec:dec:team},~\cite{worldofcode:schedule}).

\subsubsection{Trade-offs}
\begin{enumerate}[leftmargin=0.5cm]
    \item \textbf{Overnight vs breaks during the night:} The main advantage of organizing a hackathon that takes place overnight is that participants have more time to work on their projects. Working overnight can, however, take a toll in that productivity can be expected to drop during the night and the following morning. Breaks during the night limit the available working time but allow for participants to get some rest and engage with activities beyond the hackathon. To address this trade-off, organizers might consider providing the option to work during the night by, e.g., keeping the venue open but leaving it to the participants whether they would like to take a break. To avoid participants feeling social pressure to work during the night, organizers could also emphasize that it might be helpful to take a break or organize an activity that could reasonably mark the end of the hacking day such as a dinner.
    \item \textbf{Weekend vs during the week:} Organizing a hackathon during a weekend might make it more likely for people to participate since many projected participants can be expected to be busy during the week. It might not be advisable, though, to organize a hackathon for corporate employees during a weekend. Organizing a hackathon during the week might also provide access to stakeholders that will not be available during the weekend. To address this tradeoff, organizers could utilize a mixed format where the start of the hackathon is during the week and it ends during the weekend. This would allow participants to access stakeholders during the first crucial phases of a hackathon when ideas are formed.
    \item \textbf{Short vs long:} Deciding on the overall duration of a hackathon can be difficult. An event needs to be long enough for participants to be able to make progress on their projects, but it should not drag on endlessly because participants might lose interest. Moreover, short hackathons might also force teams to quickly decide on a project direction and start working early, while longer hackathons might lead to teams continuing discussions, which can compromise their artifact. Deciding too quickly might, however, also lead to teams working on solutions without understanding the problem they aim to address. Generally, it is not advisable, though, to drag an event on for too long. One of the characteristics of a hackathon, after all, is that it takes place over a limited time span. An overall hacking time of about 48 hours divided over multiple days has proved to be a good rule of thumb for in-person events. Online events typically need more time, as we will discuss below.
    \item \textbf{Time for work vs time for breaks:} Depending on the goals of the organizers, it might be advisable to organize multiple breaks during each day for participants to be able to get away from hacking and socialize. Having many breaks will, however, cut into the time participants will have for their projects and might leave participants frustrated because they did not make sufficient progress. To address this tradeoff, organizers could use breaks such as breakfast, lunch, or dinner for participants to socialize.
\end{enumerate}

\subsubsection{Modes of participation}
\begin{itemize}[leftmargin=0.5cm]
    \item \textbf{In-person:} In-person hackathons immerse participants fully in the event, with fewer external distractions compared to online formats. Organizers often structure such events as intense 24- to 48-hour~\cite{g48howitworks} sessions or spread them across several days with set working hours~\cite{oneweek}.

    \forceindent In addition to scheduling breaks, organizers can provide dedicated relaxation spaces at the venue, such as quiet rooms or lounge areas, to help combat fatigue~\cite{falk2024role}. Planned social interactions through playful activities such as karaoke and games are also good strategies~\cite{paganini2020engaging}, especially if networking is one of the goals (Section \ref{sec:dec:goal}). It is vital to strike a balance, as extended work sessions without adequate rest can lead to burnout and diminished performance.

    \forceindent For participants who wish to continue working overnight, organizers also need to ensure the venue is accessible and provide safe alternatives, such as secure overnight spaces or nearby accommodations.
    
    \item \textbf{Online:} Online hackathons are typically longer than in-person events and often last for more than 3 days with staggered working hours. This format is necessary because activities such as ideation and team formation simply take longer to organize in an online setting. Participants often also face interruptions due to work, school, or personal commitments, which would likely happen less frequently if they were physically present at a hackathon location. Moreover, participants might feel inclined to attempt to join a hackathon despite them being busy otherwise because online participation is more flexible.

    \forceindent Maintaining engagement is more difficult in an online setting. Organizers can foster engagement during extended online events by encouraging structured work sessions, such as 2-4 focused hours per day, spread over several days or weeks~\cite{braune2021interdisciplinary}. Frequent scheduled breaks are essential to mitigate screen fatigue and disengagement. Organizers should consider sending notifications and reminders on communication platforms to prompt participants to take a break. It is also important to distinguish between off-screen breaks where participants can disengage and relax and breaks which are planned as opportunities to connect~\cite{mendes2022socio,paganini2023opportunities}.

    \item \textbf{Hybrid:} Hybrid hackathons face similar issues to online events. They generally need to be longer, online participants might be interrupted and disengage due to various reasons, and time-zone differences might make it hard for teams to collaborate and for participants to join common activities such as checkpoints and presentations. Organizers should thus consider structuring schedules to allow participants to choose their preferred level of engagement -- whether fully remote, in-person, or a mix of both. Parallel schedules with separate activities, such as virtual networking for online participants and in-person games or workshops, can provide meaningful breaks for each group~\cite{mendes2022socio}. 

    \forceindent Focused work sessions of 2-4 hours with defined start and end times can ensure equity in workload across modalities, preventing in-person participants from feeling overburdened and online participants from disengaging. Events like BookDash~\cite{bookdash2024} have implemented collaboration sessions at set times to accommodate diverse time zones and participant preferences.

    \forceindent Organizers should also synchronize key activities, such as presentations or judging, across both formats. Hybrid schedules should provide remote participants with the option to follow staggered work sessions while ensuring critical moments of the hackathon, such as checkpoints and presentations (Section \ref{sec:dec:agenda}) remain synchronous for all participants, fostering a sense of shared purpose and equity.
\end{itemize}

\subsection{Ideation}
\label{sec:dec:ideation}
One of the main motivations for individuals to attend a hackathon is the prospect of working on an exciting project. It is thus crucial for organizers to think about how to support participants to come up with interesting and attainable project ideas they can work on during a hackathon. There is also evidence that some ideation approaches, such as traditional brainstorming, can help self-identified minorities feel more welcome and their ideas more accepted during the event~\cite{filippova2017diversity}.

\subsubsection{When?}
Ideation typically takes place before the hackathon, but organizers can also plan for a dedicated ideation session at the beginning of the event itself.

\subsubsection{Who?}
Participants typically propose their own ideas. Especially for ideation during the event, trained facilitators can help the participants generate ideas efficiently and harmoniously. It is also possible to guide ideation towards a certain direction that appears feasible and useful to organizers and / or connected stakeholders (section \ref{sec:dec:stake}).

\subsubsection{How?}
The most common ideation approach is for participants to develop ideas that are related to the \textbf{theme(s) of a hackathon} (section \ref{sec:dec:theme}). Hackathon themes are often intentionally broad covering areas such as civic technologies~\cite{civichackingday}, environmental sustainability~\cite{greenhack}, entrepreneurship~\cite{nolte2019touched} and others to allow for a large variety of ideas to fit under their banner. Organizers and stakeholders can also decide to narrow the scope of potential ideas by proposing specific problems (areas) that participants should address. This approach is suitable for targeted events (section \ref{sec:dec:goal}) that, e.g., aim to develop technologies for a specific company~\cite{leemet2021utilizing} or community~\cite{pe2019understanding}. It is important to leave space for participants to develop ideas that are of interest to them, to ensure their motivation to participate.

\vspace{0.15cm}
\forceindent Ideation can take place \textbf{before or during a hackathon}. For larger audiences, it might be advisable to collect ideas prior to an event using technologies such as Google Docs or GitHub issues. It is important to use technologies that projected participants are familiar with. Collecting ideas through such technologies not only allows participants to describe their ideas but also enables others to comment, provide feedback, and express interest. They also allow organizers and stakeholders to pre-screen ideas, adjust their ideation approach, and capture ideas for future use beyond the context of a particular hackathon (section \ref{sec:dec:continuity}). 

\vspace{0.15cm}
\forceindent This can reduce the logistical burden of facilitating live brainstorming sessions during the event and provide a structured starting point for collaboration. 

\vspace{0.15cm}
\forceindent Conducting a separate ideation session \textbf{at the beginning of a hackathon} using common approaches~\cite{csikszentmihalyi1996flow} such as brainstorming~\cite{brainstorming} might lead to more interaction between participants, organizers, and mentors (section \ref{sec:dec:mentor}) and foster ideation. By familiarizing themselves with submitted ideas beforehand, participants can build common ground and begin exploring the resources required to execute those ideas. This approach creates a diverse pool of potential projects, providing an opportunity for participants -- often unfamiliar with each other -- to connect during team formation (section~\ref{sec:dec:team}). It also allows organizers to guide ideation by asking targeted questions~\cite{brainstorming:question} and clustering ideas, e.g., based on participant interest. Ideation during a hackathon does take away time for hacking, though. Conducting ideation sessions during an event and sharing ideas might also prove not to be feasible if a hackathon is too large.

\vspace{0.15cm}
\forceindent Collecting a \textbf{sufficient number of interesting ideas} is crucial for a successful hackathon because ideas usually are the basis for team formation (section \ref{sec:dec:team}). Collecting many interesting ideas is, however, not the only aspect to consider during ideation. While being challenging enough to be interesting for participants to attempt and potentially continue after an event~\cite{nolte2020what}, ideas should also be attainable. This means that they need to be doable during the short duration of a hackathon, that the team that attempts them has or can quickly attain the skills required to complete a project based on that idea, and that there are sufficient resources available at the hackathon, including, for example, specialized hardware, licenses, cloud resources, or others (section \ref{sec:dec:spec}). Organizers, stakeholders, and mentors can support teams in selecting suitable ideas and help them scope their project during the hackathon. Organizers should also support this process by offering clear frameworks, such as ideation templates, and scheduling regular check-ins. The HackHPC~\cite{hackhpc} and World of Code~\cite{worldofcode} hackathons, for example, provide instructions for each checkpoint which guide participants through the process from ideation to scoping and development to ensure that teams do not get stuck ideating and discussing. 

\vspace{0.15cm}
\forceindent Finally, it is important to consider that \textbf{some ideas might be extremely popular while others do not draw much attention}. If some ideas prove extremely popular, the idea can sometimes be split into parts, or several teams can be formed to pursue the same idea. If ideas are not popular at all it should be clear to all participants that they will not be attempted during the hackathon. Participants proposing ideas thus have to be prepared to let go of their idea and potentially join a different team and work on something else.

\subsubsection{Trade-offs}
\begin{enumerate}[leftmargin=0.5cm]
    \item \textbf{Priming vs open ideation:} Leaving ideation completely open and in the hands of the participants can lead to them coming up with ideas that are not, or only marginally, related to the goal of the hackathon, or with ideas that are not doable due to other constraints imposed by the setup of the event (time, specialized resources, available skills, etc.). If the primary goal is just to have fun or some basic exposure to coding and its possibilities, this may be fine. Imposing strict limitations on the ideation process by, e.g., limiting participants to address specific challenges proposed by organizers or stakeholders can in turn negatively affect the motivation of participants to attend an event and take on the proposed challenges. To address this tradeoff, it is thus advisable to always leave room for participants to develop their own ideas even when proposing challenges.
    \item \textbf{Individual ideation vs group ideation:} This tradeoff is common for most creativity techniques. Asking participants to develop ideas individually and share them after ideation has ended typically leads to more diverse ideas since people tend to follow the direction of ideas that have already been proposed. Some participants might, however, also benefit from others sharing their ideas because it can foster their imagination. One way of dealing with this tradeoff is to take a two-step approach by asking participants to submit individual ideas prior to the hackathon and then sharing them at the beginning of the event allowing other ideas to be added. Moreover, posing multiple (potentially contradicting) ideation themes might also help participants to come up with diverse ideas.
    \item \textbf{Time for ideation vs time to hack:} Conducting the ideation at the beginning of or during a hackathon provides organizers with an opportunity to steer its direction, emphasize ideas that they perceive to be best related to their goals, and allow participants to develop additional related ideas. It does, however, also cut into the time that remains for hacking. This trade-off becomes more problematic for larger hackathons because each participant should have the chance to propose ideas to keep the morale up which might not be possible at larger events. For larger events, it is advisable to ask participants to develop -- and potentially send -- ideas before the event. Ideation prior to an event can also allow participants to familiarize themselves with the idea, create common ground, and start learning about potentially required technologies~\cite{affia2020developing,nolte2018you}.
    \item \textbf{Too large vs too small:} Ideas should be interesting and challenging but at the same time doable during the short duration of a hackathon. One approach to deal with this trade-off would be to let participants propose wild ideas first that can then be scaled down to doable projects through mentoring. In order to avoid mismatched expectations, it is important for participants to be gently encouraged to be realistic about what they can hope to accomplish during an event. Prototypes where only a few example features are implemented simply and difficult technical challenges such as analyzing substantial data sets or developing APIs are simply faked. Such compromises are common and often necessary. Organizers can provide guidance in the form of checkpoints as discussed before.
\end{enumerate}

\subsubsection{Modes of participation}

\begin{itemize}[leftmargin=0.5cm]
    \item \textbf{In-person:} For in-person events, organizers can create physical spaces conducive to ideation interactions (e.g., breakout rooms, whiteboards, and lounge areas)~\cite{mendes2022socio}. Organizers can also provide tangible resources like sticky notes, flipcharts, whiteboards, and prototypes~\cite{gama2021online}. 

    \forceindent Ideation during an in-person event can also prove to be challenging, though. Time constraints are necessary but can also result in rushed or incomplete idea development when designed too strictly. Moreover, dominant or outspoken personalities might overshadow quieter participants which can limit diverse perspectives and contributions. Organizers can address these issues by structuring in-person ideation sessions with clear guidelines and equitable participation strategies, such as rotating speaking opportunities or using moderated discussions. Organizers can also implement pre-event ideation sessions to reduce the logistical effort of in-person ideation and cater to participants who may feel overwhelmed by the intensity of in-person brainstorming. 
    
    \item \textbf{Online:} To foster ideation for an online event, organizers could consider supporting pre-event ideation activities. Digital platforms such as Miro, MURAL, or Google Jamboard~\cite{mendes2022socio} can support collaborative ideation processes, offering shared virtual spaces where participants can brainstorm and contribute ideas asynchronously at their convenience~\cite{gama2021online}. To avoid confusion, organizers should provide instructions and training for these tools before participants are expected to use them during a hackathon.
 
    \forceindent Furthermore, organizers should adopt ideation strategies designed to foster inclusivity and level the playing field for all participants. This includes creating mechanisms to actively encourage contributions from participants who may feel less confident in speaking up during virtual sessions~\cite{paganini2023opportunities,gama2021online}. For example, organizers can facilitate anonymous idea submissions, small breakout discussions, or use structured prompts to ensure equitable engagement across diverse participant groups.

    \forceindent Many online events avoid supporting ideation before and during an event and simply require participants to submit ideas when they register. This approach maximizes hacking time during an event and can help organizers pre-screen ideas, but it can also limit cross-pollination of ideas and prevent individuals from registering who do not have a specific idea before joining an event.

    \item \textbf{Hybrid:} Hybrid formats can benefit from targeted ideation sessions that cater to both modalities if ideation occurs during the hackathon event. Examples are synchronous ideation sessions (e.g., live brainstorming over Zoom) with asynchronous input (e.g., shared digital documents).

    \forceindent Organizers should be aware, thought, that in-person participants will likely have an advantage over online participants due to the immediacy of physical interactions, while remote participants may find it harder to contribute without strong facilitation. One approach to mitigate such inequalities might be to ask both in-person and online participants to submit ideas through technical means such as shared virtual whiteboards or documents (e.g., Miro, Google Docs). Parallel ideation sessions -- such as live brainstorming for in-person participants and asynchronous contributions for online participants -- can ensure all voices are heard. 
\end{itemize}

\subsection{Team formation}
\label{sec:dec:team}
Another important decision for organizers is selecting an appropriate strategy for selecting projects and forming teams. Teams are typically formed from the recruited participant pool (section \ref{sec:dec:part}), around projects of interest to them.

\subsubsection{When?}
Participant recruitment (section \ref{sec:dec:part}) and ideation (section \ref{sec:dec:ideation}) are prerequisites for the team formation process. Team formation and project selection can take place either \textbf{before the event} or \textbf{at the beginning of the event}. Each has its advantages and disadvantages, as we will discuss in the trade-offs below. Even if the intent is to choose teams and projects at the beginning of the event, organizers should expect that some participants may join the hackathon as a team with firm ideas about what they want to work on and with whom.

\subsubsection{Who?}
There are three roles involved in the team formation and project selection process. These roles are project \textbf{proposers}, \textbf{moderators}, and \textbf{joiners}. The proposer refers to someone who pitches a project idea at the event. This role can be taken by participants, organizers, or stakeholders (section \ref{sec:dec:stake}). The joiners are participants who select the project they are interested in, sometimes also selecting a role that they would like to play at the event. For example, during Microsoft’s OneWeek Hackathon~\cite{pe2022corporate}, project proposers specified roles required for their proposed projects, and other participants joined the project teams by taking one of these roles. The organizer or a dedicated person takes the role of moderator who facilitates the team formation process in order to configure project teams with skills, expertise, background, and reasonable size required to complete the projects they aim to work on. For hackathons at scale, it is important to assist the moderator with a tool that facilitates the matching of participants and projects.

\subsubsection{How?}
In order to successfully form teams with skills required to complete the projects proposed at a hackathon, it is often helpful to have a diverse participant pool (section \ref{sec:dec:part}). Organizers try to attract suitable participants as part of the participant selection and recruitment process, which has to be completed before team formation.

\vspace{0.15cm}
\forceindent Teams can be formed either by \textbf{open selection}, \textbf{assignment}, or a \textbf{hybrid} strategy. In open selection, participants select projects and roles that they want to play based on their interest from the list of all available projects and roles. In the assignment strategy, a mediator assigns projects and roles to participants. The hybrid strategy narrows down the participant’s search space by filtering out projects and roles that seem to be of lesser interest to participants or that they might be less qualified for. For assignment and hybrid strategy, it is important to gather participants’ needs and expectations beforehand to optimize team formation. Information like that is typically collected through the registration process as part of participant recruitment.

\vspace{0.15cm}
\forceindent Forming teams \textbf{before a hackathon} requires suitable online tools such as Google Docs, Google Forms, or GitHub issues. Suitable tools need to support project listing and sign-up. For example, in the STScI hack days~\cite{pe2019understanding} we observed that Google Forms were used to collect project preferences and skills. Based on this information, the organizers configured teams of 3 to 6 participants with a good mix of skills. Some hackathons, e.g., Steelhacks~\cite{steelhacks}, suggest participants to form teams of 5. These tools work well for smaller events (say, 50 or fewer participants), but a more sophisticated tool would be required for larger scale events such as Microsoft’s OneWeek Hackathon. They deployed the online tool HackBox, which allowed participants to create project proposals, sign up for projects and search for additional members with specific skills or interests (the tool is now defunct). For the HackOhio hackathon~\cite{hackohio2023}, the organizers used dedicated discord channels for pre-event introductions and matchmaking before the event.

\vspace{0.15cm}
\forceindent Another common practice is for teams to form \textbf{at the beginning of a hackathon} (section \ref{sec:dec:agenda}). This process needs to be fairly efficient so that teams will have sufficient time to actually work on their project. One common approach is to allow participants that have a project idea to pitch it in front of the other participants
%and write each idea down on a flip chart or whiteboard
(section \ref{sec:dec:ideation}). The remaining participants are then given some time to chat with the project proposers and select a project they would like to work on. It is common to aim for teams of similar size, between 3 and 6 members. It is particularly important for competitive hackathons (section \ref{sec:dec:comp}) to have teams of similar size since a large difference between team sizes could be perceived as an unfair disadvantage. Moreover, team size can also affect what a team is able to produce. If the goal of an event is innovation (section \ref{sec:dec:goal}), then larger teams might be preferred~\cite{falk2025creativity}, but larger teams typically also require additional coordination effort, which can limit the time a team has to actually work on their project. It might thus be advisable to split up large teams, and ideas that do not draw much interest are generally abandoned. In another small-scale hackathon we observed, organizers asked the participants to rank proposed projects in order of preference in GoogleDocs and participants were assigned to the project on a first-come-first-serve basis. This process might not be feasible for events at scale and using a tool like the aforementioned HackBox or Discord would be necessary even if team formation occurred at the beginning of the hackathon.

\vspace{0.15cm}
\forceindent For hackathons where teams are formed at the event, it is sometimes desirable for organizers to propose projects and either post descriptions in advance, develop brief pitches and make them available to participants as videos, or describe them at the beginning of an event. This approach is particularly useful when the goal (section \ref{sec:dec:goal}) of a hackathon is, for example, to introduce newcomers to a particular domain, tool set, or scientific community. In these cases, it is very difficult for the projected participants to develop feasible and appropriate project ideas themselves. 

\subsubsection{Trade-offs}
In the following, we describe a number of trade-offs between various strategies used to configure teams in hackathons. It is important to note here that these trade-offs are not independent, and organizers should consider balancing them when making decisions about team formation and project selection.

\begin{enumerate}[leftmargin=0.5cm]
    \item \textbf{Forming teams before vs at a hackathon:} Pitching projects and even forming teams before an event can help to get the project work underway more quickly at the hackathon itself. However, there are some costs to this approach. Projects proposed before an event, even if there is an opportunity to pitch at the event itself, are likely to be chosen since participants have become familiar with them. They will not have the benefit of being discussed at the hackathon with the potential for cross-fertilization and innovation this can provide. Moreover, if teams are chosen before an event, there is generally very little interaction during the hackathon among participants of different teams. If growing a community and forming broader social networks are important goals, it is generally advisable to pitch and discuss ideas at the hackathon itself.
    \item \textbf{Proposing projects by participants vs by organizers:} Most hackathons allow participants to define their own project ideas, and this is often a primary motivation for participants to do something fun and acquire skills they want. For some hackathons, however, participants are simply not in a good position to formulate projects that are feasible and appropriate for the theme of an event. They may lack technical skills, domain knowledge, or both. In these cases, it is desirable for organizers to propose projects that will help the participants. This approach combines well with pre-event tutorials and dedicated mentors (section \ref{sec:dec:mentor},~\cite{nolte2020support}) because teams will likely need a lot of help to make progress. Since they are not pursuing their own passion, motivation for participation needs to be carefully considered, though, which may consist of things like valuable contacts for their future profession, potential job offers, or developing skills that are in demand. Recruiting materials should lay these benefits out convincingly.
    \item \textbf{Open selection vs assignment:} Open selection of teams is common for hackathons organized around themes (section \ref{sec:dec:theme}) (e.g. particular civic issues, making use of specific data sets, etc.) and for hackathons designed just for fun or for exposure to programming, prototype development, or entrepreneurship. Open selection is beneficial in the sense that it allows participants to choose what they want to work on in contrast to a strict assignment approach that does not consider participants' motivations, needs, and expectations. However, there are some costs associated with open selection. For example, teams may not have members with the skills required to complete a desired project. This can not only lead to frustrations during the hackathon but might also negatively affect the probability of project continuation after an event has ended~\cite{nolte2020what}. %A lack of diversity may also inhibit a team’s ability to create innovative ideas and solutions and again affect project continuation after an event~\cite{nolte2020what}.
    It is often helpful to have webinars and pointers to resources prior to the actual event. This puts participants in a much better position to formulate realistic and on-target project ideas quickly. Another way to alleviate this trade-off is having a balanced -- hybrid -- approach, which could provide participants with choices that are closely related to their goals and expectations.
    \item \textbf{Large vs small teams:} Organizers should expect that even if they try to have teams with reasonable size, teams may be larger than the desirable size of 3 to 6 participants. Larger teams -- especially if they are more homogeneous in terms of their interests and skills -- are likely to create more innovative products~\cite{falk2025creativity}. At the same time, they are more likely to encounter coordination problems than smaller teams. This problem might even be more significant in teams consisting of members who have not collaborated before because they do not have common knowledge about each other’s skills and working practices~\cite{pe2022corporate}, which could lead to them not being able to generate the outcomes they want. Hackathons with open selection are more likely to suffer this problem as no moderation is applied to the team formation. To minimize this issue, the organizers should try to ensure that the teams are of a reasonable size regardless of the team formation strategy they use.
\end{enumerate}

\subsubsection{Modes of participation}
\begin{itemize}[leftmargin=0.5cm]
    \item \textbf{In-person:} In-person participation typically makes it easier for teams to form because they can walk around and talk to each other face-to-face. Participants can directly assess compatibility, skill alignment, and shared interests~\cite{pe2022corporate} in a way that is difficult to replicate in virtual settings. 

    \forceindent However, in-person team formation often occurs under tight time constraints at the start of the event, which can result in rushed decisions and mismatched teams. Large-scale events further exacerbate this challenge, as participants may find it difficult to connect with a wide range of potential teammates.

    \forceindent Shy participants or those new to hackathons may struggle to engage in organic team formation in high-pressure physical settings~\cite{paganini2023opportunities}. Organizers can facilitate structured matchmaking processes, such as skill-sharing boards or moderated discussions, to ensure all participants find compatible teammates. Icebreaker activities and scheduled networking sessions can help reduce social barriers and create a welcoming environment. However, adequate time must be allocated for team formation to prevent participants from feeling pressured or excluded.
    
    \item \textbf{Online:} Online hackathons allow for team formation across geographies. It is important to consider potential time-zone differences, though, since those can make it difficult for teams to collaborate. Participants can connect prior to an event using digital platforms like Slack or Discord. Pre-event team formation can be supported by organizers through dedicated virtual matchmaking sessions~\cite{braune2021interdisciplinary}, such as speed networking or idea pitches. Such sessions can help participants evaluate compatibility and align goals before the event starts. For example, events like EUvsVirus~\cite{bertello2022open} and EasterHack~\cite{braune2021interdisciplinary} leveraged matchmaking sessions to support pre-event team formation. It is common for organizers to ask participants to form teams prior to an event and register as a team. This approach can, however, inhibit cross-fertilization and team interaction at an event, as discussed before.

    \forceindent Organizers might also consider forming teams at the beginning of a hackathon by, e.g., providing dedicated virtual breakout rooms for each proposed idea. The HackHPC events have successfully implemented this strategy~\cite{hackhpc}. Idea proposers can wait in these rooms, and joiners can switch between virtual rooms until they have found a team that they would like to join. It is important to note, though, that compared to in-person team formation, where joiners typically have the option to walk around, switching between virtual rooms takes significantly more time. One approach to alleviate this issue is to start an event in the evening, end the first day with team formation, and ask teams to report back the next morning (section~\cite{hackhpc}). This gives participants sufficient time to see which ideas are available and make their choice without being rushed.
    
    \forceindent In an online setting, it is also more difficult to build rapport due to the lack of in-person cues, and participants with less experience or weaker communication skills may struggle to join teams~\cite{powell2021organizing,gama2018hackathon}. Mentorship and guided matchmaking are essential to support participants with less experience or weaker communication skills (section~\ref{sec:dec:mentor}) and are beneficial in such cases to support team formation~\cite{bertello2022open}. Online ice-breaking sessions, e.g., in the form of quizzes and games during synchronous interactions supported by communication channels such as Discord or Slack can be an additional alternative for team formation in an online setting~\cite{paganini2023opportunities}.

    \item \textbf{Hybrid:} Like in the case of online hackathons, hybrid events allow for team formation across geographical locations, potentially increasing diversity. Time zones again play a role in that a large disparity between team members can make it difficult to collaborate. Hackathon organizers can foster pre-event team formation, as discussed before.

    \forceindent Team formation at the beginning of a hybrid event can be particularly challenging because organizers need to consider both the needs of online and in-person participants. One approach is to collect ideas digitally, e.g., through a digital whiteboard, and project the whiteboard at the hackathon location so that both in-person and online participants can see them (section \ref{sec:dec:ideation}). Organizers can then create virtual breakout rooms based on the proposed ideas and ask at least one person who is present at the hackathon site to join the respective room. This person can be the idea proposer or someone who is interested in the idea and might want to join the team. This approach allows in-person participants to walk around to find ideas they are interested in while online participants can do the same by switching between different breakout rooms. This approach was successfully utilized during one of the World of Code hackathons~\cite{worldofcode}. It does require a large space, though, since multiple teams discussing at the same time with both online and in-person participants can be very distracting and difficult particularly for online participants. The approach also requires organizers to monitor both the in-person space and the online breakout rooms so that no participant gets lost. It is thus advisable to utilize this approach for smaller events and to make sure that multiple organizers are available to facilitate.
    
    \forceindent To make things easier, organizers sometimes enforce in-person and online-only teams. This approach still requires facilitation by multiple organizers, but it alleviates the issue of noise. While this is an understandable strategy given the previous discussion, it also forfeits one of the main advantages of running a hybrid event, which is the potential for individuals across different locations to collaborate. Organizers should thus carefully decide on one or the other option.
    
    \forceindent Shared hybrid activities, like joint icebreakers or virtual networking rooms, can again help connect participants across modalities and create more cohesive hybrid teams~\cite{gama2021online}. 

    \forceindent Organizers might also combine team formation at the event with pre-event introductions to prevent last-minute rushes and to ensure smooth integration of participants. Organizers should actively encourage participants to connect and align with potential teammates before the event, leveraging digital communication tools.
\end{itemize}

\subsection{Agenda}
\label{sec:dec:agenda}
Like any other event, hackathons need an agenda that outlines which activities will take place at which point in time. The timing and outline of activities can profoundly affect the experience of participants. Organizers thus have to carefully plan which activities they want to conduct during a hackathon for it to be satisfying and engaging.

\subsubsection{When?}
The agenda should be available at least a few days prior to a hackathon to allow participants and other stakeholders (section \ref{sec:dec:stake}) to familiarize themselves with it and make plans accordingly. Certain activities in the agenda, such as organizing speakers and awards, might require longer preparation periods and should thus be started earlier during the preparation.

\subsubsection{Who?}
Organizers typically consult with mentors (section \ref{sec:dec:stake},~\cite{nolte2020support}) and other stakeholders (section \ref{sec:dec:stake}), such as sponsors and domain experts, to decide about which activities will take place during a hackathon. Their timing then is commonly decided by the organizers to create an organic flow during the event itself.

\subsubsection{How?}
Depending on the domain the hackathon takes place in (section \ref{sec:dec:theme}), the goals (section \ref{sec:dec:goal}) that organizers aim to reach, or the type of participants they aim to attract (section \ref{sec:dec:part}), organizers might want to consider a variety of different activities during an event. In the following, we will outline common examples for activities that organizers might want to consider. This list is, however, by no means complete. Organizers can and should be creative in developing specific activities that fit their particular event.

\forceindent Hackathons typically \textbf{start with a brief welcoming address}. During this address, the organizers welcome participants and lay out the organizational details of an event, including the code of conduct (\cite{hackhpc-codeofconduct} is an example). Other organizational details should include means of reaching organizers, mentors, and other participants during a hackathon such as shared \textbf{communication channels, email lists or common document folders} as well as \textbf{links to useful resources} related to, e.g., the theme of the event or to technologies that participants might use~\cite{worldofcode:tutorial}. Organizers should also \textbf{introduce mentors} (section \ref{sec:dec:stake},~\cite{nolte2020support}), \textbf{jury and judging criteria} (in the case of a competitive event (section \ref{sec:dec:comp})), explain \textbf{checkpoints} and discuss which \textbf{outcome} is expected from each team at the end of the hackathon. Such outcomes can include but are not limited to source code or other technical artifacts and presentations, including videos and / or slides. The welcoming address can also include a \textbf{thematic keynote}, e.g., by a sponsor that provides additional background for the hackathon and sets the tone for the remainder of the event.

\vspace{0.15cm}
\forceindent Afterward, participants commonly pitch ideas (section \ref{sec:dec:ideation}) and form teams (section \ref{sec:dec:team}) before starting to work on their projects.

\vspace{0.15cm}
\forceindent During the hackathon, organizers commonly also schedule a series of \textbf{checkpoints} during which teams report their progress, discuss problems they are facing and outline their plans for the time ahead~\cite{g48cybersec}. These checkpoints should be evenly distributed along the timeline of a hackathon. It is, e.g., typical to have checkpoints at the beginning and the end of each day. They provide a great opportunity for organizers and mentors to get an overview of each team’s progress and decide which team might need additional support. Checkpoints can be organized in different ways. Some organizers may only ask team leaders to present to organizers and mentors so as not to break the teams’ rhythm. Others prefer all participants to be present during each checkpoint so that teams can share experiences and learn from each other.

\vspace{0.15cm}
\forceindent Organizers can also schedule additional \textbf{talks or training sessions} during an event~\cite{g48cybersec,affia2020developing}. These can be related to using common or specialized technologies (section \ref{sec:dec:spec}) that participants might use, provide additional domain background, or teach participants specific skills, such as how to successfully pitch their project at the end of the hackathon. Such talks or training sessions can take place once or multiple times as part of the main event. Some organizers even run them as parallel tracks over the entire duration of the hackathon~\cite{brainhack}. Such talks should be closely related to the projects that teams are working on during an event to have the desired effect~\cite{affia2020developing}.

\vspace{0.15cm}
\forceindent Depending on the goal of an event, organizers might also organize \textbf{social activities} that require participants to interact with each other beyond the teams they work in. These can include short games where participants have to form teams that are different from those they work with during the hackathon and compete for small prizes~\cite{icebreakergames}. Such games can also ease the tension of a hackathon and force participants to focus on something other than their project, which can help reduce stress and emphasize the fun aspect of a hackathon. They are particularly useful for hackathons that emphasize networking as a goal. They can, however, also be frustrating because they can break the rhythm of participants and distract them from their projects~\cite{nofun}.

\vspace{0.15cm}
\forceindent At the end of a hackathon, it is common for teams to \textbf{present their project} to the other teams, organizers, mentors, and jury (in the case of a competitive event). These presentations can take different forms depending on the outcome outlined at the beginning of the hackathon. They can be organized in the form of pitches (as common in entrepreneurial events), demos (as common in collegiate events), or project presentations (as common in civic and corporate events). In a competitive event, these presentations are followed by a deliberation of the jury and an award ceremony.

\vspace{0.15cm}
\forceindent It is generally \textbf{not advisable to plan too many activities} during a hackathon because all of them will reduce the time teams have to work on their projects (section \ref{sec:dec:dur}), which after all will be one of the main reasons for people to attend a hackathon. It is advisable, though, to conduct a thorough opening address as outlined before, schedule at least one checkpoint per day, and hold final presentations so that all teams can show what they have been working on. The other outlined activities are optional, and organizers need to decide which ones they consider useful for their specific event.

\subsubsection{Trade-offs}
\begin{enumerate}[leftmargin=0.5cm]
    \item \textbf{Input and training vs social activities:} Organizers might be inclined to provide as much input to participants as possible, especially during a hackathon that is attended by participants who are not necessarily very familiar with the theme of the event. Providing too much input during a hackathon can, however, confuse and frustrate participants because it breaks their rhythm, and they might feel inclined to change their project idea repeatedly based on the input they received. Moreover, some hackathon organizers might organize social activities for participants to network rather than work on their projects all the time. Striking a suitable balance here is crucial for a successful event. One possible way to mitigate this trade-off could be to have a thematic keynote at the beginning of an event, provide additional resources for participants to refer to during an event, and stagger social activities around common breaks such as breakfast, lunch, and dinner.
    \item \textbf{Repeat activities vs single activities:} It can be advisable to run the same talks or training sessions multiple times during an event to allow participants to attend them at the point in time that fits them best. This can, however, be difficult to organize since it requires presenters and trainers to be available during the entire duration of a hackathon. It might also be advisable to focus input at the beginning of an event so that participants can take maximum advantage of it. In addition, organizers can provide access to useful resources (e.g., instructional videos) that participants can access when needed.
    \item \textbf{Voluntary vs mandatory checkpoints:} Checkpoints are a great way for organizers and mentors to assess the progress of each team and provide targeted support if necessary. Attending checkpoints can, however, be tedious and time-consuming for teams and affect their productivity, so they might not be particularly inclined to attend. One way to deal with this tradeoff is to assign mentors to one team or a group of teams (section \ref{sec:dec:mentor},~\cite{nolte2020support}) and ask them to engage with their teams on a regular basis. This allows them to detect issues, inform the organizers, discuss strategies, and provide targeted support. This does, however, lose the advantage of familiarizing the team members with the projects and people on other teams.
\end{enumerate}

\subsubsection{Modes of participation}
\begin{itemize}[leftmargin=0.5cm]
    \item \textbf{In-person:} In-person events can allow for flexible, real-time adjustments to the agenda based on participant needs and energy levels. This spontaneity allows organizers to incorporate unplanned activities, such as networking during breaks or informal mentoring sessions, which can enhance participant interactions and foster creativity. The physical presence of participants, mentors, and organizers also creates natural opportunities for collaboration and guidance. Especially for larger events that do not take place in a single room, changing the agenda might be more challenging because organizers need to make sure that all participants are informed and have time to integrate changes into their team agenda.

    \forceindent However, physical constraints, such as room availability and venue schedules, may limit the flexibility of the agenda. Organizers must plan ahead to allocate time for key activities like checkpoints, mentoring sessions, and structured breaks. Adequate downtime between work sessions helps participants recharge and maintain focus. Networking events or informal activities during meal breaks can encourage team bonding and provide moments of relaxation, balancing the intensity of in-person hackathons.

    \item \textbf{Online:} Online hackathons require more structure compared to in-person events and are thus less flexible in terms of adjustments during an event. They are also typically longer, often spanning multiple days or weeks, not only to account for the fact that online collaboration is typically more time-consuming to organize but also to allow participants to balance personal and professional obligations while working from different time zones~\cite{paganini2023opportunities,powell2021organizing,gama2021online}. Another issue of online events is for organizers to stay in touch with teams and maintain engagement. For this, organizers should consider clear schedules that are communicated in advance and regular check-ins. Check-ins provide a good opportunity for organizers to assess team progress and spot potential issues. Organizers can also assign mentors to stay in regular contact with a limited number of teams (section \ref{sec:dec:mentor}). These mentors can then report potential issues to organizers. In addition, organizers should provide sufficient off-screen breaks to mitigate screen fatigue (section \ref{sec:dec:dur}). For example, virtual coffee breaks during EUvsVirus~\cite{bertello2022open} proved effective in maintaining engagement while allowing participants to recharge. Pre-event warm-ups, such as workshops or icebreaker sessions, can further help build excitement and foster a sense of community before the event begins~\cite{paganini2023opportunities}.
    
    \forceindent During an online event, organizers also need to consider potential time-zone differences. Activities like final presentations, workshops, and mentoring sessions should be scheduled during hours that are convenient for most participants to maximize participation. In addition, organizers could also consider asynchronous options, such as recorded sessions or flexible deadlines.

    \item \textbf{Hybrid:} Hybrid hackathons typically run over longer durations compared to in-person-only events to account for the additional time that is required to coordinate between participants who are at the event site and participants who join online. Agendas of hybrid events need to consider in-person and online participant schedules, using modality-specific activities (e.g., on-site workshops, virtual Q\&A sessions) while providing shared milestones, parallel activities, and deadlines to maintain progress~\cite{powell2021organizing,gama2021online}.

    \forceindent For example, EUvsVirus~\cite{bertello2022open} employed parallel schedules to keep virtual and in-person participants engaged. Still, aligning activities and checkpoints across modalities can be difficult, adding organizational and resource overhead, especially when participants operate in different time zones or have varied engagement levels (due to other obligations)~\cite{mendes2022socio}.

    \forceindent Regular hybrid checkpoints, using both in-person and virtual components, help ensure alignment of team goals and progress. However, without careful planning, hybrid formats risk misalignment between modalities, where remote participants may feel less integrated and focus on individual tasks rather than team activities.

    \forceindent Finally, organizers should be aware that participants might switch between in-person and online participation during an event. They might, e.g., prefer to be present for ideation and team formation and decide to leave the hackathon site to work on their part of the team project without interruption before rejoining the hackathon site for the final presentation. Organizers thus need to ensure that schedules and activities are clear and that potential changes are communicated both with in-person and online participants.
\end{itemize}

\subsection{Mentoring}
\label{sec:dec:mentor}
Mentors are the first substantial point of contact for participating teams. They provide feedback, help them when they have problems, and guide them through the hackathon process. Deciding on who to recruit as a mentor and developing a suitable mentoring strategy are thus crucial decisions for every hackathon organizer.

\subsubsection{When?}
It is important to develop a mentoring strategy and recruit suitable mentors prior to a hackathon. Since they are likely to be busy people and mentoring generally takes a substantial chunk of their time, recruiting weeks or months in advance is desirable. Mentoring itself typically takes place either over the entire duration of a hackathon or at specific points during the event. It can also continue after a hackathon has ended (section \ref{sec:dec:continuity}).

\subsubsection{Who?}
Mentoring requires the collaboration of organizers, mentors, and participants. Organizers create a mentoring strategy, recruit mentors, and support them to execute the developed strategy during and after a hackathon. Mentors support participating teams based on this strategy. The time commitment asked of the mentors should be made very clear. For example, are they expected to help participants before and/or after the event itself? Are they expected to stay for the entire event, work in shifts, or just be available at checkpoints (section \ref{sec:dec:agenda})? 

\subsubsection{How?}
Prior to a hackathon, organizers have to develop a mentoring strategy and recruit suitable individuals as mentors.

\vspace{0.15cm}
\forceindent \textbf{Mentoring strategy:} The most common strategy is for mentors to provide individual \textbf{on-demand support} during a hackathon based on the mentor’s expertise. This is appropriate when the participants are generating their own projects (section \ref{sec:dec:ideation}) and have the basic skills required to complete them. On-demand mentors typically circulate among teams and/or staff a help desk where participants can receive assistance when needed. In addition, organizers often set specific checkpoints during which mentors engage with teams, ask for their current progress, and provide targeted feedback. Alternatively, the event may have \textbf{dedicated mentors} that are assigned to an individual team~\cite{nolte2020support}. This is useful when the participants have significant skill deficiencies or don’t have sufficient domain knowledge (section \ref{sec:dec:spec}) to define projects that fit within the hackathon theme (e.g., scientific hackathons aimed at bringing neophytes into a field). For either strategy, it is crucial that mentors are accessible to teams when they need them.

\vspace{0.15cm}
\forceindent In addition, organizers might want to create mandatory \textbf{checkpoints} (section \ref{sec:dec:agenda}) during which teams present their progress to mentors and to the other participating teams. Such checkpoints allow for mentors and teams to detect deficiencies that might have remained unnoticed and provide an opportunity for broad feedback to all teams at once.

\vspace{0.15cm}
\forceindent Another aspect to consider is whether to have \textbf{individual mentors} supporting participants or to form \textbf{mentor teams} with diverse backgrounds and expertise. Individual mentors allow for a more flexible deployment while mentor teams can provide holistic feedback to participating teams on a broad range of issues. Mentor teams also provide opportunities for less experienced mentors to learn from their more experienced peers. In the case that organizers decide for mentoring teams, it is important to define them prior to or at the beginning of the event.

\vspace{0.15cm}
\forceindent Organizers also have to decide \textbf{how many mentors} to recruit for their hackathon, and how many of them to deploy at specific points during the hackathon. This decision depends on the number of participating teams, the availability of mentors, and other recruitment-related aspects we will discuss in the following. It also depends on the timing of a hackathon since mentor support is mostly needed during the early and late phases of an event. As a rule of thumb, using a minimal mentor-to-participant ratio of 1 to 10 is feasible since teams commonly have fewer than 10 participants.

\vspace{0.15cm}
\forceindent \textbf{Recruitment:} Depending on the selected strategy, the organizers have to decide how to recruit suitable mentors. Common aspects for recruitment are the expertise of individuals related to the theme or domain of the event, their technical proficiency related to the technologies that participants might use during a hackathon, their prior hackathon (mentoring) experience, and their ability to guide teams and support them to perform to the best of their abilities. A good mentor~\cite{greatmentor,perfectmentor} thus possesses a combination of domain, technical, project management, and social skills. To recruit suitable individuals, organizers also have to think about potential benefits for mentors since they will invest a lot of time and effort into mentoring. For example, mentors can sometimes be drawn from companies who are hoping to find new recruits among the participants or from faculty or postdocs looking for talented students. Inclusivity in mentoring is also crucial. Hackathons that cater to underrepresented groups, such as women-focused events, have found success in recruiting women-only mentors, which can help create a more welcoming and empowering atmosphere for participants~\cite{paganini2023opportunities}.

\vspace{0.15cm}
\forceindent After recruiting suitable individuals, organizers need to provide them with suggestions on how to mentor~\cite{g48howtomentor} teams based on the previously decided mentoring strategy. This includes potential activities before a hackathon, such as training webinars, as well as their availability during the event. It is crucial that mentors are available to teams when they need them because the tight time constraints of a hackathon do not allow teams to get stuck for long. Mentors will be particularly busy during the early and late phases of a hackathon as discussed before. During the early phases, teams commonly need help scoping their project and technical support to get started. During the late phases, everyone is scrambling to fix last-minute issues, which can also lead to increased mentor demand.

\vspace{0.15cm}
\forceindent Mentors need to be introduced to the participants either before or at the beginning of a hackathon. This introduction should include how and when participants can engage with mentors and which mentor can help them with specific topics or issues. It is important to remind mentors that a hackathon is not for them to push their own ideas. It is about helping teams to run their project their way.

\vspace{0.15cm}
\forceindent In some cases, mentoring can also continue after a hackathon has ended (section \ref{sec:dec:continuity}) to, e.g., facilitate the continuity of learning or complete the development and integration of a technical artifact. This depends, however, on the mutual interest of participants, mentors, and organizers and should be discussed, at best, before the end of a hackathon.

\subsubsection{Trade-offs}
\begin{enumerate}[leftmargin=0.5cm]
    \item \textbf{Dedicated mentors vs on-demand mentors:} On-demand mentors -- if sufficient in number and covering all the necessary skills -- can quickly address the needs of any team while dedicated mentors can build a relationship with a team and be more effective and efficient supporting them and helping them define a project and acquire skills they need. Supplying a larger hackathon with individual team mentors can, however, prove to be challenging. Moreover, dedicated mentors might be inclined to take over certain aspects of a project, which might negatively affect a team’s motivation and give them an unfair advantage, especially during a competitive hackathon (section \ref{sec:dec:comp}).
    \item \textbf{Mentor teams vs individual mentors:} One benefit of mentoring teams is that participants can get comprehensive support related to multiple aspects of their project (domain, technical advice, scoping, etc.)~\cite{affia2020developing}. They are thus particularly useful when using checkpoints because these allow for mentors to take time to address multiple potential issues the team is facing at the same time. Mentor teams also allow less experienced mentors to gain more experience while working with more experienced peers. These benefits, however, only materialize when the mentors in a team have different expertise and experiences. Moreover, forming teams requires additional coordination effort by the organizers. Individual mentors, on the other hand, can flexibly offer targeted advice for teams in need. This does, however, require teams to know who they should address for specific topics. Individual mentoring also makes better use of each individual mentor’s time but limits support to the expertise of one mentor at a time.
    \item \textbf{Mentor background:} Each mentor should, at best, be an experienced project manager and domain expert with years of technical experience in the field, lots of hackathons under her/his belt, and the ability to solve any problem a team might have. Since this is not always possible, it is important that organizers carefully select mentors with complementary backgrounds and skills. Moreover, participants and mentors need to be aware of the skills of other mentors, to be able to refer participants to suitable mentors if needed. This can be achieved by individually introducing mentors at the beginning of a hackathon, by creating short online profiles, or by, e.g., creating colored badges that indicate which kind of support each individual mentor can provide (technical, domain, etc.). Moreover, mentoring teams can mitigate this issue if they are formed as discussed before.
    \item \textbf{Participant to mentor ratio:} At first glance, it appears that more mentors are always better since more mentors can support more teams. This is, however, not true in all circumstances since, e.g., having many mentors that can help with domain-related questions and none that can help with technical problems is not desirable. Moreover, more mentors create more organizational overhead for organizers and might result in conflicting messages to participants since different mentors might provide different advice to teams based on their personal experience and background. Starting with a mentor-to-participant ratio of 1 to 10 can serve as a rule of thumb. It is, however, important that organizers still ask themselves which expertise might be required by participating teams at different points during a hackathon. At the beginning of an event, participants will mostly require help related to scoping their project, while later, the required support will likely shift stronger toward domain and technology-related questions. Inexperienced teams may also need help setting up their technical environment and tools. When deciding for mentors it is thus important to consider which expertise needs to be available to participants at which point during an event. Moreover, mentoring can also be streamlined by using the previously discussed checkpoints.
    \item \textbf{Strict guidance vs mentors decide how to engage:} While there are general mentoring~\cite{greatmentor} and hackathon mentoring~\cite{perfectmentor,g48howtomentor} guidelines available online, it is important to note that each hackathon is slightly different with different goals, organizers, participants, and mentoring requirements. Given the highly context-dependent nature of hackathons, it might be helpful under certain circumstances to advise mentors about when and how to engage with their teams. Under other circumstances organizers might also just let mentors engage with teams at any point and in any way they want. Both extremes are not feasible. Providing strict guidance would limit the ability – in particular of experienced mentors – to provide useful support. Providing no guidance might affect the overall structure of a hackathon because mentors might, e.g. be inclined to contact their teams all the time, thus affecting their rhythm~\cite{affia2020developing}. A hackathon is an intensive event during which a lot of things happen over a short period of time and mentors are crucial for an event to be successful. Spreading information about who is responsible for and knowledgeable about which topic and setting fixed checkpoints to reel everyone back in can thus help to mitigate this tradeoff. In general, however, the more competitive the event the more it is necessary, in the name of fairness, to provide relatively strict guidance and limits on what sort of assistance mentors should provide and when they should provide it.
\end{enumerate}

\subsubsection{Modes of participation}

\begin{itemize}[leftmargin=0.5cm]
    \item \textbf{In-person:} During an in-person event it is advisable for mentors to be on-site as well. This allows mentors to provide immediate support when asked. It also provides opportunities for mentors to observe how teams are collaborating and progressing which in turn can allow them to spot issues and provide support without explicitly being asked. Moreover, teams and mentors spending time together in the same space can help build connections and lower the barrier for teams to ask for help.

    \forceindent However, in-person participants can face challenges when mentors are spread thin across multiple teams, especially at large-scale events. In such cases, teams may experience delays in receiving support or feel overlooked during critical moments. In addition, mentors constantly hovering over teams might be perceived as intrusive and distracting. To mitigate these issues, organizers should schedule structured mentoring hours or team check-ins to ensure every team receives equitable support~\cite{affia2020developing,nolte2020support}.

    \forceindent While in-person mentoring is highly desirable, it is also possible to provide online mentoring during an in-person event through technical means like Discord, Slack, Zoom, Google Meet, or other conferencing and messaging platforms. The advantage of online mentoring is that experts who might not have the time or resources to attend an event in person can still serve as mentors, thus broadening the support that is available to teams. Participants, however, might find it difficult to contact online mentors because they might not be available when participants need support~\cite{nolte2020support}. Direct channels such as messengers and mentor schedules might help address this issue as discussed before. Mentors, in turn, might also find it difficult to stay in contact with teams and understand when they are struggling and need support. Moreover, using online platforms for communication might lead to participants seeking in-person help instead despite online experts being available because it can be frustrating to discuss problems and issues via video meetings or text due to the limitations of such technologies.
    
    \item \textbf{Online:} Mentoring in online events requires a more structured approach compared to in-person events where participants can just walk up to mentors and ask for help and mentors can walk around, get to know participants and provide help when they see issues arising. Organizers should thus consider scheduling virtual check-ins, regular mentoring sessions, such as team-specific check-ins or virtual office hours, and pre-pitch feedback to ensure that teams receive timely guidance~\cite{paganini2023opportunities}. These contact points need to be communicated to participants, e.g., through a centralized document or website, and it needs to be clear how mentors can be reached.

    \forceindent It is also advisable to assign mentors to specific teams so that they can track team progress over the course of an event, provide assistance if necessary, find another mentor should the required help be outside of their area of expertise~\cite{nolte2020support,mendes2022socio} and inform organizers about how the teams they are assigned to are progressing. In addition, organizers might consider mentor teams to lower the load on individual mentors, broaden support, and allow novice mentors to gain experience working with more experienced peers (\cite{hackhpc} is an example).

    \forceindent Despite careful planning, online participants might still struggle with the absence of spontaneous engagement that naturally occurs in physical settings. The absence of face-to-face cues can make it challenging for participants to build rapport with mentors or convey the nuances of their challenges effectively. For example, during EUvsVirus~\cite{bertello2022open}, mentors had to juggle multiple teams, leading to delays in feedback. Online participants must also solely rely on virtual communication through platforms like Discord, Slack, or Zoom to connect with mentors. 

    \forceindent Time zone differences also pose additional challenges for online participants~\cite{powell2021organizing,bertello2022open}. Organizers could use asynchronous tools such as shared question boards or recorded mentoring sessions to supplement synchronous activities and ensure that participants in different time zones have equal opportunities to engage with mentors.

    \item \textbf{Hybrid:} At hybrid hackathons, organizers face similar issues as those discussed before for in-person and hybrid events but in a combined form. To alleviate these issues, organizers might consider assigning modality-specific mentors to ensure parity in support. However, this can be logistically complex. In-person participants may have more frequent access to mentors, leaving remote participants feeling neglected unless proactive measures are taken.

    \forceindent Organizers should also adopt hybrid-compatible tools to facilitate mentoring across modalities. For instance, shared platforms like Slack or Discord can centralize communication, enabling mentors to interact with both online and mixed in-person and online teams. Regularly scheduled hybrid mentoring sessions, where mentors engage synchronously with participants from both modalities, can help bridge the gap~\cite{gama2021online}. For online or mixed in-person and online teams, organizers should establish virtual check-ins and provide access to shared mentoring resources to ensure they do not feel sidelined.
\end{itemize}

\subsection{Continuity planning}
\label{sec:dec:continuity}
Organizers might want to run a hackathon just for everyone to have a good time. They might, however, also want to organize one for a purpose that extends beyond the conclusion of the hackathon, for example, to kickstart a community, teach participants about new technologies, or create innovative products and services that will actually be brought to market (section \ref{sec:dec:goal}). Continuity does not come for free, though. It needs to be an integral part of the hackathon planning process.

\subsubsection{When?}
Continuity planning needs to start prior to the hackathon and continue past the end of the event itself. In order to support the continuity of hackathon outcomes, it is important that the event is embedded into a larger strategy.

\subsubsection{Who?}
Organizers are responsible for developing and deploying a suitable strategy, which includes communicating it to the participants of a hackathon. The planning and execution of that strategy needs support from stakeholders (section \ref{sec:dec:stake}) and hackathon participants. Organizers should also be aware that only a few or even none of the participants might share their continuity vision. Participants might just come for the fun, or they might have continuity plans of their own~\cite{medina2019does}.

\subsubsection{How?}
Before planning for continuity, organizers and potential stakeholders need to think about \textbf{what the outcome of a hackathon should be} and how the continuation of this outcome can be supported. When thinking about hackathon outcomes, most people will think of hackathon projects that get turned into startups. The transformation of projects into startups is, however, not the only potential outcome worthy of being continued. Participants and/or organizers might also aim to extend existing products or services, foster the reuse of technologies that were created during an event beyond the confines of the project they were created for, support community growth, or spread knowledge about certain domains and technologies. Each of these outcomes might require a different continuation strategy.

\vspace{0.15cm}
\forceindent Based on the decision of which outcome should be continued, hackathon organizers can then focus on \textbf{involving potential stakeholders} (section \ref{sec:dec:stake}). Such stakeholders can, e.g., be companies if the continuation goal is to develop a product or service, or communities with related or complementary interests if the goal is to start or grow a community. Stakeholders are vital for continuity planning since they can provide background and domain knowledge for hackathon projects, support participants to scope their projects, connect participants to key players after a hackathon has ended, and provide access to learning materials. It is also important for organizers to set suitable expectations for stakeholders and participants since there is only so much that can be achieved during the short timeframe of a hackathon. This makes it all the more important to plan for what can be done before, during (section \ref{sec:dec:agenda}), and after an event to support continuation if that is the organizers’ goal. Examples for how organizers can support continuation is to suggest for participants to scope their project (section \ref{sec:dec:ideation}) before and start learning about technologies they might want to use (section \ref{sec:dec:spec},~\cite{pe2022corporate,nolte2018you}).

\vspace{0.15cm}
\forceindent During a hackathon organizers and stakeholders should \textbf{provide an environment for participants that fosters the desired outcome}. If an important goal is for participants to have the opportunity to establish lasting social bonds, the organizers should put an emphasis on activities that allow for them to not only work on their projects in their teams but also to get in contact with other participants. This could happen in the form of games or other social activities. For participants to develop a project that can be continued afterward, they should encourage participants to form a diverse team that has the skills required to complete that project~\cite{nolte2020support}, work on a strategy on how to spread the word about their project after the hackathon~\cite{nolte2020what} or to ensure that what they develop can be easily integrated into an existing code base~\cite{nolte2018you}. For participants to develop technology that could be reused by themselves or others after an event has ended, organizers should encourage participants to form larger teams, add documentation and -- if feasible -- data to the code they develop~\cite{imam2021secret}, and add an open-source license~\cite{mahmoud2022one}. Offering prizes at an event (section \ref{sec:dec:comp}) can also be an incentive for teams to continue their projects, but they only have a short-term effect~\cite{nolte2020what}. If the goal is to sustain participants’ learning, e.g., about technologies they used during the hackathon, it might be useful to propose follow-up projects or connect them to other individuals that aim to learn about the same technology.

\vspace{0.15cm}
\forceindent \textbf{Participants and organizers might have different continuation goals} after an event~\cite{medina2019does}. Most participants might not even be interested to continue working on their project after a hackathon, or their interest might fade quickly~\cite{nolte2020what}. Since continuation requires extra work from participants, it is important to identify and provide support to those who are interested in continuation. One approach to achieve this is simply to ask participants if they are interested in continuation and, based on their response, provide support and guidance. This support can – depending on the planned outcome – come in the form of startup funding, connecting participants to relevant parties that can support them, help them find resources, or simply contacting them from time to time after the hackathon to see what happened and provide targeted support if needed.

\subsubsection{Trade-offs}
In the following, we will discuss strategies to support the continuity of different hackathon outcomes. From the descriptions, it should be clear that some outcomes might require approaches and strategies that can negatively affect other outcomes. For example, organizing social gatherings during a hackathon might foster participant networking, thus potentially contributing to connection continuity. Such gatherings, however, eat into the participants’ time to work, which can negatively affect their project outcome, thus potentially jeopardizing its continuation or the reuse of the created technologies after the hackathon.

\begin{enumerate}[leftmargin=0.5cm]
    \item \textbf{Project continuity and technology reuse:} The development of useful artifacts that continue to be developed or that get reused after a hackathon is among the most common continuity goals. This requires – as discussed before – a strong focus on the project as such. It is advisable for teams to meet (section \ref{sec:dec:team}) and refine their project idea (section \ref{sec:dec:spec}) before the event, validate it with potential stakeholders, and then use the hackathon to develop it to a stage that it can be shown to stakeholders~\cite{nolte2018you}. If the goal is technology reuse then it is also advisable to include documentation, data, and an open-source license~\cite{mahmoud2022one}. This means that the hackathon essentially serves the purpose of the team focusing on the development of a presentable prototype for their project. It also means that participants should be encouraged to find team members with diverse skill sets that fit the requirements of the project~\cite{nolte2020what} and that team members choose which aspect of a project they work on predominantly based on their current skills and not on what they want to learn about.
    \item \textbf{Connection continuity:} To support connection continuity, it is important to ensure that participants can stay in touch after a hackathon. Technical means such as a shared Slack or Discord channel, a shared Google Drive folder, or similar tools can be already used leading up to and during a hackathon. These tools enable participants to quickly get on board and start communicating and sharing information about themselves and their projects while the event is still going on. Organizers should also consider creating opportunities for participants to engage with other like-minded participants outside their project teams during and after the hackathon to foster continuous engagement after the hackathon has ended (section \ref{sec:dec:agenda}). None of these will work, however, unless the participants are substantially motivated to stay in touch, perhaps to act as sounding boards or social support, to network opportunities such as jobs, or to provide each other with needed expertise in various domains. It might also be worthwhile to introduce teams that have worked on similar projects to each other, which can serve as a proxy for connection continuity.
    \item \textbf{Learning continuity:} To foster learning continuity, it is not only important to provide learning material before, during, and after an event (section \ref{sec:dec:spec}) as well as targeted talks and mentoring during the event itself~\cite{affia2020developing}. It is also important to create continuing interests, e.g., through challenges during and after a hackathon. Moreover, it is important to ensure that participants think about their projects early~\cite{affia2020developing} and choose them based on what they want to learn rather than what they already know.
\end{enumerate}

\subsubsection{Modes of participation}
\begin{itemize}[leftmargin=0.5cm]
    \item \textbf{In-person:} In-person events naturally facilitate connections through face-to-face interactions, thus potentially creating interpersonal connections that can drive post-event collaboration. To support continuity, organizers can also consider implementing digital scaffolding. Shared virtual platforms such as Discord or Slack can act as extensions of the in-person environment, allowing teams to stay connected and continue collaborating on their projects, continue learning about topics they are interested in, and simply stay in touch. Additionally, project repositories like GitHub enable participants to document and refine their work, supporting ongoing progress.

    \forceindent Organizers can reinforce connections and platform use through pre-event and follow-up initiatives, such as smaller networking meetups or showcase events where teams can share ideas or present updates on their projects. These events can strengthen the community built during a hackathon and encourage participants to maintain engagement and collaboration over the long term.

    \item \textbf{Online:} Online hackathons by nature require organizers to set up technologies including means for asynchronous communication such as Slack and Discord for participants to connect before and collaborate during an event (section \ref{sec:dec:spec}). These technologies can then also be used to continue collaboration, knowledge exchange, and networking after an event has ended~\cite{mendes2022socio,powell2021organizing} as demonstrated in events like EUvsVirus~\cite{bertello2022open}, which used Slack and post-event matchmaking to connect participants with funders and collaborators. Moreover, online events might provide opportunities for growing a community or finding adopters of technologies developed during an event due to their global reach.

    \forceindent However, online participants may face challenges in building strong personal connections due to the absence of face-to-face interaction. This lack of connection can reduce long-term motivation and engagement~\cite{mendes2022socio,schulten2022participants}. To address this, organizers could consider incorporating structured post-event mechanisms, such as virtual feedback sessions, mentoring, and follow-up workshops. For example, the Matchathon initiative in EUvsVirus~\cite{bertello2022open} connected participants with investors and end-users, fostering sustained collaboration despite the online format. Pre-event planning can also enhance continuity by emphasizing community-building activities, such as virtual icebreakers and team-building exercises, which help participants establish rapport and allow them to share goals early.

    \forceindent Organizers should also be aware, though, that some participants simply might perceive an event as a one-time experience and are not interested in continuing their project or staying in touch with other hackathon participants and stakeholders.

    \item \textbf{Hybrid:} Most of the aspects discussed previously related to online events also hold true for hybrid events. They do, however, offer the opportunity to form in-person connections combined with a potential global audience. The main issue that organizers face is to offer similar support for in-person and online participants. It is thus advisable to set up technologies in a similar way as one would for an online event and make sure that both in-person and online participants use them. This can be achieved, for example, by running pre-hackathon sessions using technologies that are then also used to, e.g., contact organizers and mentors or share artifacts during an event. Post-event activities can then provide opportunities for participants to continue working on their projects, share artifacts with others, continue learning, and expand their networks.

    \forceindent Organizers should also be aware that forming connections is more likely to happen between in-person participants. For connections to online participants, the organizers need to provide suitable opportunities, e.g., by fostering hybrid team formation (section \ref{sec:dec:team}).
\end{itemize}

\section{Acknowledgments}
The authors gratefully acknowledge support by the Alfred P. Sloan Foundation (Grant No: G-2023-20872), the German Federal Government Commissioner for Culture and the Media, Microsoft Research, Microsoft Garage, the Science Gateways Community, the Space Telescope Science Institute, HackHPC, Omnibond, Garage48, and the Oak Ridge National Laboratory. We would also like to thank Christian Bird, Amy Cannon, Irene-Angelica Chounta, Tapajit Dey, Kevin Ellet, Jeanette Falk, Linda Bailey Hayden, John Holmen, Timothy Holston, Ahmed Samir Imam Mahmoud, Rajesh Kalyanam, Thomas Maillart, Raimundas Matulevičius, Audris Mockus, Lavinia Paganini, Sudhakar Pamidighantam, Suzanne Parete-Koon, Je'aime Powell, Steve Scallen, Cleo Schulten, Alexander Serebrenik, Daniel Spikol, Kathryn Traxler, Nancy Wilkins-Diehr, Boyd Wilson, Mona Wong, and the organizers, mentors and participants of the various hackathons we studied, co-organized and supported.

% Balancing columns in a ref list is a bit of a pain because you
% either use a hack like flushend or balance, or manually insert
% a column break.  http://www.tex.ac.uk/cgi-bin/texfaq2html?label=balance
% multicols doesn't work because we're already in two-column mode,
% and flushend isn't awesome, so I choose balance.  See this
% for more info: http://cs.brown.edu/system/software/latex/doc/balance.pdf
%
% Note that in a perfect world balance wants to be in the first
% column of the last page.
%
% If balance doesn't work for you, you can remove that and
% hard-code a column break into the bbl file right before you
% submit:
%
% http://stackoverflow.com/questions/2149854/how-to-manually-equalize-columns-
% in-an-ieee-paper-if-using-bibtex
%
% Or, just remove \balance and give up on balancing the last page.
%
\balance{}

% REFERENCES FORMAT
% References must be the same font size as other body text.
\bibliographystyle{SIGCHI-Reference-Format}
\bibliography{references}


\begin{thebibliography}{00}

%%% the Web address.

\ifx \showCODEN    \undefined \def \showCODEN     #1{\unskip}     \fi
\ifx \showDOI      \undefined \def \showDOI       #1{{\tt DOI:}\penalty0{#1}\ }
  \fi
\ifx \showISBNx    \undefined \def \showISBNx     #1{\unskip}     \fi
\ifx \showISBNxiii \undefined \def \showISBNxiii  #1{\unskip}     \fi
\ifx \showISSN     \undefined \def \showISSN      #1{\unskip}     \fi
\ifx \showLCCN     \undefined \def \showLCCN      #1{\unskip}     \fi
\ifx \shownote     \undefined \def \shownote      #1{#1}          \fi
\ifx \showarticletitle \undefined \def \showarticletitle #1{#1}   \fi
\ifx \showURL      \undefined \def \showURL       #1{#1}          \fi

\bibitem{affia2020developing}
{Abasi-Amefon~Obot Affia}, {Alexander Nolte}, {and} {Raimundas
  Matulevi{\v{c}}ius}. 2020.
\newblock \showarticletitle{Developing and Evaluating a Hackathon Approach to
  Foster Security Learning}. In {\em Collaboration Technologies and Social
  Computing}. Springer.
\newblock


\bibitem{affia2022integrating}
{Abasi-amefon~Obot Affia}, {Alexander Nolte}, {and} {Raimundas
  Matulevi{\v{c}}ius}. 2022.
\newblock \showarticletitle{Integrating hackathons into an online cybersecurity
  course}. In {\em Proceedings of the ACM/IEEE 44th International Conference on
  Software Engineering: Software Engineering Education and Training}. 134--145.
\newblock


\bibitem{affia2025hybrid}
{Abasi-amefon Affia-Jomants} {and} {Alexander Nolte}. 2025.
\newblock Research instruments for hybrid hackathon organizers and researchers.
\newblock
\newblock
\showDOI{%
\url{http://dx.doi.org/10.5281/zenodo.14736328}}


\bibitem{hackweektoolkit}
{{Arendt, Anthony and Huppenkothen, Daniela}}. 2020a.
\newblock HackWeek Toolkit.
\newblock
\newblock
\showURL{%
\url{https://uwescience.github.io/HackWeek-Toolkit/}}
\newblock
\shownote{[Online; accessed 03-March-2025].}


\bibitem{hackweektoolkit:targetaudience}
{{Arendt, Anthony and Huppenkothen, Daniela}}. 2020b.
\newblock HackWeek Toolkit - Target Audience and Scoping to Specific
  Communities.
\newblock
\newblock
\showURL{%
\url{https://uwescience.github.io/HackWeek-Toolkit/\#Objectives/Objectives-and-Goals/\#target-audience-and-scoping-to-specific-communities}}
\newblock
\shownote{[Online; accessed 03-March-2025].}


\bibitem{baccarne2014urban}
{Bastiaan Baccarne}, {Peter Mechant}, {Dimitri Schuurma}, {Lieven De~Marez},
  {and} {Pieter Colpaert}. 2014.
\newblock \showarticletitle{Urban socio-technical innovations with and by
  citizens}.
\newblock {\em Interdisciplinary Studies Journal\/} {3}, 4 (2014), 143.
\newblock


\bibitem{berger2017karachi}
{Eric Berger}. 2017.
\newblock \showarticletitle{Karachi Hackathon Takes on Emergency Medicine
  Challenges: Solutions Pitched for Resource-Poor Environments}.
\newblock {\em Annals of Emergency Medicine\/} {69}, 3 (2017), A17--A20.
\newblock


\bibitem{bertello2022open}
{Alberto Bertello}, {Marcel~LAM Bogers}, {and} {Paola De~Bernardi}. 2022.
\newblock \showarticletitle{Open innovation in the face of the COVID-19 grand
  challenge: insights from the Pan-European hackathon ‘EUvsVirus’}.
\newblock {\em R\&d Management\/} {52}, 2 (2022), 178--192.
\newblock


\bibitem{bookdash2024}
{{{BookDash 2024}}}. 2024.
\newblock The Turing Way Book Dashes.
\newblock
\newblock
\showURL{%
\url{https://the-turing-way.netlify.app/community-handbook/bookdash}}
\newblock
\shownote{[Online; accessed 03-March-2025].}


\bibitem{braune2021interdisciplinary}
{Katarina Braune}, {Pablo-David Rojas}, {Joscha Hofferbert}, {Alvaro
  Valera~Sosa}, {Anastasiya Lebedev}, {Felix Balzer}, {Sylvia Thun}, {Sascha
  Lieber}, {Valerie Kirchberger}, {and} {Akira-Sebastian Poncette}. 2021.
\newblock \showarticletitle{Interdisciplinary online hackathons as an approach
  to combat the COVID-19 pandemic: case study}.
\newblock {\em Journal of medical Internet research\/} {23}, 2 (2021), e25283.
\newblock


\bibitem{briscoe2014digital}
{Gerard Briscoe}. 2014.
\newblock \showarticletitle{Digital innovation: The hackathon phenomenon}.
\newblock {\em Creativeworks London\/}  {6} (2014), 1--13.
\newblock


\bibitem{chounta2023re}
{Irene-Angelica Chounta}, {H~Ulrich Hoppe}, {Alexander Nolte}, {and} {Daniel
  Spikol}. 2023.
\newblock \showarticletitle{Re-inventing project-based learning: Hackathons,
  datathons, devcamps as learning expeditions}. In {\em Frontiers in
  Education}, Vol.~8. Frontiers Media SA, 1182264.
\newblock


\bibitem{civichackingday}
{{Code for America}}. 2019.
\newblock National Day of Civic Hacking.
\newblock
\newblock
\showURL{%
\url{https://www.codeforamerica.org/events/national-day-of-civic-hacking-2019}}
\newblock
\shownote{[Online; accessed 03-March-2025].}


\bibitem{codeforpgh}
{{Code for Pittsburgh}}. 2020.
\newblock Code for Pittsburgh Meetups.
\newblock
\newblock
\showURL{%
\url{https://www.meetup.com/codeforpgh/}}
\newblock
\shownote{[Online; accessed 03-March-2025].}


\bibitem{codecademy}
{{Codecademy}}. 2020.
\newblock Learn Python 3.
\newblock
\newblock
\showURL{%
\url{https://www.codecademy.com/learn/learn-python-3}}
\newblock
\shownote{[Online; accessed 03-March-2025].}


\bibitem{csikszentmihalyi1996flow}
{Mihaly Csikszentmihalyi}. 1996.
\newblock {\em Flow and the psychology of discovery and invention}. Vol.~56.
\newblock New York: Harper Collins.
\newblock


\bibitem{biohack}
{{Database Center for Life Science}}. 2008.
\newblock BioHackathon 2008.
\newblock
\newblock
\showURL{%
\url{http://hackathon.dbcls.jp/}}
\newblock
\shownote{[Online; accessed 03-March-2025].}


\bibitem{facultyhack2024}
{{{FacultyHack organizers}}}. 2024.
\newblock FacultyHack.
\newblock
\newblock
\showURL{%
\url{https://hackhpc.github.io/facultyhack-gateways24/}}
\newblock
\shownote{[Online; accessed 25-February-2025].}


\bibitem{falk2025creativity}
{Jeanette Falk}, {Yiyi Chen}, {Janet Rafner}, {Mike Zhang}, {Johannes Bjerva},
  {and} {Alexander Nolte}. 2025.
\newblock \showarticletitle{How Do Hackathons Foster Creativity? Towards AI
  Collaborative Evaluation of Creativity at Scale}.
\newblock {\em arXiv preprint arXiv:2503.04290\/} (2025).
\newblock


\bibitem{falk2024future}
{Jeanette Falk}, {Alexander Nolte}, {Daniela Huppenkothen}, {Marion Weinzierl},
  {Kiev Gama}, {Daniel Spikol}, {Erik Tollerud}, {Neil~Chue Hong}, {Ines
  Kn{\"a}pper}, {and} {Linda~Bailey Hayden}. 2024.
\newblock \showarticletitle{The future of hackathon research and practice}.
\newblock {\em IEEE Access\/} (2024).
\newblock


\bibitem{falk2024role}
{Jeanette Falk}, {Daniel Spikol}, {Diana Meda}, {and} {Mark Melnykowycz}. 2024.
\newblock \showarticletitle{The Role of Slowing Down in Fast-Paced Game Jams}.
  In {\em Proceedings of the 8th International Conference on Game Jams,
  Hackathons and Game Creation Events}. 33--40.
\newblock


\bibitem{falk2022supporting}
{Jeanette Falk} {and} {Faith Young}. 2022.
\newblock \showarticletitle{Supporting fast design: The potential of hackathons
  for co-creative systems}. In {\em Proceedings of the 14th Conference on
  Creativity and Cognition}. 515--519.
\newblock


\bibitem{falk202010}
{Jeanette Falk~Olesen} {and} {Kim Halskov}. 2020.
\newblock \showarticletitle{10 Years of Research With and On Hackathons}. In
  {\em Proceedings of the 2020 ACM on Designing Interactive Systems
  Conference}. 1073--1088.
\newblock


\bibitem{filippova2017diversity}
{Anna Filippova}, {Erik Trainer}, {and} {James~D Herbsleb}. 2017.
\newblock \showarticletitle{From diversity by numbers to diversity as process:
  supporting inclusiveness in software development teams with brainstorming}.
  In {\em 2017 IEEE/ACM 39th International Conference on Software Engineering
  (ICSE)}. IEEE, 152--163.
\newblock


\bibitem{gama2018hackathon}
{Kiev Gama}, {Breno Alencar}, {Filipe Calegario}, {Andr{\'e} Neves}, {and}
  {Pedro Alessio}. 2018.
\newblock \showarticletitle{A Hackathon Methodology for Undergraduate Course
  Projects}. In {\em 2018 IEEE Frontiers in Education Conference (FIE)}. IEEE,
  1--9.
\newblock


\bibitem{gama2023developers}
{Kiev Gama}, {George Valen{\c{c}}a}, {Pedro Alessio}, {Rafael Formiga},
  {Andr{\'e} Neves}, {and} {Nycolas Lacerda}. 2023a.
\newblock \showarticletitle{The developers’ design thinking toolbox in
  hackathons: a study on the recurring design methods in software development
  marathons}.
\newblock {\em International Journal of Human--Computer Interaction\/} {39}, 12
  (2023), 2269--2291.
\newblock


\bibitem{gama2023comfort}
{Kiev Gama}, {Carlos Zimmerle}, {and} {Lavinia Paganini}. 2023b.
\newblock \showarticletitle{The Comfort of Distance: Student Choices and Soft
  Skill Development during a Hybrid Hackathon in Post-Pandemic Learning}. In
  {\em Proceedings of the XXXVII Brazilian Symposium on Software Engineering}.
  368--377.
\newblock


\bibitem{gama2021online}
{Kiev Gama}, {Carlos Zimmerle}, {and} {Pedro Rossi}. 2021.
\newblock \showarticletitle{Online hackathons as an engaging tool to promote
  group work in emergency remote learning}. In {\em Proceedings of the 26th ACM
  Conference on Innovation and Technology in Computer Science Education V. 1}.
  345--351.
\newblock


\bibitem{g48howtomentor}
{{Garage48}}. 2017.
\newblock How to mentor teams at a hackathon.
\newblock
\newblock
\showURL{%
\url{https://garage48.org/blog/how-to-mentor-teams-at-a-hackathon}}
\newblock
\shownote{[Online; accessed 03-March-2025].}


\bibitem{g48cybersec}
{{Garage48}}. 2019.
\newblock Garage48 Cyber Security 2019.
\newblock
\newblock
\showURL{%
\url{http://garage48.org/events/garage48-cyber-security-2019}}
\newblock
\shownote{[Online; accessed 03-March-2025].}


\bibitem{g48defence}
{{Garage48}}. 2020a.
\newblock Garage48 Defence Makeathon.
\newblock
\newblock
\showURL{%
\url{https://garage48.org/events/defence-makeathon-2020}}
\newblock
\shownote{[Online; accessed 03-March-2025].}


\bibitem{g48howitworks}
{{Garage48}}. 2020b.
\newblock How does it work?
\newblock
\newblock
\showURL{%
\url{http://garage48.org/how-it-works}}
\newblock
\shownote{[Online; accessed 03-March-2025].}


\bibitem{mlh:prizes2}
{{Gottfried, Jon}}. 2014.
\newblock Are Hackathon Prizes the Worst Thing Since Moldy Sliced Bread?
\newblock
\newblock
\showURL{%
\url{https://news.mlh.io/are-hackathon-prizes-the-worst-thing-since-moldy-sliced-bread-04-18-2014}}
\newblock
\shownote{[Online; accessed 03-March-2025].}


\bibitem{hackbeyondthecode2024}
{{{Hack beyond the code organizers}}}. 2024.
\newblock Hack Beyond the Code.
\newblock
\newblock
\showURL{%
\url{https://hackailiteracy.github.io/}}
\newblock
\shownote{[Online; accessed 03-March-2025].}


\bibitem{hackohio2023}
{{{HackOHI/O 11}}}. 2023.
\newblock Ohio State's Premier Hackathon.
\newblock
\newblock
\showURL{%
\url{https://hack.osu.edu/2023/}}
\newblock
\shownote{[Online; accessed 03-March-2025].}


\bibitem{brainstorming}
{{Heston, Klare}}. 2020.
\newblock How to Brainstorm.
\newblock
\newblock
\showURL{%
\url{https://www.wikihow.com/Brainstorm}}
\newblock
\shownote{[Online; accessed 03-March-2025].}


\bibitem{holmen2024facultyhack}
{John~K Holmen}, {Je'Aime Powell}, {Alexander Nolte}, {Elijah MacCarthy},
  {Charlie Dey}, {Ver{\'o}nica~G Vergara~Larrea}, {Suzanne Parete-Koon}, {and}
  {Linda Hayden}. 2024.
\newblock \showarticletitle{FacultyHack Events: Faculty-Focused Hackathons for
  High-Performance Computing Curriculum Development}. In {\em Proceedings of
  the 8th International Conference on Game Jams, Hackathons and Game Creation
  Events}. 67--71.
\newblock


\bibitem{huppenkothen2018hack}
{Daniela Huppenkothen}, {Anthony Arendt}, {David~W Hogg}, {Karthik Ram},
  {Jacob~T VanderPlas}, {and} {Ariel Rokem}. 2018.
\newblock \showarticletitle{Hack weeks as a model for data science education
  and collaboration}.
\newblock {\em Proceedings of the National Academy of Sciences\/} {115}, 36
  (2018), 8872--8877.
\newblock


\bibitem{imam2021secret}
{Ahmed Imam}, {Tapajit Dey}, {Alexander Nolte}, {Audris Mockus}, {and} {James~D
  Herbsleb}. 2021.
\newblock \showarticletitle{The Secret Life of Hackathon Code Where does it
  come from and where does it go?}. In {\em 2021 IEEE/ACM 18th International
  Conference on Mining Software Repositories (MSR)}. IEEE, 68--79.
\newblock


\bibitem{irani2015hackathons}
{Lilly Irani}. 2015.
\newblock \showarticletitle{Hackathons and the making of entrepreneurial
  citizenship}.
\newblock {\em Science, Technology, \& Human Values\/} {40}, 5 (2015),
  799--824.
\newblock


\bibitem{greatmentor}
{{Kerpen, Carrie}}. 2018.
\newblock If You Want To Be A Great Mentor Do These 5 Things.
\newblock
\newblock
\showURL{%
\url{https://www.forbes.com/sites/carriekerpen/2018/06/18/5-things-great-mentors-do/}}
\newblock
\shownote{[Online; accessed 03-March-2025].}


\bibitem{kraus2022coworking}
{Sascha Kraus}, {Ricarda~B Bouncken}, {Lars G{\"o}rmar}, {Maria~H
  Gonz{\'a}lez-Serrano}, {and} {Ferran Calabuig}. 2022.
\newblock \showarticletitle{Coworking spaces and makerspaces: Mapping the state
  of research}.
\newblock {\em Journal of Innovation \& Knowledge\/} {7}, 1 (2022), 100161.
\newblock


\bibitem{leemet2021utilizing}
{Alar Leemet}, {Fredrik Milani}, {and} {Alexander Nolte}. 2021.
\newblock \showarticletitle{Utilizing hackathons to foster sustainable product
  innovation-the case of a corporate hackathon series}. In {\em 2021 IEEE/ACM
  13th International Workshop on Cooperative and Human Aspects of Software
  Engineering (CHASE)}. IEEE, 51--60.
\newblock


\bibitem{liu2022understanding}
{Szu-Yu Liu}, {Brian~A Smith}, {Rajan Vaish}, {and} {Andr{\'e}s
  Monroy-Hern{\'a}ndez}. 2022.
\newblock \showarticletitle{Understanding the role of context in creating
  enjoyable co-located interactions}.
\newblock {\em Proceedings of the ACM on Human-Computer Interaction\/} {6},
  CSCW1 (2022), 1--26.
\newblock


\bibitem{mahmoud2022one}
{Ahmed Samir~Imam Mahmoud}, {Tapajit Dey}, {Alexander Nolte}, {Audris Mockus},
  {and} {James~D Herbsleb}. 2022.
\newblock \showarticletitle{One-off events? An empirical study of hackathon
  code creation and reuse}.
\newblock {\em Empirical software engineering\/} {27}, 7 (2022), 167.
\newblock


\bibitem{mahmoud2023exploratory}
{Ahmed Samir~Imam Mahmoud}, {Alexander Nolte}, {and} {Dietmar Pfahl}. 2023.
\newblock \showarticletitle{An Exploratory Study on the Evidence of Hackathons'
  Role in Solving OSS Newcomers' Challenges}.
\newblock {\em arXiv preprint arXiv:2305.09546\/} (2023).
\newblock


\bibitem{maillart2024computational}
{Thomas Maillart}, {Lucia Gomez}, {Ewa Lombard}, {Alexander Nolte}, {and}
  {Francesco Pisano}. 2024.
\newblock \showarticletitle{Computational diplomacy: how ‘hackathons for
  good’feed a participatory future for multilateralism in the digital age}.
\newblock {\em Philosophical Transactions A\/} {382}, 2285 (2024), 20240103.
\newblock


\bibitem{mlh:judges}
{{Major League Hacking}}. 2019a.
\newblock Judging \& Submissions.
\newblock
\newblock
\showURL{%
\url{https://guide.mlh.io/digital-hackathons/judging-and-submissions}}
\newblock
\shownote{[Online; accessed 03-March-2025].}


\bibitem{mlh:prizes1}
{{Major League Hacking}}. 2019b.
\newblock Ordering Swags \& Prizes.
\newblock
\newblock
\showURL{%
\url{https://guide.mlh.io/digital-hackathons/event-logistics/ordering-swags-and-prizes}}
\newblock
\shownote{[Online; accessed 03-March-2025].}


\bibitem{mlh:sponsorship}
{{Major League Hacking}}. 2020a.
\newblock Getting Sponsorship.
\newblock
\newblock
\showURL{%
\url{https://guide.mlh.io/digital-hackathons/getting-sponsorship}}
\newblock
\shownote{[Online; accessed 03-March-2025].}


\bibitem{mlh:promote}
{{Major League Hacking}}. 2020b.
\newblock How to Promote Your Event.
\newblock
\newblock
\showURL{%
\url{https://guide.mlh.io/digital-hackathons/marketing-your-event/how-to-promote-your-event}}
\newblock
\shownote{[Online; accessed 03-March-2025].}


\bibitem{medina2019does}
{Maria~Angelica Medina~Angarita} {and} {Alexander Nolte}. 2019.
\newblock \showarticletitle{Does it matter why we hack? - Exploring the impact
  of goal alignment in hackathons}. In {\em Proceedings of 17th European
  Conference on Computer-Supported Cooperative Work}. European Society for
  Socially Embedded Technologies (EUSSET).
\newblock


\bibitem{mendes2022socio}
{Wendy Mendes}, {Albert Richard}, {T{\"a}he-Kai Tillo}, {Gustavo Pinto}, {Kiev
  Gama}, {and} {Alexander Nolte}. 2022.
\newblock \showarticletitle{Socio-technical constraints and affordances of
  virtual collaboration-a study of four online hackathons}.
\newblock {\em Proceedings of the ACM on Human-Computer Interaction\/} {6},
  CSCW2 (2022), 1--32.
\newblock


\bibitem{oneweek}
{{Microsoft Garage}}. 2017.
\newblock oneweek hackathon 2017.
\newblock
\newblock
\showURL{%
\url{https://news.microsoft.com/life/one-week-microsoftlife/}}
\newblock
\shownote{[Online; accessed 03-March-2025].}


\bibitem{moller2014community}
{Steffen M{\"o}ller}, {Enis Afgan}, {Michael Banck}, {Raoul~JP Bonnal},
  {Timothy Booth}, {John Chilton}, {Peter~JA Cock}, {Markus Gumbel}, {Nomi
  Harris}, {Richard Holland}, {and} {others}. 2014.
\newblock \showarticletitle{Community-driven development for computational
  biology at Sprints, Hackathons and Codefests}.
\newblock {\em BMC bioinformatics\/} {15}, 14 (2014), S7.
\newblock


\bibitem{morrison2020challenges}
{Sarah Morrison-Smith} {and} {Jaime Ruiz}. 2020.
\newblock \showarticletitle{Challenges and barriers in virtual teams: a
  literature review}.
\newblock {\em SN Applied Sciences\/} {2}, 6 (2020), 1--33.
\newblock


\bibitem{nolte2019touched}
{Alexander Nolte}. 2019.
\newblock \showarticletitle{Touched by the Hackathon: a study on the connection
  between Hackathon participants and start-up founders}. In {\em Proceedings of
  the 2nd ACM SIGSOFT International Workshop on Software-Intensive Business:
  Start-ups, Platforms, and Ecosystems}. 31--36.
\newblock


\bibitem{nolte2020what}
{Alexander Nolte}, {Irene-Angelica Chounta}, {and} {James~D Herbsleb}. 2020.
\newblock \showarticletitle{What Happens to All These Hackathon Projects? -
  Identifying Factors to Promote Hackathon Project Continuation}.
\newblock {\em Proceedings of the ACM on Human-Computer Interaction\/} {4},
  CSCW2 (2020), 1--26.
\newblock


\bibitem{nolte2025survey}
{Alexander Nolte}, {Anna Filippova}, {and} {James~D. Herbsleb}. 2025.
\newblock A survey instrument for hackathon organizers and researchers.
\newblock
\newblock
\showDOI{%
\url{http://dx.doi.org/10.5281/zenodo.14705828}}


\bibitem{nolte2020support}
{Alexander Nolte}, {Linda~Bailey Hayden}, {and} {James~D Herbsleb}. 2020a.
\newblock \showarticletitle{How to Support Newcomers in Scientific Hackathons -
  An Action Research Study on Expert Mentoring}.
\newblock {\em Proceedings of the ACM on Human-Computer Interaction\/} {4},
  CSCW1 (2020), 1--23.
\newblock


\bibitem{nolte2020organize}
{Alexander Nolte}, {Ei~Pa~Pa Pe-Than}, {Abasi-amefon~Obot Affia}, {Chalalai
  Chaihirunkarn}, {Anna Filippova}, {Arun Kalyanasundaram}, {Maria
  Angelica~Medina Angarita}, {Erik Trainer}, {and} {James~D Herbsleb}. 2020b.
\newblock \showarticletitle{How to organize a hackathon--A planning kit}.
\newblock {\em arXiv preprint arXiv:2008.08025\/} (2020).
\newblock


\bibitem{nolte2018you}
{Alexander Nolte}, {Ei~Pa~Pa Pe-Than}, {Anna Filippova}, {Christian Bird},
  {Steve Scallen}, {and} {James~D Herbsleb}. 2018.
\newblock \showarticletitle{You Hacked and Now What? -Exploring Outcomes of a
  Corporate Hackathon}.
\newblock {\em Proceedings of the ACM on Human-Computer Interaction\/} {2},
  CSCW (2018), 1--23.
\newblock


\bibitem{openbio}
{{Open Bioinformatics foundation}}. 2018.
\newblock OBF hackathons.
\newblock
\newblock
\showURL{%
\url{https://www.open-bio.org/}}
\newblock
\shownote{[Online; accessed 03-March-2025].}


\bibitem{paganini2020engaging}
{Lavinia Paganini} {and} {Kiev Gama}. 2020.
\newblock \showarticletitle{Engaging women’s participation in hackathons: A
  qualitative study with participants of a female-focused hackathon}. In {\em
  Proceedings of the 5th International Conference on Game Jams, Hackathons and
  Game Creation Events}. 8--15.
\newblock


\bibitem{paganini2023opportunities}
{Lav{\'\i}nia Paganini}, {Kiev Gama}, {Alexander Nolte}, {and} {Alexander
  Serebrenik}. 2023.
\newblock \showarticletitle{Opportunities and constraints of women-focused
  online hackathons}. In {\em 2023 IEEE/ACM 4th Workshop on Gender Equity,
  Diversity, and Inclusion in Software Engineering (GEICSE)}. IEEE, 33--40.
\newblock


\bibitem{pe2019understanding}
{Ei~Pa~Pa Pe-Than} {and} {James~D Herbsleb}. 2019.
\newblock \showarticletitle{Understanding Hackathons for Science:
  Collaboration, Affordances, and Outcomes}. In {\em International Conference
  on Information}. Springer, 27--37.
\newblock


\bibitem{pe2022corporate}
{Ei~Pa~Pa Pe-Than}, {Alexander Nolte}, {Anna Filippova}, {Christian Bird},
  {Steve Scallen}, {and} {James Herbsleb}. 2022.
\newblock \showarticletitle{Corporate hackathons, how and why? A multiple case
  study of motivation, projects proposal and selection, goal setting,
  coordination, and outcomes}.
\newblock {\em Human--Computer Interaction\/} {37}, 4 (2022), 281--313.
\newblock


\bibitem{pe2019designing}
{Ei~Pa~Pa Pe-Than}, {Alexander Nolte}, {Anna Filippova}, {Christian Bird},
  {Steve Scallen}, {and} {James~D Herbsleb}. 2019.
\newblock \showarticletitle{Designing Corporate Hackathons With a Purpose: The
  Future of Software Development}.
\newblock {\em IEEE Software\/} {36}, 1 (2019), 15--22.
\newblock


\bibitem{icebreakergames}
{{Pietroluongo, Lindsay}}. 2023.
\newblock 11 Icebreaker Games for Work That Your Team Will Love.
\newblock
\newblock
\showURL{%
\url{https://www.elegantthemes.com/blog/business/great-icebreaker-games-your-employees-will-actually-enjoy}}
\newblock
\shownote{[Online; accessed 03-March-2025].}


\bibitem{porras2019code}
{Jari Porras}, {Antti Knutas}, {Jouni Ikonen}, {Ari Happonen}, {Jayden
  Khakurel}, {and} {Antti Herala}. 2019.
\newblock \showarticletitle{Code camps and hackathons in education-literature
  review and lessons learned}. In {\em Proceedings of the 52nd Hawaii
  International Conference on System Sciences}.
\newblock


\bibitem{porter2017reappropriating}
{Emily Porter}, {Chris Bopp}, {Elizabeth Gerber}, {and} {Amy Voida}. 2017.
\newblock \showarticletitle{Reappropriating Hackathons: The Production Work of
  the CHI4Good Day of Service}. In {\em Proceedings of the 2017 CHI Conference
  on Human Factors in Computing Systems}. ACM, 810--814.
\newblock


\bibitem{powell2021organizing}
{Jeaime Powell}, {Linda Bailey~Hayden}, {Amy Cannon}, {Boyd Wilson}, {and}
  {Alexander Nolte}. 2021.
\newblock \showarticletitle{Organizing online hackathons for newcomers to a
  scientific community--Lessons learned from two events}. In {\em Proceedings
  of the 6th Annual International Conference on Game Jams, Hackathons, and Game
  Creation Events}. 78--82.
\newblock


\bibitem{prado2020trans}
{Rafa Prado}, {Wendy Mendes}, {Kiev~S Gama}, {and} {Gustavo Pinto}. 2020.
\newblock \showarticletitle{How trans-inclusive are hackathons?}
\newblock {\em IEEE Software\/} {38}, 2 (2020), 26--31.
\newblock


\bibitem{perfectmentor}
{{Savisaari, Olli}}. 2017.
\newblock How to be the perfect mentor at a hackathon?
\newblock
\newblock
\showURL{%
\url{https://medium.com/@o.savisaari/709e0ab2d032}}
\newblock
\shownote{[Online; accessed 03-March-2025].}


\bibitem{schulten2024we}
{Cleo Schulten} {and} {Irene-Angelica Chounta}. 2024.
\newblock \showarticletitle{How do we learn in and from Hackathons? A
  systematic literature review}.
\newblock {\em Education and Information Technologies\/} (2024), 1--32.
\newblock


\bibitem{schulten2022participants}
{Cleo Schulten}, {Alexander Nolte}, {Daniel Spikol}, {and} {Irene-Angelica
  Chounta}. 2022.
\newblock \showarticletitle{How do participants collaborate during an online
  hackathon? An empirical, quantitative study of communication traces}.
\newblock {\em Frontiers in Computer Science\/}  {4} (2022), 983164.
\newblock


\bibitem{schulten2024hack}
{Cleo Schulten}, {Li Yuan}, {Kiev Gama}, {Wayne Holmes}, {Alexander Nolte},
  {Tore Hoel}, {and} {Irene-Angelica Chounta}. 2024.
\newblock \showarticletitle{Hack Beyond the Code: Building a Toolbox of
  Human-Centred Strategies for AI Literacy}. In {\em International Conference
  on Artificial Intelligence in Education}. Springer, 467--472.
\newblock


\bibitem{stakeholderanalysis}
{{Schuurman, Robbin}}. 2019.
\newblock Creating a Stakeholder Analysis: How do you do that?
\newblock
\newblock
\showURL{%
\url{https://bit.ly/33xZguf}}
\newblock
\shownote{[Online; accessed 03-March-2025].}


\bibitem{taylor2018everybody}
{Nick Taylor} {and} {Loraine Clarke}. 2018.
\newblock \showarticletitle{Everybody's Hacking: Participation and the
  Mainstreaming of Hackathons}. In {\em Proceedings of the 2018 CHI Conference
  on Human Factors in Computing Systems}. ACM, 172.
\newblock


\bibitem{hackplaybook}
{{Thackery, L.}} 2017.
\newblock Hackathon playbook.
\newblock
\newblock
\showURL{%
\url{https://hackathon-planning-kit.org/files/hackathon-playbook-external.pdf}}
\newblock
\shownote{[Online; accessed 03-March-2025].}


\bibitem{astrohackweek}
{{The AstroHackWeek organizers}}. 2020.
\newblock Astro hack week 2020.
\newblock
\newblock
\showURL{%
\url{http://astrohackweek.org/2020/}}
\newblock
\shownote{[Online; accessed 03-March-2025].}


\bibitem{chihacknight}
{{The Chi Hack Night organizers}}. 2020.
\newblock Chi Hack Night.
\newblock
\newblock
\showURL{%
\url{https://chihacknight.org/projects.html}}
\newblock
\shownote{[Online; accessed 03-March-2025].}


\bibitem{debcamp}
{{The DebCamp organizers}}. 2019.
\newblock DebCamp.
\newblock
\newblock
\showURL{%
\url{https://wiki.debian.org/DebCamp}}
\newblock
\shownote{[Online; accessed 03-March-2025].}


\bibitem{entrofy}
{{The entrofy organizers}}. 2017.
\newblock entrofy.
\newblock
\newblock
\showURL{%
\url{https://github.com/dhuppenkothen/entrofy}}
\newblock
\shownote{[Online; accessed 03-March-2025].}


\bibitem{geohackweek}
{{The Geohackweek organizers}}. 2019.
\newblock Geohackweek 2019.
\newblock
\newblock
\showURL{%
\url{https://geohackweek.github.io/}}
\newblock
\shownote{[Online; accessed 03-March-2025].}


\bibitem{globalhack}
{{The global hack organizers}}. 2020.
\newblock The global hack.
\newblock
\newblock
\showURL{%
\url{https://garage48.org/events/the-global-hack}}
\newblock
\shownote{[Online; accessed 12-August-2025].}


\bibitem{greenhack}
{{The Green Hackathon organizers}}. 2019.
\newblock Green Hackathon.
\newblock
\newblock
\showURL{%
\url{http://www.greenhackathon.com/}}
\newblock
\shownote{[Online; accessed 03-March-2025].}


\bibitem{hackhpc}
{{The HackHPC organizers}}. 2020a.
\newblock HackHPC.
\newblock
\newblock
\showURL{%
\url{http://hackhpc.org/}}
\newblock
\shownote{[Online; accessed 03-March-2025].}


\bibitem{hackhpc-codeofconduct}
{{The HackHPC organizers}}. 2020b.
\newblock HackHPC Code of Conduct.
\newblock
\newblock
\showURL{%
\url{http://hackhpc.org/codeofconduct/}}
\newblock
\shownote{[Online; accessed 03-March-2025].}


\bibitem{hpcinthecity}
{{The HackHPC organizers}}. 2020c.
\newblock HPC in the city.
\newblock
\newblock
\showURL{%
\url{https://hackhpc.github.io/HPCintheCity20/}}
\newblock
\shownote{[Online; accessed 03-March-2025].}


\bibitem{brainhack}
{{The OHBM Brainhack organizers}}. 2018.
\newblock OHBM Brainhack 2018.
\newblock
\newblock
\showURL{%
\url{https://ohbm.github.io/hackathon2018/}}
\newblock
\shownote{[Online; accessed 03-March-2025].}


\bibitem{sheinnovates}
{{The SheInnovates organizers}}. 2020.
\newblock SheInnovates.
\newblock
\newblock
\showURL{%
\url{http://sheinnovates.us/}}
\newblock
\shownote{[Online; accessed 03-March-2025].}


\bibitem{steelhacks}
{{The SteelHacks organizers}}. 2020.
\newblock SteelHacks.
\newblock
\newblock
\showURL{%
\url{http://steelhacks.com/}}
\newblock
\shownote{[Online; accessed 03-March-2025].}


\bibitem{worldofcode}
{{The world of code hackathon organizers}}. 2019a.
\newblock World of code hackathon.
\newblock
\newblock
\showURL{%
\url{https://github.com/woc-hack}}
\newblock
\shownote{[Online; accessed 03-March-2025].}


\bibitem{worldofcode:schedule}
{{The world of code hackathon organizers}}. 2019b.
\newblock World of code hackathon - schedule.
\newblock
\newblock
\showURL{%
\url{https://github.com/woc-hack/schedule}}
\newblock
\shownote{[Online; accessed 03-March-2025].}


\bibitem{worldofcode:tutorial}
{{The world of code hackathon organizers}}. 2019c.
\newblock World of code hackathon - tutorial.
\newblock
\newblock
\showURL{%
\url{https://github.com/woc-hack/tutorial}}
\newblock
\shownote{[Online; accessed 03-March-2025].}


\bibitem{trainer2014community}
{Erik~H Trainer}, {Chalalai Chaihirunkarn}, {Arun Kalyanasundaram}, {and}
  {James~D Herbsleb}. 2014.
\newblock \showarticletitle{Community code engagements: summer of code \&
  hackathons for community building in scientific software}. In {\em
  Proceedings of the 18th International Conference on Supporting Group Work}.
  ACM, 111--121.
\newblock


\bibitem{brainstorming:question}
{{Visser, Friso}}. 2018.
\newblock 7 tricks to build the perfect brainstorming question.
\newblock
\newblock
\showURL{%
\url{https://bit.ly/30zytw1}}
\newblock
\shownote{[Online; accessed 03-March-2025].}


\bibitem{nofun}
{{vj, satish}}. 2016.
\newblock Please don’t organize ‘fun’ activities at hackathons.
\newblock
\newblock
\showURL{%
\url{https://bit.ly/2EOdPj4}}
\newblock
\shownote{[Online; accessed 03-March-2025].}


\bibitem{zapico2013hacking}
{Jorge~Luis Zapico}, {Daniel Pargman}, {Hannes Ebner}, {and} {Elina Eriksson}.
  2013.
\newblock \showarticletitle{Hacking sustainability: Broadening participation
  through green hackathons}. In {\em Fourth International Symposium on End-User
  Development. June 10-13, 2013, IT University of Copenhagen, Denmark}.
\newblock


\end{thebibliography}

\end{document}